\documentclass{ieeeaccess}
\usepackage{cite}
\usepackage{amsmath,amssymb,amsfonts}
\usepackage{algorithmic}
\usepackage{graphicx}
\usepackage{textcomp}

\usepackage[utf8]{inputenc}

\usepackage[normalem]{ulem}
\useunder{\uline}{\ul}{}
\usepackage{diagbox}

\usepackage{array}
\newcolumntype{L}[1]{>{\raggedright\let\newline\\\arraybackslash\hspace{0pt}}m{#1}}
\newcolumntype{C}[1]{>{\centering\let\newline\\\arraybackslash\hspace{0pt}}m{#1}}
\newcolumntype{R}[1]{>{\raggedleft\let\newline\\\arraybackslash\hspace{0pt}}m{#1}}

\usepackage{enumitem}

\def\BibTeX{{\rm B\kern-.05em{\sc i\kern-.025em b}\kern-.08em
    T\kern-.1667em\lower.7ex\hbox{E}\kern-.125emX}}

\begin{document}
\history{© 2019 IEEE. Personal use of this material is permitted. Permission from IEEE must be obtained for all other uses, in any current or future media, including reprinting/republishing this material for advertising or promotional purposes, creating new collective works, for resale or redistribution to servers or lists, or reuse of any copyrighted component of this work in other works.}
\doi{10.1109/ACCESS.2019.DOI}

\title{A Survey on Data Plane Flexibility and Programmability in Software-Defined Networking}
\author{
	\uppercase{Enio~Kaljic},~\IEEEmembership{Member, IEEE},
	\uppercase{Almir~Maric},~\IEEEmembership{Member, IEEE},
	\uppercase{Pamela~Njemcevic}, and
	\uppercase{Mesud~Hadzialic$\dagger$},~\IEEEmembership{Member, IEEE}	
}

\address{Department of Telecommunications, Faculty of Electrical Engineering, University of Sarajevo, Bosnia and Herzegovina\\
$\dagger$ deceased on December 16, 2018.}

\tfootnote{This work was supported by the Government of Federation of Bosnia and Herzegovina, Federal Ministry of Education and Science under Grant 05-39-2565-1/18.}

\markboth
{Kaljic \headeretal: A Survey on Data Plane Flexibility and Programmability in Software-Defined Networking}
{Kaljic \headeretal: A Survey on Data Plane Flexibility and Programmability in Software-Defined Networking}

\corresp{Corresponding author: Enio Kaljic (enio.kaljic@etf.unsa.ba).}

\begin{abstract}
Software-defined networking (SDN) attracts the attention of the research community in recent years, as evidenced by a large number of survey and review papers. The architecture of SDN clearly recognizes three planes: application, control, and data plane. The application plane executes network applications; control plane regulates the rules for the entire network based on the requests generated by network applications; and based on the set rules, the controller configures the switches in the data plane. The role of the switch in the data plane is to simply forward packets based on the instructions given by the controller. By analyzing SDN-related research papers, it is observed that research, from the very beginning, is insufficiently focused on the data plane. Therefore, this paper gives a comprehensive overview of the data plane survey with particular emphasis on the problem of programmability and flexibility. The first part of the survey is dedicated to the evaluation of actual data plane architectures through several definitions and aspects of data plane flexibility and programmability. Then, an overview of SDN-related research was presented with the aim of identifying key factors influencing the gradual deviation from the original data plane architectures given with ForCES and OpenFlow specifications, which we called the data plane evolution. By establishing a correlation between the treated problem and the problem-solving approaches, the limitations of ForCES and OpenFlow data plane architectures were identified. Based on identified limitations, a generalization of approaches to addressing the problem of data plane flexibility and programmability is made. By examining generalized approaches, open issues have been identified, establishing the grounds for future research directions proposal.
\end{abstract}

\begin{keywords}
Data plane, 
data plane abstractions, 
data plane architectures, 
data plane flexibility, 
data plane implementations, 
data plane languages, 
data plane programmability, 
deeply programmable networks, 
description languages, 
energy consumption, 
energy efficiency, 
hardware-based implementations, 
measurement, 
monitoring, 
network virtualization, 
network functions virtualization, 
networking technologies, 
performance, 
programming languages, 
quality of service, 
reliability, 
security, 
software-based implementations, 
software-defined networking, 
stateful data plane
\end{keywords}

\titlepgskip=-15pt

\IEEEoverridecommandlockouts

\IEEEpubid{\makebox[\columnwidth]{\hfill 978-1-4799-4937-3/14/\$31.00~\copyright{}2014 IEEE} \hspace{\columnsep}\makebox[\columnwidth]{ }}

\maketitle

\section{Introduction}
\label{sec:introduction}
	\Figure[t!](topskip=0pt, botskip=0pt, midskip=0pt)[clip, trim=2cm 10.8cm 2cm 10.8cm, width=1.0\textwidth]{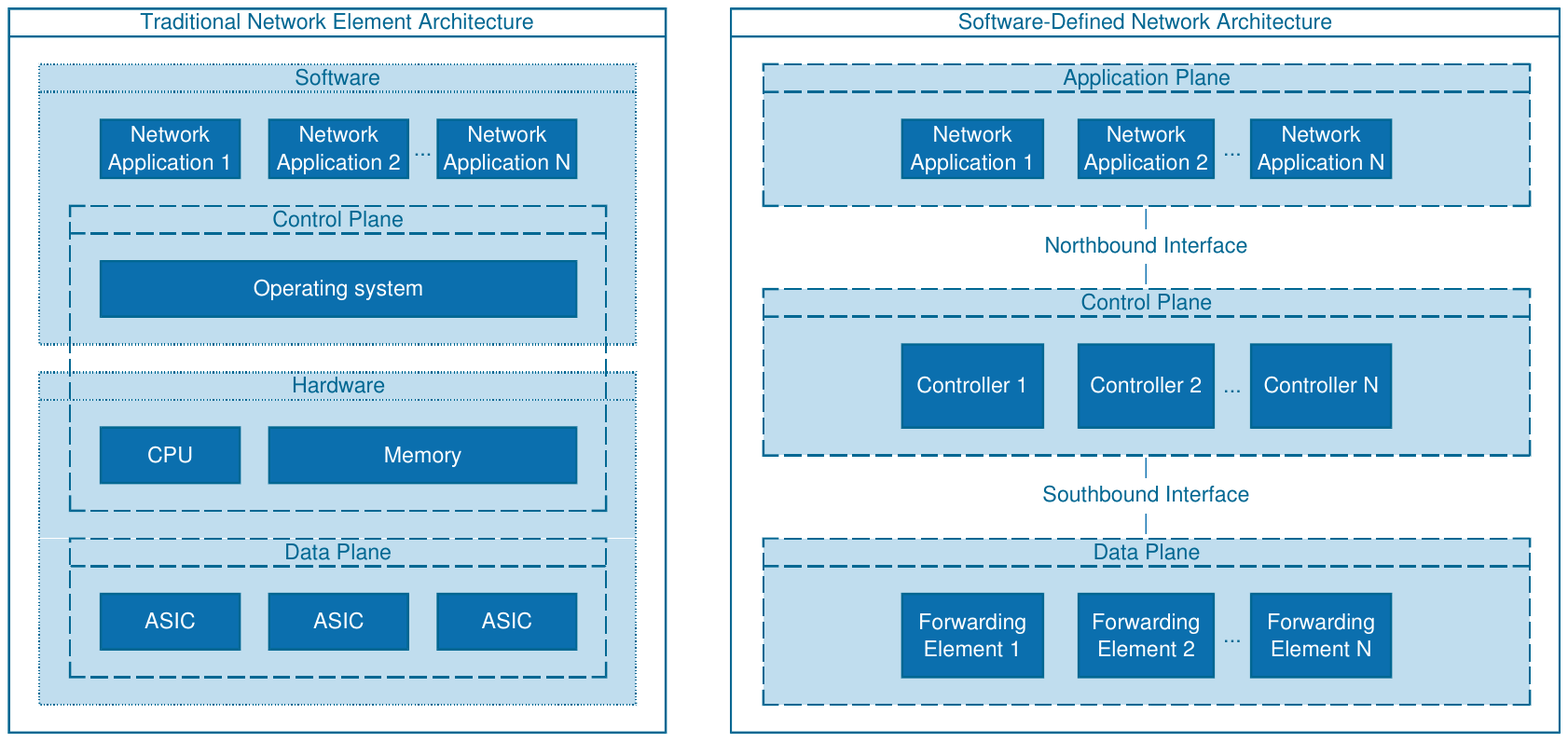}
	{Migration from traditional to software-defined network architecture\label{fig:sdn}}
	
	\PARstart{T}RADITIONAL communication networks are constructed from a large number of network devices that perform various tasks such as switching, routing, maintaining quality of service, monitoring and management, ensuring security and reliability, etc. To respond to these challenges, complex network algorithms and protocols are implemented on these network devices, which in the majority of cases are proprietary and implemented in the form of closed code. Maintenance and management of such networks is achieved by particular configuration of network devices through interfaces that vary from a vendor to a vendor. Standardization of network interfaces and protocols, aimed at unifying the network ecosystem and increasing the degree of interoperability, is a complex and time-consuming process. This approach has slowed down innovation and increased complexity and cost of maintenance and network management. In response to this problem, researchers in the 1990's \cite{Leslie1991,Merwe1998} have applied the analogy of relatively simple reprogramming of classical computers to computer networks, thus establishing the basis for the development of active networks \cite{Tennenhouse1997}. The active network concept is based on the fact that the packet carries the program instead of raw data (ie., smart packets \cite{Campbell1999}). Network devices, when they receive a smart packet, execute the program that it carries, and in accordance with the data plane design carry out different actions on the packet. In this way, network devices create an environment that responds to what the packet carries instead of passively transmitting packet payload from one node to another.
	
	In the early 2000s, the idea of network programmability, which comes from active networks, is articulated by separating the control and data plane, thus creating the concept of software-defined networking (SDN) \cite{Feamster2014}. Figure~\ref{fig:sdn} illustrates the transition from a traditional network architecture to a software-defined network. In the backbone of conventional network architecture, there is a networking device which performs all control and data plane tasks using a hard separable combination of software and hardware. On the other hand, in the SDN, the entire network intelligence is centralized in the application and control plane, where the application plane is composed of different network applications, and one or more controllers make the control plane. Network applications are performing routing algorithms, quality of service (QoS) mechanisms, control and network management mechanisms, etc., and are generating rules, according to which the network traffic should be treated. Generated rules are delivered to the control plane via a specially defined northbound interface. Based on these rules, controllers make specific forwarding rules and, according to them, configure packet switches via a southbound interface. In the end, network devices (routers and switches), in the data plane, perform a simple forwarding of packets based on a quick lookup of the forwarding tables.
	
	The standardization of the first SDN architecture was started through the Forwarding and Control Element Separation (ForCES) requirements specification by the IETF in 2003 in RFC3654 \cite{Khosravi2003} and a year later confirmed in RFC3746 \cite{Yang2004}. According to the ForCES specification given in RFC3746, the network element (NE) consists of several control elements (CE) in the control plane and the forwarding elements (FE) in the data plane. Since ForCES was not designed with a long-term vision to implement the SDN architecture, only with the emergence of OpenFlow \cite{McKeown2008}, the significance and usefulness of SDN architecture has arisen. OpenFlow is based on an Ethernet switch with an internal flow table and a standardized interface for adding and deleting records in the flow table. With understanding the need for standardization of communication interfaces and protocols, IETF expands the ForCES specification with RFC5810 \cite{Dong2010}. Although both ForCES and OpenFlow follow the same idea of the control and data plane separation, they are technically different from the aspects of architecture and forwarding model, as analyzed in \cite{Wang2012}.
	
	\Figure[t!](topskip=0pt, botskip=0pt, midskip=0pt)[clip, trim=0.5cm 10cm 0.5cm 10cm, width=1.0\textwidth]{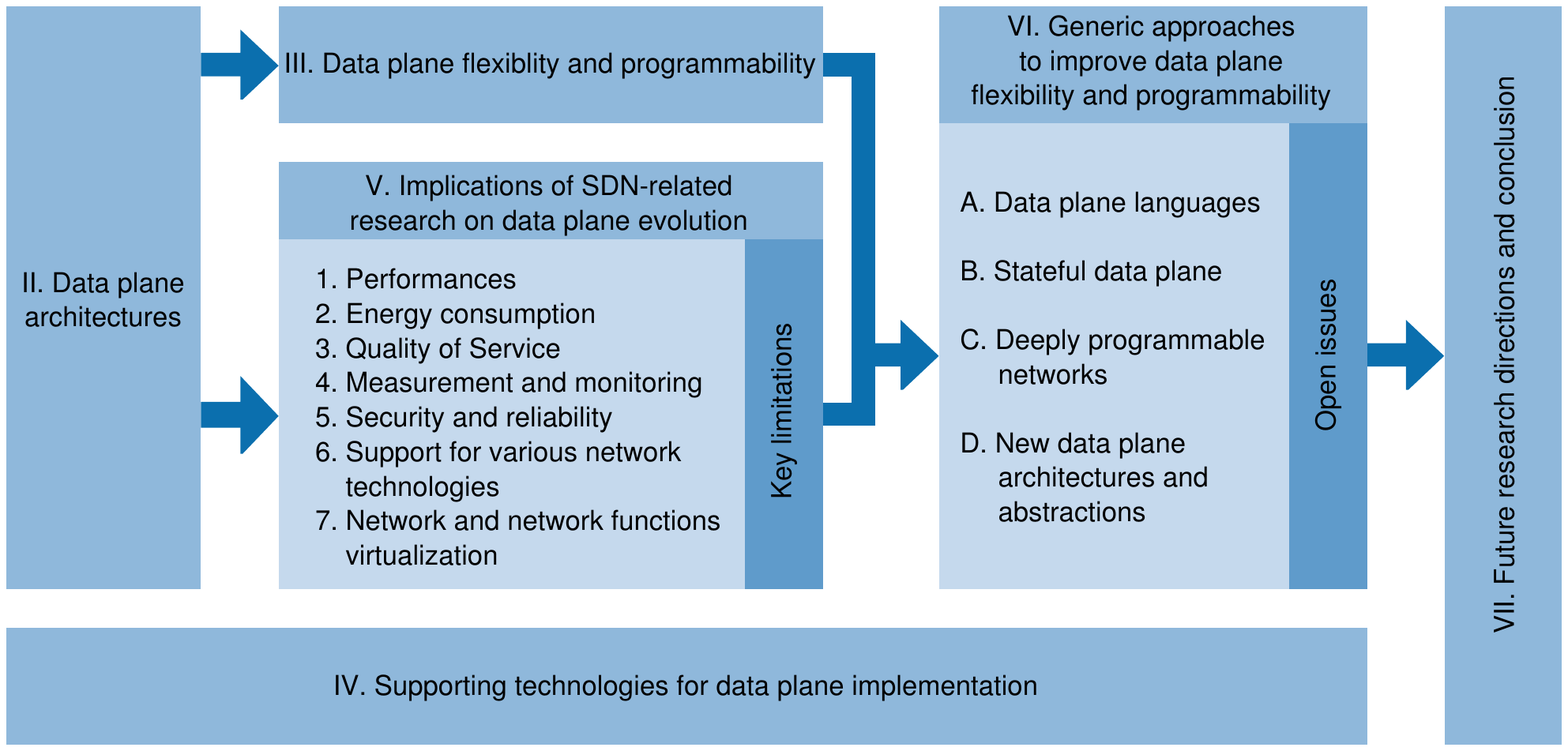}
	{Survey methodology\label{fig:methodology}}
		
	The development of SDN attracts the attention of the research community in recent years, as evidenced by the large number of review and survey papers. General overviews of SDN research are provided in \cite{Kreutz2015,Xia2015,Farhady2015,Masoudi2016,Cox2017}, while in a significant number of papers, targeted reviews were made by the issue addressed within the SDN. Thus, the development of traffic engineering in OpenFlow based networks is shown in \cite{Akyildiz2014}, while a review of new challenges in the development of SDN-based traffic engineering is given in \cite{Akyildiz2016}. Special attention was given to the review of QoS models and mechanisms in \cite{Karakus2017}. In \cite{Benzekki20161}, problems and solutions related to scalability, reliability, security, and performance of SDN were analyzed, while in \cite{Bolla2011,Tuysuz2017}, issues of energy efficiency and energy consumption in SDN were discussed. An overview of the research experimentation platforms is given in \cite{Anan2016}.
	
	By analyzing the survey papers mentioned above, it has been noted that the SDN research has been focusing on the control and application plane programmability from the very beginning. For the data plane, ever since the inception of the SDN idea, it was considered that it should follow two basic principles:
	\begin{enumerate}
		\item{simplicity that is seen in the process of packets forwarding in data plane, and}
		\item{the generality that is indicated in the independence of the SDN architecture from the technology through which the network is implemented.}
	\end{enumerate}
	Since the main challenges of SDN have been in the control plane, that led to the neglect of data plane development. Therefore, the focus of this paper is on the data plane, especially in term of its flexibility and programmability.
		
	Considering the different definitions of the data plane flexibility \cite{Kellerer2016,Kellerer2018,He2019} and programmability \cite{Campbell1999,Farhad2014,Nakao20151,Zilberman2015,Dargahi2017,Bifulco2018}, we advocate that flexibility means the possibility of a data plane to timely respond to new conditions in the network, and programmability as a method by which flexibility is achieved. Under new conditions in the network, we include changes in requirements, constraints, and data plane state.
	
	\begin{table}[h!t!]
		\centering
		\caption{List of abbreviations}
		\label{abbreviations}
		\begin{tabular}{|L{1.5cm}|L{6cm}|}
			\hline
			Abbreviation	& Full phrase \\ \hline
			ARP				& Address Resolution Protocol \\ \hline
			ATCA            & Advanced Telecommunications Computing Architecture \\ \hline
			BSV             & Bluespec System Verilog \\ \hline
			CE				& Control Element \\ \hline
			CPU             & Central Processing Unit \\ \hline
			DCN             & Data Center Network \\ \hline
			DDoS            & Distributed Denial-of-Service \\ \hline
			DDR             & Double Data Rate \\ \hline
			DOCSIS          & Data Over Cable Service Interface Specification \\ \hline
			DPDK            & Data Plane Development Kit \\ \hline
			DPI             & Deep Packet Inspection \\ \hline
			DPN             & Deeply Programmable Network \\ \hline
			DRAM            & Dynamic Random-Access Memory \\ \hline
			FE				& Forwarding Element \\ \hline
			FIFO            & First-In-First-Out \\ \hline
			ForCES			& Forwarding and Control Element Separation \\ \hline
			ForTER          & ForCES-based Router \\ \hline
			FP              & Forwarding Processor \\ \hline
			FPGA			& Field-Programmable Gate Array \\ \hline
			FSM             & Finite State Machine \\ \hline
			GPON            & Gigabit Passive Optical Network \\ \hline
			HAL             & Hardware Abstraction Layer \\ \hline
			IoT             & Internet of Things \\ \hline
			IP				& Internet Protocol \\ \hline
			I/O             & Input/Output \\ \hline
			LFB				& Logical Functional Block \\ \hline
			MAC             & Medium Access Control \\ \hline
			MCF             & Multi-Core Fiber \\ \hline
			MPLS            & MultiProtocol Label Switching \\ \hline
			NAT             & Network Address Translation \\ \hline
			NE				& Network Element \\ \hline
			NFV             & Network Function Virtualization \\ \hline
			NIC             & Network Interface Card \\ \hline
			OCS             & Optical Circuit Switching \\ \hline
			OPP             & Open Packet Processor \\ \hline
			OPS             & Optical Packet Switching \\ \hline
			OVS             & Open vSwitch \\ \hline
			PCI(e)          & Peripheral Component Interconnect (Express)\\ \hline
			PE              & Processing Element \\ \hline
			PHY             & Physical Layer \\ \hline
			POF             & Protocol-oblivious Forwarding \\ \hline
			ROADM           & Reconfigurable Optical Add-Drop Multiplexer \\ \hline
			QoS				& Quality of Service \\ \hline
			RISC            & Reduced Instruction Set Computer \\ \hline
			RLDRAM          & Reduced Latency DRAM \\ \hline
			RMT             & Reconfigurable Matching Table \\ \hline
			SFP             & Small Form-factor Pluggable \\ \hline
			SMF             & Single-Mode Fiber \\ \hline
			SoC				& System on a Chip \\ \hline
			SoFPGA          & System on FPGA \\ \hline
			SDN				& Software-Defined Network(ing) \\ \hline
			SRAM			& Static Random-Access Memory \\ \hline
			TCAM            & Ternary Content-Addressable Memory \\ \hline
			TCP				& Transport Control Protocol \\ \hline
			ToR             & Top of the Rack \\ \hline
			UDP				& User Datagram Protocol \\ \hline
			VM              & Virtual Machine \\ \hline
			VNF             & Virtual Network Function \\ \hline
			XFSM            & eXtended FSM \\ \hline
		\end{tabular}
	\end{table}
	
	Although, specific issues of data plane flexibility and programmability were addressed in above-mentioned papers, it is important to note that there is no comprehensive survey of data plane research from the aspect of its flexibility and programmability. It is also important to emphasize that although there are review papers that dealt with software-defined wireless networks \cite{Cho2014,Macedo2015,Sun2015}, data plane of the wireless network is out of the scope of this paper because it is being implemented using the Software-Defined Radio (SDR) techniques.
	
	Consequently, the aim of this paper is a survey of data plane research in a wired SDN, which appropriately addresses the problems of programmability and flexibility, and establishes preconditions for the advancement of its development through a proposal of future research directions.
	
	To accomplish this goal, several tasks have been carried out according to the methodology presented in Figure~\ref{fig:methodology}, which are at the same time the outline of this paper.
	At first, an overview of the data plane architecture in ForCES and the OpenFlow-based SDN, reflecting on the historical context of development and the differences between these two models is given in Section~\ref{sec:architectures}. Afterwards, in Section~\ref{sec:flexibility} is given an overview of the definitions of network flexibility and programmability and some general considerations of flexibility in other domains. Then, a review of the constraints of ForCES and OpenFlow-based data plane architectures, through the considered definitions and aspects of flexibility and programmability, is presented. Given that a lot of the data plane research, discussed in Sections \ref{sec:architectures}, \ref{sec:implications}, and \ref{sec:programmability}, is established on the experimental evaluation, in Section~\ref{sec:implementations} is given an overview of hardware- and software-based technologies which served as good support for data plane implementation. In Section~\ref{sec:implications} is given an overview of SDN-related research whose results have implied a data plane evolution. Under the data plane evolution, we indicate a gradual deviation from the original data plane architectures given with ForCES and OpenFlow specifications, resulting in the need to address the problem of programmability and the flexibility of the data plane in a more generic way. By reviewing the research which had addressed different problems in seven categories, as shown in Figure~\ref{fig:methodology}, we observed several common problem-solving approaches. Then, by establishing the correlation between treated problems and problem-solving approaches, we identified the key limitations of ForCES and OpenFlow data plane architectures. Based on identified key limitations in Section~\ref{sec:implications} and discussed aspects of flexiblity and programmability in Section~\ref{sec:flexibility}, we generalized approaches to improving the data plane flexibility through four methods for improvement of the data plane programmability. Based on critical review of generic approaches to improve data plane flexibility and related open issues, future research directions are proposed in Section~\ref{sec:futurework}.
	
	Table~\ref{abbreviations} shows the list of abbreviations in alphabetical order which are used more than once throughout the paper or outside the same paragraph.
	
\section{Data Plane Architectures}
\label{sec:architectures}
	An overview of the data plane architecture in ForCES and the OpenFlow based SDN is given in this section. In addition, some architectures inspired by ForCES have been reviewed, and at the end of the section, a review of the differences between the ForCES and OpenFlow data plane architectures is provided.
	
	The first proposal of the data plane architecture was specified by RFC5812 \cite{Halpern2010}, according to which the resources within ForCES FE are represented by logical functional blocks (LFBs), as illustrated in Figure~\ref{fig:forces}. LFB is a logical representation of a single packet processing functionality. The data paths through which the packets pass are formed by the interconnection of multiple LFBs, and they enable complex tasks execution during packet processing.
	Definitions and implementations of 22 LFBs in accordance with ForCES specification are presented in \cite{Dong2007}. In addition to the definition of LFBs, the definition of eight frame types, 43 data types, and 14 metadata types are given. Metadata are associated with packets when traveling between LFBs within the network element. The proposed definitions covered basic Internet protocol version 4 (IPv4) packet processing functionality, while the definitions of advanced elements required to support QoS functionality or firewall functionality were beyond the scope of the above-mentioned paper. 
	The design and implementation of the ForCES protocol are presented in \cite{Ramirez2017}. The introduced implementation enables the creation of new LFB test topologies within the network element structure, which makes it a useful tool in the research of the ForCES data plane. 
	The LFB chain composition method, which allows the LFB series to combine in the LFB chain according to the application, is presented in \cite{Rong20161}. LFB chaining method is divided into three types: (1) sequential chaining, (2) chaining of branches and (3) a hybrid chaining.
	In the mentioned work, a method for sequential chaining was proposed, and a work frame for that method was implemented. The framework consists of an LFB chain matching agent, which generates an appropriate LFB chain based on the incoming request. Agent performs the matching process in three steps: (1) mapping requests to a set of LFBs that can respond to a given request, (2) combining LFBs into one or more chains, and (3) selection of the best chain from a set of chains.
	An overview of research in the field of development and application of the ForCES specification is given in \cite{Haleplidis20152}. It has been noted that ForCES provides a significant support in various areas where distributed packet processing on network elements is required. The possibility of realization of custom-defined LFBs makes the ForCES model very flexible and powerful. Also, the vision of using ForCES in the unification of all network technologies, such as optical networks, wireless networks and Internet of Things (IoT), is presented. Although ForCES was presented as a promising model, it was concluded that its application is not at a satisfactory level due to several factors of which the most significant are: (1) lack of model adoption in the network industry, and (2) little use in academic environment due to lack of support for experimental work in the form of usable open source code.
	
	\Figure[t!](topskip=0pt, botskip=0pt, midskip=0pt)[clip, trim=3.2cm 11.4cm 3.2cm 11.4cm, width=0.99\columnwidth]{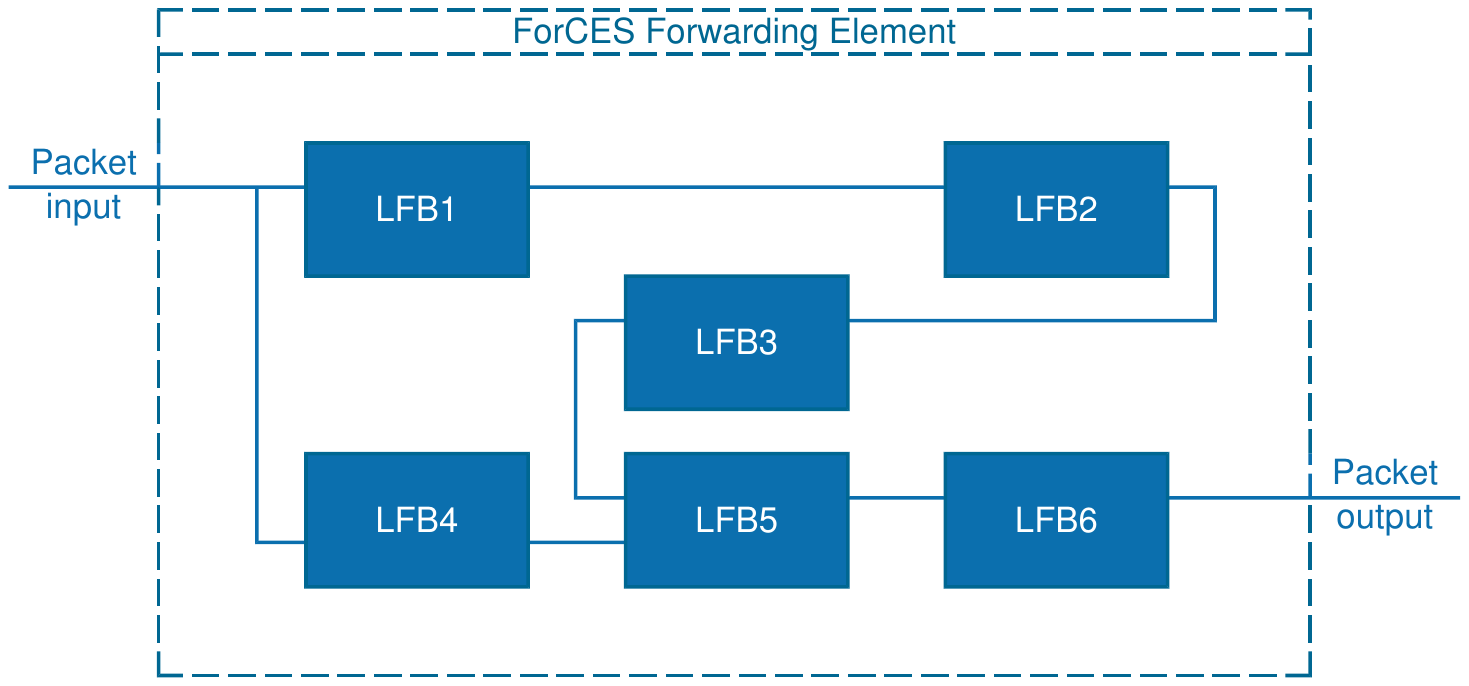}
	{ForCES Forwarding Element architecture\label{fig:forces}}
	
	\Figure[t!](topskip=0pt, botskip=0pt, midskip=0pt)[clip, trim=3.2cm 12.8cm 3.2cm 12.8cm, width=0.99\columnwidth]{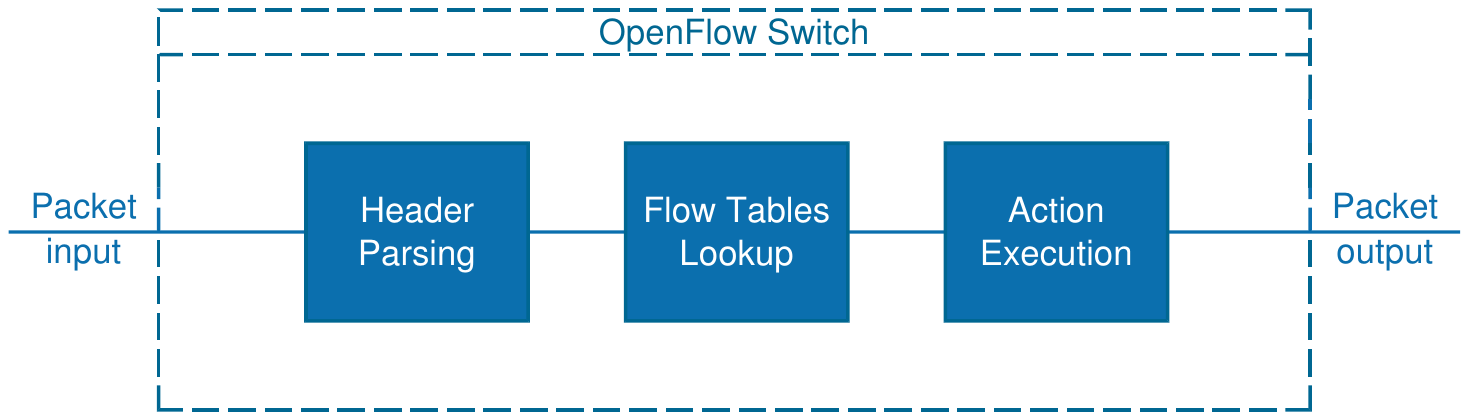}
	{OpenFlow switch architecture\label{fig:openflow}}
	
	Inspired by ForCES, two architectures based on the idea of separating the control and data plane, NEon and Ethane, are presented in \cite{Schuba2005,Casado2007}. NEon \cite{Schuba2005}, in accordance with the SDN principles, consists of two planes:	(1) control plane policy manager, and (2) programmable rule enforcement device (PRED). PRED has been realized as a high-performance programmable machine for packets classification and actions execution. It consists of logical functional blocks aggregation that enables the implementation of different network services. Logical functional blocks are providing packet processing functionalities such as flow identification, packet classification, and action processing. Action processing was achieved by using programmable dispatch machines that allow packet data manipulation.
	Ethane \cite{Casado2007} is an enterprise network architecture, consisting of: (1) a centralized controller that defines network policies for all packets, and (2) a group of simple Ethane switches. The Ethane switch contains a secure channel to the controller and a simple flow table. Packets arriving at the Ethane switch are forwarded based on the records in the flow table. In the absence of an appropriate record, the packet is forwarded to the controller together with information about where the packet came from. In this case, the controller has the task of defining the forwarding rule for that packet and updating the flow table on the switch. The records from the Ethane flow table contain: (1) header according to which the matching is performed with the headers of incoming packets, (2) action that tells the switch to what to do with the packet, and (3) additional data related to the flow (various counters). Header fields cover transport control protocol (TCP), user datagram protocol (UDP), IP and Ethernet protocol, and physical port information. Supported actions are: (1) forwarding the packets to the corresponding interface, (2) update of byte and packet counters, and (3) setting the flows activity bits. Beside listed, additional actions are possible such as placing packets in different queues or changing packet headers. In Ethane network architecture, all switches do not have to be Ethane switches, enabling a gradual migration from classic networks to Ethane-based networks.
	
	On the other hand, the OpenFlow data plane \cite{McKeown2008} consists of fixed architecture switches made up of: (1) flow table containing flow records with associated actions, (2) a secure channel to the OpenFlow controller, and (3) OpenFlow protocol that provides standardized communication between switches and controllers. OpenFlow switch working principle, illustrated in Figure~\ref{fig:openflow}, is based on simple forwarding of the packet between the ports based on the records in the flow table defined by the remote control process, i.e. the controller. Each record in the flow table contains three fields: (1) the packet header which defines the flow, (2) an action that specifies how packets are processed, and (3) statistics related to the flow (e.g., packet and byte count). OpenFlow switch has to support at least the following three actions: (1) forwarding a packet to a specific port or multiple ports, (2) packet encapsulation and forwarding to the controller, and (3) packet discarding. The encapsulation of the packet and sending to the controller is performed only for the first packet of the new flow, i.e., when for the incoming packet there are no records present in the flow table. Then, the controller generates the forwarding rule for the new flow and delivers it to the switch, allowing the switch to autonomously handle all subsequent packets from that flow.
	
	An overview and the analysis of the differences between ForCES and OpenFlow regarding architecture and forwarding model are given in \cite{Wang2012}. ForCES architecture implies separating the control and data plane within a single network element. This is achieved on two levels: (1) the first level implies communication separation in the sense that CE and FE are using a standard protocol instead of proprietary interfaces, and (2) physical separation that allows CEs and FEs to be executed in physically separated devices, and together form a network element. Interfaces between two ForCES NEs are the same as between standard switches and routers, so ForCES NEs can be transparently connected to existing conventional network elements. The ForCES control functions are still based on the use of distributed protocols. 
	However, OpenFlow separates the control and data planes in such a way that the data plane consists of more simple switches and the control plane of the whole network is made up from one centralized controller. The OpenFlow architecture supports two types of switches:
	\begin{enumerate}
		\item{"pure" OpenFlow switch - contains only a data plane based on flow tables, and provides an interface to a logically centralized controller. Logically centralized controller performs all control tasks such as: (1) collecting data on network operation and making decisions according to management logic, (2) installing rules on switches' flow tables through the OpenFlow protocol, and (3) providing an open application programming interface (API) to user applications.}
		\item{hybrid switch - also supports autonomous operation as a conventional Ethernet switch (eg., in the absence of OpenFlow controller).}
	\end{enumerate}
	The forwarding model in the ForCES data plane is based on packet processing through the LFB composition described by the directed graph. Each LFB defines a straightforward operation which is performed on the packet that passes through it. Typical examples of LFBs are: IPv4 Longest Prefix Matching, address resolution protocol (ARP), Internet control message protocol (ICMP), Queue, Scheduler, etc. However, with OpenFlow, the forwarding model is based on the flow tables manipulations. OpenFlow switches handle packets with flow granularity. Therefore, some standard network functionality that is run on the packet level (e.g., ARP) is very difficult to implement with OpenFlow. On the other hand, using the ForCES architecture, it is possible to define LFBs whose work principle is similar to OpenFlow flow tables, which confirms the flexibility of ForCES forwarding model.	
	Thus, the possibility of extending OpenFlow architecture with ForCES concepts is explored in \cite{Haleplidis2012}. By comparing the data plane of the OpenFlow and ForCES architectures, it has been observed that ForCES elements can describe some aspects of OpenFlow: (1) suitable LFB components can describe packet header fields lookup, counters, and actions, (2) unique attributes of LFBs can describe the set of supported actions and the mode of their execution, and (3) directed graph of LFBs can describe OpenFlow pipeline. In regard to the observation, the middleware based on the ForCES wrapper around the OpenFlow switch, which would allow switch control via the ForCES control element or OpenFlow controller, was proposed.

\section{Data plane flexibility and programmability}
\label{sec:flexibility}
	An overview of network flexibility definitions and general considerations of flexibility in other domains, which can be useful in valuing data plane flexibility, is given in this section. Then, an overview of the definition of programmability and the connection between the programmability and the flexibility of the data plane is presented. Finally, a review of the limitations of ForCES and OpenFlow based data plane architectures, from the perspective of the described definitions of flexibility and programmability, is given.
	
	\subsection{Definition of flexibility}
	Since there is no common approach to the definition of network or data plane flexibility, various definitions have been used and proposed in many papers. While some observed flexibility through the structure and design of the system \cite{Kellerer2016,Kellerer2018,He2019,Broniatowski2016,Nilchiani2007,Broniatowski2017}, others tied the definition of flexibility to the resilience of the system \cite{Omer20092,Zobel2011,Filippini2014}.

	\subsubsection{Flexibility in SDN}
	Although the benefit of SDN paradigm is in the development flexibility of new control logic, in \cite{Zilberman2015} is emphasized the importance of reconfiguration flexibility which allows the addition of support for new protocols and flexibility of the data plane structure. According to papers \cite{Kellerer2016} and \cite{Kellerer2018} which have dealt with the flexibility of softwarized networks, which include SDN, flexibility is defined as the ability of the network to adapt its resources, such as flows and topology, to changes in the requirements placed to the network. Adaptation implies changing the network topology, configuration and position of network functions. Given that there is no generally accepted definition of flexibility, the following aspects of flexibility are proposed in \cite{Kellerer2016}:
	\begin{itemize}
		\item{flow steering,}
		\item{function placement,}
		\item{function scaling,}
		\item{function operation.}
	\end{itemize}
	From the aspect of flow steering, an element, which supports both packet forwarding and copying, is more flexible regarding an element which supports only packet forwarding. Flexibility from the aspect of function placement reflects in the ability to dynamically change the position of functions during network operation, while the granularity of the resource allocation between the functions affects the aspect of scalability. The configurability and programmability of the network element functionalities are covered through the aspect of the function operation. Additionally, an aspect of topology adaptation, in term of adding or deleting links and nodes, is introduced in \cite{Kellerer2018}.
	
	Network flexibility was also a subject of research presented in \cite{He2019}, where it is defined as "timely support of changes in the network requirements." Under changes in network requirements, they include traffic variation, user mobility, network lease, network upgrades, and failure mitigation. In the aforementioned papers, they have considered two cases in which the network can support these requirements: 
	\begin{enumerate}
		\item{a network design which allows meeting the requirements without adaptation, and}
		\item{an adaptation of topology, flows, resources, and functions.}
	\end{enumerate}
	In the second case, adaptation should be carried out within the given time constraints.
	
	\subsubsection{Flexilibity in other domains}
	Flexibility can be observed through the inclusion of different options or design alternatives in the system structure, which was the basic idea of \cite{Broniatowski2016}. According to the proposed approach, the system can be designed using nodes and links, where each node represents the decision in the design process, and the link indicates logical or temporal dependence. According to cited paper, the system flexibility manifests in the total number of paths from the source node to the destination node, where each path represents a sequence of decisions leading to the fulfillment of the set requirement. Similarly, in \cite{Nilchiani2007} flexibility is defined via the ratio of the number of different paths and the total number of nodes.

	The impact of layer abstraction and modular decomposition on the flexibility of the system is explored in \cite{Broniatowski2017}. The analysis was carried out for four design strategies:
	\begin{enumerate}
		\item{integral system design in which one form shares functions, including unused functions,}
		\item{modular system design in which multiple forms are mapped to multiple forms by a one-to-one principle,}
		\item{layered system design in which all functions are included even not used (latent functions), and}
		\item{synergistic system design which includes latent functions but also allows adding new ones as needed.}
	\end{enumerate}
	Using the simulation techniques in \cite{Broniatowski2017} has been shown that both system design strategies contribute to the flexibility of the system, and their contribution is additive, making the synergistic system design the most flexible.
	
	In addition to the flexibility definitions which are related to the structure and design of the system, in \cite{Omer20092,Zobel2011,Filippini2014} flexibility is observed through resilience, defined as the ability of the system to recover quickly from external or internal disruptions and return to the state of equilibrium. In this context, the disruption can be modeled as a new requirement set to the system, and the ability of the system to respond to new requirements and continue with the correct functioning as a system resilience.
	
	\subsection{Definition of programmability}
	Data plane programmability has been in the focus of research presented in \cite{Campbell1999,Farhad2014,Nakao20151,Zilberman2015,Dargahi2017,Bifulco2018}. Programmability has been observed in \cite{Campbell1999} as a significant characteristic of the network through the level of programmability indicating the method, the granularity and the time scale in which new functionalities can be introduced into the network infrastructure. 
	
	In \cite{Farhad2014} and \cite{Nakao20151} researchers advocate that data plane programmability reflects in its depth and the way of its implementation. The depth of programmability they see through management capabilities of processes below the level of packet forwarding, which includes caching, transcoding, support for new protocols, and so on. Regarding the method of data plane implementation, a data plane which is implemented entirely in the software is programmable, and one implemented in hardware is non-programmable. Contrary to this, in \cite{Zilberman2015} researchers believe that the data plane programmability can be achieved with the use of reconfigurable hardware and convenient programming languages. Regardless of the implementation method, in \cite{Dargahi2017} it is deemed that data plane programmability can be seen in stateful flow processing. 
	
	A comprehensive definition of data plane programmability is given in \cite{Bifulco2018}, according to which programmability implies the switch capability to expose the packet processing logic to the control plane to support systematic, fast and comprehensive reconfiguration.
	
	From the considered definitions of flexibility and programmability, we see data plane programmability as a key factor in achieving flexibility from the aspects of adaptation of topology, flows, functions and resources.
	
	\subsection{Flexibility and programmability of ForCES and OpenFlow}
	According to the specification given in RFC5812 \cite{Halpern2010}, the data plane of ForCES consists of FEs presented by interconnected LFBs. Therefore, with the adequate support for the programmability of individual LFBs and their arrangement into arbitrary functional topologies, the data plane of ForCES can be viewed as highly flexible from the aspect of adaptation of functions.
	
	On the other hand, the data plane of OpenFlow is fixed pipeline structure whose forwarding model is based on the lookup of flow tables and execution of associated actions. Programmability of the data plane of OpenFlow is limited to the level of table flow content, which restricts its flexibility solely to the aspect of flows adaptation.
	
	Since neither ForCES nor OpenFlow based data plane architectures are flexible enough in term of considered aspects, a significant number of research has gone in the direction of data plane evolution to adequately respond to various functional requirements. Section~\ref{sec:implications} is dedicated to the review of that research with the aim of identifying the key limitations of ForCES and OpenFlow based data plane architecture in term of flexibility and programmability.

\section{Supporting technologies for data plane implementations}
\label{sec:implementations}
	Given that a lot of research, discussed in Sections \ref{sec:implications} and \ref{sec:programmability}, is based on experimental verification of proposed data plane architectures, this section presents an overview of hardware- and software-based technologies which served as a good platform for their implementations. At first, hardware architectures were used to implement packet switching nodes in the SDN, but slightly later software architectures were also used due to processing power limitations of conventional computer systems no longer being an issue. The processing power of today's computer systems has reached a significant level which enables the implementation of packet switching nodes whose performance is comparable to hardware-based implementations, placing the software-based implementations in an equally prominent position.
	
\subsection{Hardware-based implementations}
	By reviewing the research that dealt with hardware-based implementation of packet switching nodes and its application in SDN's data plane, we perceived the following categories:
	\begin{enumerate}
		\item{field-programmable gate array (FPGA) based implementations,}
		\item{system on a chip (SoC) based implementations,}
		\item{network processor (NP) based implementations,}
	\end{enumerate}

	\subsubsection{FPGA-based implementations}
	
	\textit{\textbf{NetFPGA}} --- the first FPGA-based platform, specially designed to teach network equipment development, is presented in \cite{Casado20051,Casado20052}. The first version of the NetFPGA board contains three Altera EP20K400 APEX devices, three 1MByte static random-access memory (SRAM) chips and 8-port Ethernet controller. One of three FPGA chips, called Control FPGA (CFPGA), is pre-programmed and connects two user FPGA chips (UFPGA) to an Ethernet controller. All communication on the board takes place via Ethernet ports. Although the board does not have a central processing unit (CPU), its operation is possible thanks to the virtual network system, the software executed on the computer where the card is embedded. The software can access the hardware registers, using a dedicated Ethernet frame with the CFPGA being responsible for its decoding and execution.
	
	The development of new versions of the NetFPGA board proceeded because of the limitations of the first version such as: impractical size of the board which can be only fitted into specially designed computer chassis, low speed - the first version had eight 10Mbps Ethernet ports, and lack of processor. Thus, in \cite{Watson2006,Lockwood2007}, \mbox{NetFPGA-v2} and \mbox{NetFPGA-v2.1} boards are presented. NetFPGA-v2 is made as a 32-bit full-length peripheral component interconnect (PCI) board running at 33MHz. The board is equipped with a Xilinx Spartan chip through which PCI communication is supported and the Xilinx V2P30 chip to which the user design is programmed. The UFPGA has two 512Kx36 SRAMs and is connected to the Marvell Quad 10/100/1000 Ethernet physical layer (PHY) chip through the standard gigabit media-independent interface (GMII). \mbox{NetFPGA-v2.1} brings two additional DDR3 synchronous dynamic random-access memory (SDRAM) chips that work asynchronously with the UFPGA chip. The standard NetFPGA library contains a skeleton of Verilog design that instantiates four Gigabit Ethernet Media Access Controllers (GMAC) and interfaces to SRAM and DDR2 memory. User design is implemented as a pipeline following the standard request-grant first-in-first-out (FIFO) protocol. The pipeline consists of input modules, user filter, and the output module. Input modules are connected to four Gigabit Ethernet network interfaces and host processors via a PCI interface. The user filter performs tasks such as decapsulation, decryption, and other user-defined functions. The output module performs an output port lookup to determine the port to which the packet must be forwarded. For example, the Ethernet switch or IP router logic are mainly implemented in the output module of the pipeline.
	
	The first application of FPGA technology in the SDN's data plane is the Ethane switch implementation \cite{Luo2007}. The data plane of the switch is implemented on the NetFPGA-1G board as a pipeline with two exact match flow tables, one for the packets to be forwarded and the other for the packets to be discarded. Packets that do not match the records in flow tables are forwarded to the software responsible for maintaining the flow table through record addition and deletion.
	
	Considering the growing need for fast prototyping platforms in the forthcoming period, a 40Gbps PCI Express card with a Xilinx Virtex-5 chip, called NetFPGA-10G, is presented in \cite{Blott2010}. NetFPGA-10G has four 10Gbps Ethernet interfaces in SFP+ form that are connected via additional PHY transceivers to the FPGA, and RLDRAMII and QDRII memory controllers for additional SRAM and DRAM memory. The Open Component Portability Infrastructure (OpenCPI) interface is used to connect the NetFPGA-10G card via PCIe to a computer. The AMBA4 AXI-Stream interface is used for packet transmission within the reference design and the AMBA4 AXI-Lite interface for signaling.
	
	NetFPGA-1G-CML and NetFPGA-SUME, featured in \cite{Zilberman2014,Cao2015}, are the successors of NetFPGA-1G and NetFPGA-10G platforms based on the 7th generation of Xilinx FPGAs.
	NetFPGA-1G-CML board is based on the Xilinx Kintex-7 FPGA. Improvements compared to NetFPGA-1G are reflected in:
	\begin{itemize}
		\item{three times more of FPGA logical elements,}
		\item{four times more of Block RAM capacity,}
		\item{512 MB DDR3 instead of 64 MB DDR2,}
		\item{additional 4.5 MB QDRII+ memory,}
		\item{4x Gen. 2 PCIe instead of PCI.}
	\end{itemize}
	It is compatible with Stanford's 10G architecture design, which allows relatively easy portability of designs from NetFPGA-10G platforms. NetFPGA-SUME is based on the Xilinx Virtex-7 960T FPGA chip containing 30 serial 13.1Gbps transceivers through which the FPGA chip is connected to four 10Gbps SFP+ Ethernet interfaces, PCIe, and two expansion connectors through which multiple NetFPGA-SUME boards can be interconnected. Therefore, no additional physical layer controller is required for the implementation of 10G Ethernet applications, which is the most significant improvement compared to the NetFPGA-10G board. Through the expansion connector, it is possible to implement a 100Gbps switch, and by the interconnection of multiple cards, it is possible to produce a 300Gbps non-blocking switch, which makes this platform suitable for the research of high-throughput networks.
	
	The multi-purpose highly-programmable network platform based on FPGA technology, called C-GEP, is presented in \cite{Varga2015}. C-GEP enables flexible implementation of network nodes that support different application types such as SDN switches, media gateways, traffic generators, deep packet inspection (DPI), etc. All of these applications are possible over 1, 10, 40 and 100 Gbps traffic. The Virtex-6 FPGA chip on the C-GEP motherboard performs packet forwarding tasks, while the embedded COM Express PC is responsible for management and SDN control functionality implementation. Installation of the appropriate firmware defines the architecture of the network device implemented on C-GEP.
	
	\textit{\textbf{OpenFlow switch implementation on NetFPGA}} --- which can store more than 32,000 flow records and performs at the speed of four NetFPGA 1G ports, is described in \cite{Naous20081}. The switch is made of software and hardware components. The software component of the switch is from the user space responsible for communication to the OpenFlow controller, and from the kernel space to the table flow maintenance, packet processing, and statistics update. The hardware component of the OpenFlow switches implements a different output port lookup as compared to the reference router and uses dynamic random-access memory (DRAM) for output queues.
	
	A good foundation for the future implementations of SDN's data plane using the NetFPGA platform was provided by the reference architecture of the packet switching node, presented in \cite{Gibb2008,Naous20082}. The idea is based on the fact that network hardware is generally implemented as a pipeline through which packets flow and are processed in different stages of the pipeline. Thus, the API that enables the configuration of modular architecture and the transfer of packets from one component to the other is proposed. Components of the pipeline are modular and can be reused in other projects. For example, an IPv4 router was built using the reference pipeline, with five modules: (1) medium access control (MAC) layer reception queues, (2) input arbiter, (3) output port lookup, (4) output queues, and (5) MAC layer transmission queues. Other examples of network devices that are built using the reference design are 4-port network interface card (NIC), 4-port Ethernet learning switch, OpenFlow switch, etc.
	
	OpenFlow switch implementations on NetFPGA-10G, ML605 and DE4 platforms, which have demonstrated the portability and flexibility of the proposed architecture, are presented in \cite{Khan2013}. The switch design is described using the Bluespec System Verilog (BSV) language. Particular attention is devoted to solving the challenges of portability and flexibility through high modularity and configurability. For the design of the switch pipeline, the following modules were used: (1) flow table records composer, (2) flow table controller, (3) action processor, (4) arbiter i (5) switch controller interface. All modules are designed in such a way that machining functionality is independent of the platform (e.g., type of memory used, type of network or PCIe interface).
	
	\textit{\textbf{Network-attached FPGA concept}} --- which enables the distribution of hardware design to multiple physical resources (e.g., FPGA devices), is presented in \cite{Gibb2010}. The proposed architecture, called OpenPipes, follows the basic principles of system design according to which the system consists of multiple modules: (1) processing modules, (2) flexible interconnections, and (3) a controller that configures interconnections, and manages the location and configuration of process modules. OpenFlow was used to implement interconnection architecture. Processing modules can be implemented in hardware or software and can be relocated while the system is operating, enabling real-time experimentation or migration from the old implementation platform to the new one.
	
	Another example of network-attached FPGA concept is presented in \cite{Weerasinghe2015}. A direct connection of FPGAs to a data center network (DCN) using integrated network cards is suggested. The FPGA is divided into three parts: (1) users logic, which implements custom applications, (2) network service layer which connects FPGA with DCN, and (3) management layer which performs resource management tasks. The integration of the proposed architecture into the cloud is envisaged using the new OpenStack service.
	
	\textit{\textbf{Network-on-chip enhanced FPGA}} --- was used to develop a new programmable packet processor \cite{Bitar2015}. The new form of a packet processor, providing a high degree of flexibility and throughput of 400 and 800 Gbps, has been developed by interconnecting multiple protocol-specific processing modules. Instead of using the match tables that support the entire set of protocols, in the proposed design, packets are sent to the suitable modules depending on the protocol specified in the packet headers. Each processing module determines the actions that will be taken for that protocol, and which is the next processing module in the packet processing chain. Reconfigurable nature of the FPGA provides complete freedom in adapting and supplementing the processing modules and effectively brings programmability directly into the data plane.
	
	\subsubsection{SoC-based implementations}
	
	In response to the code re-use problem in FPGA-based networking hardware, a new flexible legacy design support for SoC and system on FPGA (SoFPGA) platforms was proposed in \cite{Antichi2013}. The proposed architecture consists of the data plane and control plane bridges, in which the data plane bridge encapsulates the old design and integrates it into a new one. Integration of NetFPGA-1G Output Port Lookup module into NetFPGA-10G design is given as an example. In this case, the data plane bridge is situated in the pipeline between the input arbiter and the output queues and is connected using the AMBA4 AXI-Stream interface. The control plane bridge is connected via the AMBA4 AXI interconnection interface to the data plane bridge. In this way, the control plane bridge has access to the internal data registers of the data plane.
	
	\textit{\textbf{OpenFlow switch implementation on SoC}} --- is presented in \cite{Chengchen2014}. The programmable platform ONetSwitch is based on the Xilinx Zynq-7045 SoC, which features a dual-core ARM Cortex-A9 Processor System and a Kintex-7 FPGA Programmable Logic (PL) within the same chip. One side of the Zynq PL is connected to four 1Gbps Ethernet and four 10Gbps SFP+ interfaces, and the other side is connected with Zynq processor system. The switch data plane is based on a hybrid software-hardware solution. The flow table lookup is performed in hardware, and in if there is no matching in hardware, the software is utilized. An algorithm, for flow table distribution into hardware and software, uses switching performance as an optimization criterion and is implemented within the hardware abstraction layer (HAL).
	
	The hardware-software co-design of OpenFlow switch using SoC platforms is presented in \cite{Zhou2014}. The architecture of the proposed switch consists of a software agent and a hardware-based data plane. The agent is implemented as a Linux application running on an ARM processor and performs following tasks: sending packets to data plane, updating/deleting flow records, reading counters, and accepting packets from the data plane. The data plane is implemented as a modular pipeline on the programmable logic of SoC. Packets can enter the pipeline, consisting of the packet parser and lookup table, through the OpenFlow agent or physical interface. The link between the OpenFlow agent and the data plane is realized through the AMBA4 AXI-Stream interface, the packet transfer, and the AMBA4 AXI-Lite interface, for signaling.
	
	\subsubsection{NP-based implementations}
	
	On the other hand, network processors have been used in a considerably smaller scope for implementation of SDN's data plane than reconfigurable hardware, but it is still important to review some of the most significant research.
	
	ServerSwitch design, motivated by the cognizance that commodity Ethernet switching chips have become programmable, is proposed in \cite{Lu2011}. The switching Ethernet chip is used for adaptive packet transfer while the server CPU is used to control and process traffic in the data plane. The prototype is implemented on the ServerSwitch card that uses the Broadcom switching chip, and is connected to the server via the PCIe 4x interface. Also, a software stack for card management and traffic control and processing in the data plane is implemented. The software stack contains the kernel component through which the card driver is implemented and the application component that provides the API to the driver.
	
	\textit{\textbf{ForCES router implementation}} --- based on Intel IXP network processor is presented in \cite{Wang2007}. Router architecture consists of CE and FE, where CE implements management and control planes, and FE implements management and data planes. The data plane is split into a fast and a slow path. Fast path is in charge of packet processing at line speed, while routing, network management and packet exceptions management are executed in a slow path. LFBs, that build up the data plane, are implemented using multiple microblocks in a flat plane. Microblocks differ from vendor to vendor and perform the single function. The FE prototype is implemented on the Intel IXP network processor, where the fast path is directly mapped to the MicroEngine layer, while the slow path is running on the Intel XScale layer. Multiple FEs were interconnected using a switch based on the Advanced Telecommunications Computing Architecture (ATCA). The presented implementation of ForTER was used for the realization of LFBs defined in \cite{Dong2007}.
	
	\textit{\textbf{OpenFlow switch implementation}} --- based on ATCA platform with architecture consisting of a data path (i.e., fast path) for packet forwarding and a control plane (i.e., slow path) for management and signaling, is presented in \cite{Rostami2012}. Data plane elements are implemented using the Broadcom Ethernet chipset on the AT8404 card that supports header parsing, packet classification, and frame forwarding by header field content. The packets which can not be directly forwarded are encapsulated and delegated to the embedded processor.
	
	\subsubsection*{Remark on flexibility and programmability}
	
	Data plane implementations based on network processors have low flexibility from aspects of function placement and operation. The network processor programmability is generally limited to configuring the parameters of the data plane functions such as queue capacity, scheduling mechanism, packet header filter, etc. On the other hand, in the reviewed papers it has been shown that the use of reconfigurable hardware and suitable hardware description languages can achieve a high level of programmability, which positively affects the data plane flexibility.	NetFPGA project, which stands out of all the reviewed approaches, simplified the process of implementing innovative network hardware using FPGA technology through the great support of the research community. A step further in the hardware programmability was made by SoC technology, which enabled the implementation of hybrid data plane architectures composed of a fast hardware path and a highly programmable software path.

\subsection{Software-based implementations}
	Only after increasing the processing power of conventional computer systems, the software-based implementation of SDN's data plane became attractive both to researchers and industry. However, development of the idea about a software-based implementation of a programmable packet switching node, which later served as a useful tool for SDN's data plane implementation, began several years ago. By reviewing the research dealing with software-based implementations, we have noted the following categories:
	\begin{enumerate}
		\item{pure software-based implementations,}
		\item{implementations based on virtualization techniques,}
		\item{implementations supported by hardware-based acceleration.}
	\end{enumerate}

	\subsubsection{Pure software-based implementations}
	
	\textit{\textbf{Click}} --- a new software architecture for the realization of configurable and flexible routers is presented in \cite{Kohler2000}. Each Click router is built from the packet processing module, so-called elements. The elements are performing simple router functions such as queue processing, packet classification, and providing an interface to network devices. For a description of the router configuration, a declarative language that supports user-defined abstractions has been proposed. The declarative language, which is also readable to humans, is suitable for machine processing and can be easily translated into a directed graph, where elements are represented by nodes and a packet transmission lines between elements with branches. 
	
	Similar to the idea of Click, the extensible open router platform (XORP) is presented in \cite{Handley2003}. Its design addresses four key objectives: (1) features, (2) extensibility, (3) performance i (4) robustness. XORP was conceived both as a stable platform and as a research tool that would allow smooth transfer of new ideas from the lab environment to the production network. It consists of two subsystems: (1) a high-level subsystem that performs routing protocols and other management processes in the user space, and (2) a low-level subsystem that manages data plane processes. Data plane is realized using the Click modular router, but it can also be implemented at the UNIX kernel level by exploiting the forwarding engine abstraction layer that abstracts the implementation-specific data to routing processes.
	
	To increase the performance of the Click software router, the RouteBricks architecture, which enables parallel process execution on multiple processor cores within one or more servers, is proposed in \cite{Dobrescu2009}. The design is based on the Click Router extension with the support of allocating specific elements of the Click router to particular processor cores. To achieve this, the 10G network adapter driver is additionally extended with support for multiple queues and support for NIC-driven batching. This ensured that one packet is processed on only one processor core, and the number of input/output (I/O) transactions is also reduced. By implementing a prototype RouteBricks router, named RB4, made up of four servers connected by 10Gbps links in a full-mesh topology, it has been shown that a bandwidth of up to 40Gbps can be achieved.
	
	\textit{\textbf{OpenFlow switch implementation}} --- based on Click modular platform is described in \cite{Mundada2009}. To create a hybrid model that allows packet- and flow-based processing, an OpenFlowClick element has been added within the Click router, that enables rule tables management through the OpenFlow protocol. OpenFlowClick runs as a Click kernel module, and uses the \textit{secchan} and \textit{dpctl} tools to communicate with the OpenFlow controller. Within the OpenFlowClick element, the data path module performs a rule checking and packet forwarding, and the control plane module manages the forwarding rules table according to the controller commands. Linear and hash tables are used for the implementation of wildcard and exact matching.
	
	Another software-based implementation of the OpenFlow switch is presented in \cite{Fernandes2014}. The switch architecture is based on the NetBee library and consists of: (1) ports that use (2) NetBee Link components for switch and network interfaces connection, (3) NetPDL \cite{Risso2006} description of OpenFlow 1.3 protocol formatted in XML, (4) NetBee XML protocol description parser, (5) flow table, (6) rules grouping table, (7) meter table, (8) oflib library for OpenFlow messages conversion to internal format and vice versa, and (9) secure channel to the OpenFlow controller.
	
	A cost-effective alternative to SDN implementation using Raspberry Pi single-board computers and the OVS software switch is proposed in \cite{Kim2014}. Although the new architectural aspects of the data plane SDN are not presented in this paper, it has been shown that single-board computers can be used as a platform for execution of software switches. The performance of the prototype implemented is similar to that achieved by using a hardware switch based on the NetFPGA-1G platform.
		
	In addition to the implementation of OpenFlow switches that are published in research papers, here we list some other open-source implementations of OpenFlow switches and supporting libraries:
	\begin{itemize}
		\item{OpenFlow v1.0 reference implementation\cite{OFREFIMPL} - written in C,}
		\item{Indigo \cite{INDIGO} - support for physical and hypervisor-based switches written in C,}
		\item{Pantou \cite{PANTOU} - port of OpenFlow implementation for OpenWRT wireless platform written in C,}
		\item{OpenFaucet \cite{OPENFAUCET} - implementation of v1.0 written in Python,}
		\item{OpenFlow Java \cite{OFJAVA} - OpenFlow stack written in Java,}
		\item{oflib-node \cite{OFLIBNODE} - implementation of v1.0 and v1.1 protocols in the form of libraries for Node.js.}
	\end{itemize}

	\textit{\textbf{Combination of emulation and simulation}} --- to support the realization of large network experiments is proposed in \cite{Fernandes2018}. The proposed architecture, inspired by the idea of SDN, separates the control and data planes so that the control plane is emulated and the data plane is simulated. Within the simulated data plane, there are common elements of a network simulator based on discrete-event simulations (DES):
	\begin{itemize}
		\item{event queues,}
		\item{queue scheduler,}
		\item{event processor,}
		\item{network status and statistics register,}
		\item{topology containing a simulated logic of network nodes.}
	\end{itemize}
	The following events are held in the queues: (1) start of an application, (2) flow arrival, (3) flow departure, (4) arrival of message from a control plane, and (5) departure of message to the control plane. The flows consist of aggregated packets with common headers. The flows are separated into incoming and outgoing, to take into account the traffic losses. In a presented example of the SDN implementation, switches and end nodes are implementing the OpenFlow protocol within DES, while controllers are running as real and independent software instances.
	
	\subsubsection{Implementations based on virtualization techniques}

	\textit{\textbf{Machine virtualization technique}} --- has been used for the first implementation of virtualized data plane, called Mininet \cite{Lantz2010}. Mininet is the networking virtualization environment based on Linux virtual machines running on standard platforms such as VMware, Xen, and VirtualBox. It allows the creation of virtual networks by setting up and interconnecting the host processes within the network namespace. For host interconnection virtual Ethernet (\textit{veth}) is used. The user can use various building elements for implementation of the SDN-based virtual network: (1) links made as \textit{veth} pairs, (2) hosts realized as shell processes, (3) switches implemented as OpenFlow software switches, and (4) controllers that can be run anywhere in a real or simulated network.
	
	The idea of scaling the Click modular router performance by increasing the number of instances of Click routers running within miniature virtual machines is presented in \cite{Ahmed2012}. These miniature virtual machines are called ClickOS, and are running on a Xen hypervisor. Xen hypervisor has a shared network \textit{netback} driver, which communicates with hardware on the one side, and through shared memory with the ClickOS \textit{netfront} driver on the other side. The task of the \textit{netback} driver is to forward packets between from network adapter to shared memory, and vice versa, over a virtual network interface. On the other hand, the \textit{netfront} driver is scheduling packets from shared memory to the transceiving interfaces of Click router (FromClickOS and ToClickOS), and vice versa. This builds a bridge between the Click router and NICs while preserving all gains of virtualization.
	
	\textit{\textbf{Network/NIC virtualization technique}} --- can be used for implementation of generic high throughput bus or in a concrete case for connecting virtual machines. In that sense, a virtual local Ethernet (VALE) is proposed in \cite{Rizzo20121}. VALE, using the \textit{netmap} API \cite{Rizzo20121} as a communication mechanism, exposes ports to hypervisors and processes. The core \textit{netmap} is based on shared memory, which represents the interface between network hardware and packet processing applications. Within that memory, packet transmission and reception buffers are assigned to each network interface, and two circular arrays called \textit{netmap} rings for storage of metadata about transmission and reception buffers.
	
	The software switch mSwitch \cite{Honda2015} has simultaneously responded to several shortcomings noted in the previous solutions by utilization of techniques for network interfaces virtualization and the separation of switching from packet processing. mSwitch provides: (1) flexible data plane, (2) efficient processor utilization, (3) high throughput, (4) high packet processing intensity, and (5) high port density. The central principle of the proposed architecture is the division of the data plane into the switch fabric which is responsible for packet switching and the switch logic, which is the modular part of the switch, responsible for the packet processing. This allowed a high throughput, while maintaining the high level of programmability of the packet processing functions.
	
	Open vSwitch (OVS), a multilayer virtual switch presented in \cite{Pfaff2015}, is intended for networking in virtual production environments and supports most hypervisor platforms. In the OVS architecture, the packet forwarding is performed using two components:
	\begin{enumerate}
		\item{\textit{ovs-vswitchd} daemon in the user space which is identical for all operating systems,}
		\item{high-perfomance datapath kernel module written for the target operating system.}
	\end{enumerate}
	The datapath kernel module is responsible for receiving packets from a NIC or a virtual machine, and its processing according to the instructions given by the \mbox{\textit{ovs-vswitchd}} module. In the case that there are no defined processing rules for the specific packet in the kernel module, this packet is forwarded to the \mbox{\textit{ovs-vswitchd}} module, which then makes the decision on further processing and returns it together with the packet. When the OVS is used as an SDN switch, then the agent side of the OpenFlow protocols is running inside the \mbox{\textit{ovs-vswitchd}} module.
	
 	\subsubsection{Implementations supported by hardware-based acceleration}
 	
 	The problem of the limited performance of the pure software-based implementation of packet switching nodes has been addressed in some research by utilization of hardware-based acceleration. 
 	
 	\textit{\textbf{Graphics processing unit (GPU) based acceleration framework}} --- has been used for development of a software router PacketShader as presented in \cite{Han2010}. The challenge of maintaining high forwarding rate while preserving sufficient processing power for different routing applications is solved as follows: (1) the I/O engine for fast and efficient packet processing is implemented, (2) routing table lookup and IPsec encryption are offloaded from the main processor to GPUs. I/O engine functions, which are implemented at the kernel level, are used for kernel-level packet handling operations. The remaining packet processing operations are executing in the multi-threading application in user space with the help of three callback functions: (1) pre-shader, (2) shader, and (3) post-shader. The pre-shader function performs fetching of packet parts from the receiving queues to the GPU. The shader function performs processing of the packet within the GPU kernel, and the post-shader function delivers processed parts of the packet to the destination ports. In this way, an efficient pipeline for packet processing by FIFO principles has been established. By implementing the prototype of the router, it has been shown that high throughput such as 40Gbps can be achieved.
			
	\textit{\textbf{NIC-accelerated Ethernet switch implementation}} --- called CuckooSwitch \cite{Zhou2013}, combines the hash-based forwarding information base (FIB) design with Intel's Data Plane Development Kit (DPDK) \cite{DPDK} platform that performs inbound/outbound packet transmission between hardware and user-space threads. The packet processing pipeline has three stages: (1) reception of the packet via the NIC and its storage in the reception queue using direct memory access, (2) processing of packets from queues using a worker thread in the user space, and (3) scheduling of processed packets into output queues associated with appropriate output ports. The number of input queues has been chosen to correspond to the number of worker threads, where each worker thread is allocated to one processor core. In this way, competition and synchronization overhead have been avoided. In the second stage of the pipeline, the FIB search is performed based on the destination MAC address of the Ethernet frame. FIB supports dynamic updating of rules in real time and reading of records from multiple concurrent worker threads, eliminating the need for storage of numerous copies of FIB content. By implementing a prototype, in \cite{Zhou2013} it has been demonstrated that using a standard server with eight 10Gbps Ethernet interfaces, processing power of 92.22 Mpps can be achieved for 64B packets.
	
	\textit{\textbf{NIC-accelerated OpenFlow switch implementation}} --- based on the Intel DPDK library is presented in \cite{Pongracz20131}. By using the DPDK library, implementation costs have been reduced in terms of the packet I/O overheads, DRAM buffering, interrupt processing, kernel structure overheads, and copying data when changing the context from the kernel space to the user space and vice versa. DPDK enables, through Direct I/O mechanism, direct data transfer between the program running in user space and input/output cards.
	
	\subsubsection*{Remark on flexibility and programmability}
	
	Pure software-based data plane implementations are characterized by excellent flexibility due to high level of programmability and configurability of forwarding functions. For example, software architecture of the modular router Click allows the realization of arbitrary data plane structures using fully programmable packet processing modules. In the reviewed papers it has been shown that high flexibility, in terms of the granularity of dynamic resource management, may be achieved by parallel execution of instances of software implementations on multi-core processors. Virtualization-based techniques additional contribute to flexibility, as they enable efficient scaling of forwarding functions. The use of hardware-based acceleration enables the increase of performance of software-based implementations without losing inherent flexibility. A large number of software-based implementations of the OpenFlow switch indicates that the software-based approach to the realization of the SDN's data plane is powerful, especially if their application is planned in modern data centers which often use virtualization techniques.

\section{Implications of SDN-related research on data plane evolution}
\label{sec:implications}
	The previous sections focused on standard SDN's data plane architecture, such as ForCES and OpenFlow, and their implementation using software and hardware technologies. On the other hand, an overview of SDN-related research whose results have implied the data plane evolution, is given in this section. Under the data plane evolution, we indicate a gradual deviation from the original data plane architectures given with ForCES and OpenFlow specifications, resulting in the need to address the problem of programmability and flexibility of the data plane in a more generic way. By reviewing these studies, we have found that the treated problems can be classified into the following categories:
	\begin{enumerate}
		\item{performance,}
		\item{energy consumption,}
		\item{quality of service (QoS),}
		\item{measurement and monitoring,}
		\item{security and reliability,}
		\item{support for various network technologies,}
		\item{network and network functions virtualization.}
	\end{enumerate}
	Therefore, this section is organized in accordance with the above mentioned categories. Within each category, reviewed research is organized according to the problem-solving approach. Given that many problem-solving approaches are common to several categories of problems, here is given a brief overview of identified problem-solving approaches.
	
	\textit{\textbf{Data plane programming}} --- implies the introduction of programmed packet processing into the data plane. The machine structure, which executes program instructions on packets, is mainly fixed and predefined.
	
	\textit{\textbf{Stateful packet processing}} --- allows packet processing which is aware of the state of the data plane. A lot of research examined in this section has shown that the stateless nature of the OpenFlow switch does not provide adequate support for processing of packets coming from a stateful protocol (e.g., TCP, FTP) or the implementation of some mechanisms such as a stateful firewall. Stateful packet processing is most often achieved by introducing finite automata into the data plane.
	
	\textit{\textbf{Reconfigurable architectures}} --- most commonly based on reconfigurable technology (e.g., FPGA), allow the implementation of a variable data plane structure. That has been used in some research to solve a specific problem from the above categories by increasing the flexibility of the data plane.
	
	\textit{\textbf{Physical layer management}} --- is a frequently used technique for managing the energy consumption of the network device's physical interface.
	
	\textit{\textbf{New structures of flow tables}} --- are introducing, in addition to basic information such as packet headers of a specific flow, additional data to support the treatment of issues related to data plane performances, energy consumption, QoS, etc.
	
	\textit{\textbf{New mechanisms for flow tables lookup}} --- often accompany structural changes of flow tables. In some cases, new mechanisms are based on the enhancement of existing OpenFlow flow table lookup mechanisms.
	
	\textit{\textbf{New packet classification mechanisms}} --- allow packet classification based on the header of higher-layer protocols. In this way, they enable the implementation of advanced mechanisms such as DPI.
	
	\textit{\textbf{Hybrid architectures}} --- are data plane architectures implemented using a combination of hardware and software technologies or using different types of hardware technologies (e.g., FPGA and CPU).
	Such architectures enable efficient distribution of packet processing tasks according to the affinities of the particular technology.
	
	\textit{\textbf{New data plane architectures}} --- are proposed in a significant number of research as an answer to the limited flexibility of OpenFlow and ForCES data plane architectures.
	
	\textit{\textbf{Hardware abstraction layer (HAL)}} --- is a method for providing support for non-IP networking technologies in such a way that specificities of target technology are abstracted to the data plane through suitable processes and interfaces independent of underlying network technology.
	
	At the end of the section, we have established a correlation between treated problems and problem-solving approaches, which is the first step towards the generalization of approaches to increase the programmability and flexibility of the SDN's data plane.
	
\subsection{Performance}
	The performance improvement problem of SDN has been addressed by introducing various changes to the data plane. The focus of \cite{Keinanen2009,Antichi2011,Curtis2011,Ferkouss2011,Qu2013,Martinello2014,Perez20141,Perez20142,Yan2014,Yanbiao2014,Kalyaev2015} was on changes in the flow table structure and the introduction of new flow table lookup or packet classification mechanisms. On the other hand, in \cite{Ciesla2009,Luo2009,Ram2010,Tanyingyong2010,Tanyingyong2011,Gao2012,Lu2012,Katta2014,Vencioneck2014,Dang2015,Bifulco2015,Sanvito2017}, the focus was on changing data plane architecture by using hybrid software-hardware architectures or combinations of different types of hardware, by introducing reconfigurability into the data plane, and by introducing stateful packet processing,
	
	\textit{\textbf{New structure of flow table}} --- which supports load balancing based on regular expressions, is proposed in \cite{Antichi2011}. Half-SRAM was used on NetFPGA because in load balancing scenarios, it is necessary to keep a large number of records in flow tables. The hardware plane takes care of the longest prefix matching (LPM) in the dFA structure, and the software plane manages the data structure by inserting/removing the rules in/from the forwarding table.
	
	The implementation of the packet switch with Bloom-filter based forwarding, called zFilter, is described in \cite{Keinanen2009}. The proposed forwarding mechanism is based on the identification of links instead of nodes. Packet switching nodes do not need to maintain any status except link identifier information for each interface. The forwarding information is constructed based on the aforementioned link identifiers and is transmitted in the header of the packet as the Bloom filter structure. Based on the presence of the link identifier in the Bloom filter structure carried in the packet, each packet switching node decides to which interface a packet should be forwarded.
	
	The new method of the hardware-based organization of the forwarding tables in SDN switches is presented in \cite{Kalyaev2015}. Given the advantages of parallel processing on the FPGA chip compared to serial processing on network processors or general purpose processors, the proposed solution uses all 512 bits of the header in a wildcard-based lookup.
	
	\textit{\textbf{New mechanism for flow table lookup}} --- called DevoFlow, which reduces the number of interactions on the switch-to-controller interface and the number of records in ternary content-addressable memory (TCAM), through the aggresive use of wildcard rules, is introduced in \cite{Curtis2011}. The proposed modification is based on the introduction of two new mechanisms for devolving the control from the controller to the switch: rules cloning and local actions. Rules cloning allows the creation of new rules for micro-flows based on templates, which rule search diminishes to direct matching, thus reducing the use of TCAM. Local actions allow prediction of rules and their establishment without contacting the controller. Another contribution of DevoFlow is lessening the need for statistics transmission for less dynamic flows through the use of three mechanisms for efficient statistics collection: (1) sampling, (2) triggering, and (3) approximate counting.
	
	The classic routing table lookup is modeled as the problem of the LPM and is divided into three main categories: 
	\begin{enumerate}
		\item{TCAM-based solutions which provide a deterministic and quick lookup,}
		\item{hash-based solutions which provide a quick lookup with simple table update mechanisms,}
		\item{trie-based solutions.}
	\end{enumerate}
	Instead of optimizing classical models, the brand new model, the Split Routing Lookup Model, is proposed in \cite{Yanbiao2014}. The basic idea is to divide the original flow table into two smaller, perform LPM lookup over them, and to aggregate two results into one. This results in savings in chip resources, and increased performance by introducing parallelism in the lookup process.
	
	KeyFlow, proposed in \cite{Martinello2014}, is a new approach to building a flexible network-fabric based model. The flow table lookup mechanism in the forwarding engine was replaced by simple operations based on the residual number system. Principally, the proposed model is based on source routing. Unlike Multiprotocol Label Switching (MPLS) based solutions, where there is a need for intensive communication between the core and the controller to establish end-to-end connections, all possible routes are ready for use in KeyFlow and only the appropriate route identifier for the ingress packet has to be allocated. In this way, round-trip time reduction was achieved by more than 50 percent.
	
	The caching system for the SDN based on the wildcard rules, called \textit{CAching in Buckets} (CAB), is presented in \cite{Yan2014}. The basic idea of the CAB is the partitioning of the field into logical structures, called \textit{bucket}, and bucket caching along with the associated rules. The CAB switch is implemented as a two-stage pipeline consisting of bucket filters and flow tables. In the first stage, the matching of the packets is done in all buckets, and in the second stage, the flow table lookup is performed. Packets that do not belong to a single bucket are encapsulated and sent to the controller. Logically, the bucket filter and the flow table are two tables that can be implemented using two separate TCAM memory or one multiple-lookup TCAM memory. The CAB solves the problem of the rules dependence and achieves efficient utilization of network bandwidth by reducing communication between the switch and the controller.
	
	\textit{\textbf{New packet classification mechanism}} --- based on resursive flow classification is presented in \cite{Ferkouss2011}.	Extended recursive flow classification (ExRFC), using the combination of SRAM and TCAM memory, improves the parallelization of the classification processes and exploits the hardware resources of most hardware platforms better.

	The 2-dimensional pipeline for packet classification on the FPGA, which consists of multiple self-configurable processing elements (PE), is presented in \cite{Qu2013}. The proposed architecture achieves scalability, regarding the size of the rule set, while maintaining a high-throughput packet classification in the OpenFlow switch. The modular PE can perform matching by the scope and by the prefix, making it suitable for different types of packet headers. Connecting PEs in the 2-dimensional pipeline is possible in two directions:
	\begin{enumerate}
		\item{horizontal propagation of the bit vectors of the PE output registers in a pipeline fashion,}
		\item{vertical propagation of the packet header bits (PE input registers) in a pipeline fashion.}
	\end{enumerate}
	A complex packet classification operations can be realized, by using PEs striding and clustering problem-solving techniques, whereby the clock signal frequency is not limited by the length of the packet header and the size of the rule set.
	
	The new hardware solution for the configurable packet classification is presented in \cite{Perez20141,Perez20142}. The performance of different classification algorithms was analyzed through two approaches: (1) a multi-field lookup, and (2) a single-field lookup. It has been shown that a parallel combination of different lookup algorithms based on one packet header field has achieved better performance than using a lookup algorithm based on multiple packet header fields. Therefore, the design of a hardware-based classifier architecture is proposed, which achieves optimal lookup performance by running the best set of algorithms for a given type of record in the flow table. Sharing memory resources between different lookup algorithms has resulted in efficient memory utilization. The SDN controller selects the best combination of search algorithms, and following the decision, configures the appropriate memory blocks on the hardware platform of the switch. The packet classification process is performed within a four-stage pipeline:
	\begin{enumerate}
		\item{splitting of packet header into multiple fields over which individual lookup algorithms will be performed,}
		\item{parallel execution of lookups, using a selected set of algorithms,}
		\item{combining lookup results into a single tag with the highest priority,}
		\item{reading the highest priority match rule based on the tag from the preceding stage.}
	\end{enumerate}
	The proposed architecture can follow the evolution of SDN applications by the simple extension of the existing set of lookup algorithms with a new one.
	
	\textit{\textbf{Hybrid architecture}} --- consisting of hardware and software components, was used in \cite{Ciesla2009} for the implementation of the uniform resource locator (URL) extraction mechanism from the hypertext transfer protocol (HTTP). The hardware component is an extension of the IPv4 reference router on the NetFPGA platform, and is implemented by modifying the \textit{Output Port Lookup} module in the reference design. Its task is to extract the HTTP GET request and send it to the software component. The software component extracts the URL from the HTTP GET request and updates the internal database. This paper presents an approach to the performance improvement of the deep packet inspection by applying the hybrid architecture of the packet switching node.
	
	The OpenFlow switch reference design, accelerated using multi-core network processors, is presented in \cite{Luo2009}. In the proposed design, 16 micro-machines (programmable cores) of the network processor are programmed to perform tasks of receiving packets, sending packets, processing packets, managing queues and serving orders, communicating via PCI bus, etc. The implementation of the given design consists of the software on the host and the network processor (NP) acceleration card. On the host side, the OpenFlow software communicates with the NP accelerator card via the PCIe bus using the kernel module. The experiments carried out showed a reduction in packet delay by 20\% compared to the conventional OpenFlow switch design.
	
	The sNICh architecture, which is a combination of the NIC and a data center switching accelerator, is proposed in \cite{Ram2010}. sNICh uses acceleration hardware in the form of a PCIe card, for offloading the server concerning the packet forwarding. The data and control planes in the sNICh architecture are separate, where the data plane is implemented in the NIC hardware, and the control plane is implemented within the sNICh backend. In addition to standard NIC functionality, sNICh also realizes a flow-based switch whose flow table is implemented using TCAM. In the data plane, a copy engine and memory-to-memory direct memory access engine for direct access to shared system memory are implemented, to improve performance in communication between virtual machines (VM) within the data center.
	
	Inspired by the idea of the packet processing offloading from the processor to the NIC, the architectural design that improves the flow table lookup performance on a PC-based OpenFlow switch is proposed in \cite{Tanyingyong2010,Tanyingyong2011}. The proposed solution is based on flow caching and placing the lookup process in the fast path of the switch. The acceleration of OpenFlow flow table lookup process is achieved by using the classification features of Intel 82599 10GbE controller found on modern 10GbE NICs.
	
	The use of the CPU in switches to handle not only control traffic but data plane traffic as well is presented in \cite{Lu2012}. A powerful processor has been added to a commodity switch and connected by a high-bandwidth link to the application-specific integrated circuit (ASIC), which can be programmed to redirect a portion of the traffic to the processor. In this way, the design limitations of the switches, regarding the forwarding table size and the depth of the packet buffers, have been overcome. Additionally, adding a CPU to the network device has increased the data plane programmability.
	
	CacheFlow, a new splicing technique of a large number of unpopular forwarding rules, from dependency chains, to a small number of new rules, aiming for the cache pollution avoidance, is presented in \cite{Katta2014}. It combines high-throughput hardware switches with large flow tables of the software-based switches. Hardware switches implement cache on TCAM, and in the case of cache miss, software switches are engaged, thus avoiding unnecessary and time-consuming communication with the controller.
	
	Following the trend of adding powerful processors to conventional switches, in \cite{Bifulco2015} is argued that a hybrid software-hardware switch can reduce the time needed to install rules in the flow table. Accordingly, ShadowSwitch (sWs) was proposed as the prototype of the OpenFlow switch which implements this design. Besides the hardware switch (HwSw), a high-performance software layer (SwSw), which performs packet forwarding, was introduced in the fast path of sWs architecture. The control logic (sSwLogic), whose task is to manage the record installation in the flow table, is placed in a slow path of the switch. The prototype is implemented using the commodity hardware-based OpenFlow switch and the OVS instance which runs on the server.	
	
	\textit{\textbf{Reconfigurable architecture}} --- based on a virtual emulated network on a chip, called Diorama network, is proposed in \cite{Gao2012}. The prototype is implemented on the dynamic reconfigurable processor DAPDNA-2, which consists of a high-performance digital application processor based on reduced instruction set computer (RISC) architecture and distributed network architecture (DNA). DNA connects 376 PEs and is used to construct emulated nets on the chip, where each PE emulates different functions of the actual router or link. Test results have shown that the prototype can perform the shortest path calculation 19 times faster than conventional solutions.
	
	FlexForward, which enables flexible adaptation of forwarding algorithms in software-defined DCN, is proposed in \cite{Vencioneck2014}. Reconfiguration of forwarding mechanism on switches is supported by the OpenFlow protocol and implemented by OVS extension. Forwarding mechanisms are implemented between flow extraction and flow table lookup processes. Performance improvements have been achieved by introducing an additional feature to skip the flow table lookup process. Supported forwarding mechanisms are:
	\begin{itemize}
		\item{regular OpenFlow - used when FlexForward is switched on, as long as the OpenFlow switch does not get command to change the forwarding mechanism,}
		\item{hypecube topology \cite{Bhuyan1984} - used in the server-centric DCN,}
		\item{KeyFlow - used in arbitrary topology.}
	\end{itemize}
	
	\textit{\textbf{Stateful packet processing}} --- enables offloading of simple networking processes, such as filtering and counting, from the control plane to the network switches. By using the open packet processor (OPP) in \cite{Sanvito2017}, a stateful DPI application was implemented as a pipeline consisting of two tables: one for selecting an output interface and the other for implementing DPI functionality. OPP will be described in more detail in the Section~\ref{sec:programmability}.
	
	The use of Paxos protocols for distributed and programmable network systems to improve SDN performance is demonstrated in \cite{Dang2015}. For Paxos protocol implementation, it was necessary to add support for the roles of coordinator and acceptor in the OpenFlow switch logic. According to the Paxos protocol, the coordinator must support the generation of a unique round number and a monotonically incrementing sequential number. The acceptor switch must support the storage and stateful comparison of the received random number with the appropriate field in the packet header, and maintenance of the local state by the protocol. It is proposed to implement these functionalities at the edge of the network using programmable NICs.
	
\subsection{Energy consumption}
	The issue of energy efficiency and energy consumption in the SDN has also been addressed in several ways. Some have focused on reducing energy consumption in memory for storing flow table and introducing new flow table lookup mechanisms. Others have decided to reduce the energy consumption of network's physical layer by introducing the support for physical layer management and by abstracting underlying hardware.
	
	\textit{\textbf{New structure of flow table}} --- which stores flow identifiers (Flow-ID), represented by a smaller number of bits, instead of standard flow records, is proposed in \cite{Kannan2013}. The proposed Compact TCAM summarizes the information about the flow to the size needed to identify unique flows. It has been experimentally demonstrated that it is possible to save energy by about 2.5 times the standard layer 2 switches, and up to 80\% compared to OpenFlow switches.
	
	\textit{\textbf{New flow table lookup mechanism}} --- based on flow tagging, called Tag-In-Tag, is proposed in \cite{Banerjee2014}. Tag-In-Tag is based on a generally-perceived phenomenon in networks:
	\begin{itemize}
		\item{flows travel by paths,}
		\item{paths form a deterministic set, i.e., all source and destination pairs are known in advance,}
		\item{multiple flows can travel along the same path.}
	\end{itemize}
	Based on that, the proposed mechanism denotes a packet with two tags: PATH TAG used for packet routing, and FLOW TAG that associates the packet with the corresponding flow. PATH TAG encapsulates within FLOW TAG. Within the switch, the data path is divided into two TCAMs, one for the incoming and the other for the outgoing packets. On edge switches, TCAM for incoming packets contains full headers, and TCAM for outgoing packets holds only PATH TAG and FLOW TAG. In core switches there is only TCAM that contains PATH TAG and FLOW TAG. In the end, the lookup is performed over a small number of bits in parallel, thereby achieving a TCAM energy savings of up to 80\%.
	
	Motivated by the limitations of commercial switches at the time, regarding latency and energy consumption, improvements that simultaneously reduce energy consumption and switch latency, while retaining the flexibility offered by the OpenFlow, are proposed in \cite{Congdon2014}. The proposed enhancements exploit local time stationarity of network traffic and predict the affiliation of the packet to the flow for each port of the switch. Thus, they try to avoid traditional flow-based or TCAM lookups, which simultaneously reduces both energy consumption and latency of forwarding.
	
	\textit{\textbf{Physical layer management}} --- as an extension for OpenFlow switch with support for different power saving modes is presented in \cite{Tran2012}. It includes the definition of new messages within the OpenFlow standard and the design of a separate controller that manages port powering. The prototype is implemented by combining a NetFPGA platform and a specially designed hardware controller.
	
	\textit{\textbf{Hardware abstraction layer}} --- for a unified and straightforward representation of power management features in heterogeneous data plane hardware, called Green Abstraction Layer (GAL), is presented in \cite{Bolla2013}. On the side of network devices, manufacturers can implement power management primitives (PMP) by introducing specific hardware-level elements such as controllable clock generators, manageable power supplies, sleeping transistors, etc. On the network management side, the GAL abstracts the PMPs and enables simple power management through higher layer protocols.
	
\subsection{Quality of service}
	\textit{\textbf{New structure of flow table}} --- supporting fine-grained tracking of flows with a separate flow classification, is proposed in \cite{NamSeok2013}. The flow table is divided into three different tables: the flow state table, the forwarding rules table, and the QoS rules table. The forwarding and QoS information are searched in the rule tables and linked to the record from the flow state table on the arrival of the first packet from a new flow. The arrival of the next packet from the same flow only requires a referral to the flow state table without the need to search through the forwarding rules table and the QoS rules table again. To assure performance at the micro-flows and the aggregate flows levels, a flow-based packet scheduling algorithm was developed. The proposed architecture was implemented using the Cavium OCTEON CN5650 multi-core processor, whose 12 cores were assigned as follows: one core for the OpenFlow agent, eight cores for flow-based packet processing, one core for future functionalities, one core for QoS coprocessing, and last core for the server. All cores share data from the forwarding rules, QoS rules and flow state tables, and queues.
	
	\textit{\textbf{New data plane architecture}} --- is introduced in \cite{Sonkoly2012} to support QoS experimentation in OpenFlow testbed Ofelia \cite{Melazzi2012}. The proposed QoS framework did not restrict the configuration of queues regarding minimum and maximum speed, as was the case with OpenFlow. Configuration of queues is provided through additional control protocols such as OF-Config \cite{OFCONFIG} and NETCONF \cite{Enns2011}.
	
	B4, introduced in \cite{Jain2013}, is Google's approach to SDN-based DCN implementation with QoS support. The proposed SDN architecture is based on OpenFlow, but due to the limitation of existing switch architecture concerning low-level behavior management, a custom hardware platform was used. B4 switches consist of more commodity switching chips organized in two-stage Clos topology. Each chip in the first stage (input stage) of the Clos topology is configured to forward incoming packets to a second stage (backbone) except when the packet terminates on the same chip. Chips from the second stage pass the packets to the chip from the first stage depending on the packet destination. The specially developed OpenFlow agent running inside the B4 switches mediates in the configuration of the forwarding table in this non-standard pipeline by translating the OpenFlow messages into the corresponding chipset driver commands.
	
	Since the OpenFlow switches implemented in the OVS did not support the queue configuration at that time, the QueuePusher architecture based on the OVSDB standard supported in the OVS was designed in \cite{Palma2014}. Since OVSDB is not part of the existing OpenFlow controllers, QueuePusher has been created as an extension of the existing Floodlight controller interface to simplify the procedure for creating queues within the OpenFlow switch. Although there were no direct changes to the SDN's data plane, it is remarked that there was an OpenFlow protocol constraint regarding the configuration of the queues which belong to the SDN's data plane.
	
	The API for configuration of priority queues on switch ports, based on an extension of the interface between the SDN controller and the switch with OVSDB protocol support, is proposed in \cite{Caba2015}. Unlike the solution proposed in \cite{Palma2014}, the QoS abstraction model of OVS switch is defined here. The model is stored within the OVS database and accessed through the OVSDB protocol. A QoS object which specifies the maximum speed that can be shared between the priority queues is defined on each OVS switch port. The QoS configuration module inside the controller does not keep information about the state of switch queues, but only maps the QoS configuration to the switch port. Retaining information about the state of queues within switches makes it easier to maintain data consistency.
	
	\textit{\textbf{Hybrid architecture}} --- based on the combination of FPGAs and commodity switching hardware, is proposed in \cite{Sivaraman2013}. Extending SDN flexibility by making queueing and scheduling decisions inside a fast path of the switch is achieved by adding small FPGAs, with well-defined interfaces to packet queues on the switch, in a fast path of the switch. As proof of the proposed concept, two scheduling schemes have been described and implemented: CoDel and RED.
	
	With the addition of the FPGA controller used for the deterministic guaranteed-rate (GR) services, the relocation of the network inntelligence from the control plane to the data plane is presented in \cite{Szymanski2015}. The data plane within the developed solution consists of eight FPGA controllers and packet switches of minimal complexity implemented on the Altera Cyclone IV FPGA. The FPGA controller allows the provision of deterministic GR services via IP routers, MPLS, Ethernet, InfiniBand and Fiber Channel switches. By using optoelectronic packet switches implemented on a single chip, it is possible to achieve an aggregated capacity of 100 Tbps.
	
	\textit{\textbf{Stateful packet processing}} --- was used in \cite{Wang2014} to create the autonomous QoS management mechanism in the SDN (AQSDN). AQSDN enables the independent configuration of QoS features in the switch using the OpenFlow and OF-Config protocols. Autonomy is ensured by implementing the packet context-aware QoS (PCaQoS) model inside the data plane, which enables the switch to know the context of the packet and to respond locally. The PCaQoS model consists of two components: packet context-aware packet marker (PCaPM) and packet context-aware queue management (PCaQM). PCaPM is based on a multi-color DSCP-based packet marking, while PCaQM coordinates these activities at the integrated flow level. The system prototype is implemented by extending the software switch Ofsoftswitch13 \cite{Fernandes2014}.
	
	\textit{\textbf{Hardware abstraction layer}} --- for representation of physical network layer in the hierarchically autonomous QoS model, is proposed in \cite{Wang2015}. In the proposed model, a data plane is precisely cut into multiple virtual networks with the capabilities of the dynamic allocation of resources. For the implementation of the proposed architecture, the following mechanisms have been used: 
	\begin{itemize}
		\item{redistribution of virtual resources based on the context,}
		\item{network virtualization,}
		\item{autonomous network structure.}
	\end{itemize}
	Redistribution of virtual resources is achieved by cutting overall physical network infrastructure resources according to the needs of a single service or application. Network virtualization is enabled by introducing a flow structure with an extendable packet matching mechanism. Autonomous network infrastructure is ensured by introducing control loops in the control plane which are responsible for the settings configuration on physical network elements.
	
\subsection{Measurement and monitoring}
	\textit{\textbf{New structure of flow table}} --- to support sampling of packets covered by wildcard flow records in OpenFlow switch, is proposed in \cite{Wette2013}. In the mentioned paper is was noted that traffic intensity between controllers and switches can be significant in networks such as DCN, which may result in slower forwarding of new flows. One way to reduce control traffic is the usage of wildcard records for the creation of default routes in the network. Since switches do not keep track of the flows covered by wildcard records, the controller has no more information about individual flows.
	
	\textit{\textbf{New data plane architecture}} --- which, in addition to data and control planes, introduces a history plane to support the packet history logging regarding its entire journey through the network is presented in \cite{Handigol2014}. The proposed history plane includes NetSight servers and the coordinator. In addition to basic SDN tasks, controllers are responsible for configuring packet history filters (PHF) on switches. PHFs are described in a language based on regular expressions, and specify the path, switch state, and packet header fields for the history of the packet of interest. When the packet passes through the switch, a postcard is generated based on the PHF trigger and delivered to the NetSight server. A postcard represents a packet summary that contains elements essential to tracking its journey through the network. NetSight servers collect postcards, process them and store them in compressed lists, and on the request submit compressed packet history to the coordinator. The coordinator is responsible for processing the history of the packet and generating useful information for the network monitoring applications.
	
	Concerning the evolution of OpenFlow-based SDN founded on the principles of network devices generality, distributed control plane and simple packet processing, in \cite{Zuo2014} they argue that three principles need to be satisfied to maintain vertical scalability:
	\begin{enumerate}
		\item{control functionalities should remain in the control plane domain, other than those which promote the efficiency of packet processing and adapt to the hardware and software requirements of the data plane,}
		\item{control functionalities cannot change the basic data plane processes,}
		\item{the collection of statistics in the data plane should not affect the accuracy and validity of the measurement and should not cause an increase in the control plane load.}
	\end{enumerate}
	Following these principles, it is proposed to offload the control plane from the control messages, by introducing a statistical server that closely cooperates with network devices in the data plane and submits collected statistics only at the request of the controller.
	
	\textit{\textbf{Stateful packet processing}} --- was used in StreaMon \cite{Bonola20172} to separate the program logic of the traffic analysis applications from elementary primitives implemented in the network device probes. StreaMon abstracts the measurement process through three phases:
	\begin{enumerate}
		\item{identification of the entity being monitored,}
		\item{measurement by applying efficiently-implemented primitives to the configurable fields of the packet header,}
		\item{making decisions using extended finite-state machines (XFSM).}
	\end{enumerate}
	The implementation of this abstraction is enabled by using a stream processing engine consisting of four layers: (1) a layer of events that parse the recorded packets, (2) a layer of metrics applied to parsed packets, (3) a layer of features in which different statistics are derived from the calculated metrics, and (4) a layer of decision in which the measurement application logic is executed.
	
	\textit{\textbf{Hybrid architecture}} --- which enables the introduction of software-defined counters by connecting ASIC to the general purpose processor and the cost-effective DRAM, and replacing the classic counters with small buffers, is proposed in \cite{Mogul2012}. The principle of work is the following. With the arrival of the packet, instead of incrementing the counter, ASIC creates a record of that event and adds it to the buffer. The buffer is divided into several blocks which, when filled, are transmitted to a general purpose processor. A general purpose processor based on the block content is updating the counters located in the DRAM.
	
	Similar to previous research, the software-defined measurement architecture, OpenSketch, which separates data plane measurement from the control plane, is proposed in \cite{Yu2013}. In data plane, OpenSketch provides a simple three-stage pipeline: hashing, filtering, and counting. A measurement library has been created, which enables the automatic configuration of the data plane pipeline and the switch memory allocation for each measurement task. The prototype is implemented on the NetFPGA platform, by inserting packet header parsing, hashing and lookup modules, and SRAM-based counters in the reference switch pipeline.
	
	\textit{\textbf{Data plane programming}} --- as a solution for network monitoring is proposed in \cite{Jeyakumar2013}. Within the proposed solution, there is a simple programmable interface which allows end nodes in the network to query and calculate over the switch memory using tiny packet programs (TPP). The TPPs are embedded in the packet headers and contain several instructions for reading, writing or performing arithmetic operations over SRAM data or processor registers. The pipeline of the proposed solution is based on an ASIC containing a TPP processor (TCPU) between L2/L3/TCAM tables and the memory of the output queues. TCPU is based on a RISC processor, which executes instructions in five stages: 
	\begin{enumerate}
		\item{fetching,}
		\item{decoding,}
		\item{executign,}
		\item{reading from memory,}
		\item{writing to memory.}
	\end{enumerate}
	Examples of supported statistics which can be obtained from the switch memory are: counters associated with L2/L3 flow tables, link utilization, number of received/sent/dropped bytes, queue size, etc.
	
\subsection{Security and reliability}
	Although the security \cite{Choi2010,Benzekki20162,Fiessler2016,Han2016} and reliability \cite{Kempf2012,Capone2015,Cascone20172,Petrucci2017} issues of the network are of different categories, they are commonly concerned with detection, isolation, and rapid resolution of problems that can degrade the performance or completely impair normal network operation. The SDN's data plane is attractive as a place where security and reliability issues are solved because it is the first target for potential threats to security or network reliability.
	
	Despite the fact that there was no change in the switch architecture, in \cite{Benzekki20162} it is shown that by relocating relatively straightforward access control operations to a data plane, the load of the SDN controller can be reduced, thereby increasing the scalability and security of the entire network.
	
	\textit{\textbf{New data plane architecture}} --- which enables fast detection of the impairments along the entire path, is proposed in \cite{Kempf2012}. It is shown that the data plane recovery can be achieved in less than 50 ms by the relocation of the connection monitoring from the control plane to the data plane. To do this, generators of monitoring messages, whose absence on the destination switch side can point to problems in the connection, were added to the OpenFlow switch architecture.
	
	\textit{\textbf{New packet classification mechanism}} --- as an OpenFlow extension to support detection and blocking of the distributed denial-of-service (DDoS) attack through content-oriented networking, is introduced in \cite{Choi2010}. In proposed extension an OpenFlow switch can respond to the requested content based on URL in requests. To achieve this, interceptor of the packet with the appropriate URL and the rate limiter were added to the hardware architecture of the OpenFlow switch.
	
	\textit{\textbf{Hybrid architecture}} --- for packet classification, called HyPaFilter, is proposed in \cite{Fiessler2016}. HyPaFilter exploits the advantages of parallel and massive hardware-based packet processing and the large inspection capabilities of software-based packet filters. It partitions user-defined policies for packet processing into simple parts executed on specialized hardware and complex parts run in the software. The firewall prototype is implemented by a combination of the NetFPGA-SUME hardware platform and \textit{netfilter/iptables} software on the Linux-based system. Packets which come to the proposed switching node are handled primarily on the hardware, and only in case of need for complex operation execution are redirected to the software.
	
	\textit{\textbf{Stateful packet processing}} --- was applied in \cite{Capone2015} for solving the problem of network failure. The ability of OpenState \cite{Bianchi20141} to respond to packet-level events has been utilized to determine the fast path recovery mechanisms for the relocation of flows affected by the network failure.
	
	The connection monitoring workspace, called StateMon, is proposed and implemented in \cite{Han2016}. To keep the data plane as simple as possible, only the table of an open connection, based on OpenFlow match-action abstraction, is added to the end of the switch pipeline. The rest of the logic, which is in charge of maintaining the global status table and the state management tables, is implemented in the controller. Due to the OpenFlow communication limitations, the switch has been extended with a new protocol for the open connections table programming. Using StateMon, the stateful firewall and the port knocking application were implemented.
	
	Because the current abstractions of the SDN's data plane, for the detection of network failures, did not provide the ability for fine tuning the detection mechanism in the switches, the new data plane design, called SPIDER, is proposed in \cite{Cascone20172}. SPIDER is a pipeline similar to OpenFlow which allows: (a) failure detection mechanisms based on periodic link checking on switches, (b) rapid redirection of traffic flows even in the case of remote failures, regardless of the availability of controllers. A model of flow structure based on the stateful data plane abstraction such as OpenState and P4, and the corresponding behavior model, are presented.
	
	The feasibility of using the abstraction of a programmable data plane in the \textit{iptables} offloading from the server processor to the smart network card is investigated in \cite{Petrucci2017}. On a smart network card, based on the NetFPGA-SUME platform, an open packet processor based on the XFSM has been implemented.
		
\subsection{Support for various network technologies}
	Since the inception of SDN idea, the research focused on Ethernet and IP network technologies. However, a significant need for the expansion of the SDN, to support other network technologies such as optical circuit switching (OCS), optical packet switching (OPS), gigabit passive optical networks (GPON), data over cable service (DOCSIS) and so on, has been noted. Some of the research focused on introducing reconfigurability into the data plane architecture, especially in the domain of the physical layer, while others opted for data plane abstraction to make it independent of lower-layer technology.
	
	\textit{\textbf{New structure of flow table}} --- is proposed in \cite{Kempf2011} as an extension of OpenFlow v1.0 to support MPLS technology. The proposed extension allows an OpenFlow switch without the ability to route IP traffic to forward MPLS traffic. This is accomplished by having three packet header modification operations implemented in the switching node of the data plane:
	\begin{enumerate}
		\item{push - adding new labels to the MPLS label stack,}
		\item{pop - removing labels from the MPLS label stack,}
		\item{swap - replacing the label at the top of MPLS label stack with a new label.}
	\end{enumerate}
	The prototype is implemented on the NetFPGA platform.
	
	An extension of OpenFlow v1.1 architecture to support switch management in multi-technological transport layers is presented in \cite{Shirazipour2012}. The circuit flow table, which is not used for search, but contains information about existing connections, is proposed. This enabled support for hybrid switches that perform both the circuit and the packet switching. The proposed extension can be used for a smooth migration to fully packet-optical integrated nodes where, for example, packet routers would contain reconfigurable optical add-drop multiplexers (ROADM).
	
	\textit{\textbf{Reconfigurable architecture}} --- called Open Transport Switch (OTS), which enables packet-optical cross-connections (XCON) and allocation of bandwidth on the optical element, is proposed in \cite{Sadasivarao2013}. OTS consists of the following building elements: (1) a discovery agent which is in charge of detecting and registering resources, (2) a control agent which is in charge of surveillance and propagation of alarms and notification to the controller, and (3) the data plane agent which is responsible for programming the data path of the network element. Data plane entities can be time slots, XCONs, or MPLS labels. OTS is integrated into the SDN framework using the OpenFlow protocol.
	
	The first demonstration of a fully-programmable space division multiplexing (SDM) optical network consisting of three architecture on demand (AoD) nodes interconnected by multi-core fibers (MCF) is presented in \cite{Amaya2013}. AoD nodes dynamically implement a node architecture based on the traffic requirements and consist of an optical backplane that interconnects MCF/single mode fiber (SMF) inputs, modules such as a spectral selective switch (SSS) or EDFA (erbium-doped fiber amplifiers) and MCF/SMF outputs. Later, a programmable FPGA-based optical switch and interface card (SIC) replacing traditional NIC and allowing direct interconnection of servers via optical top-of-the-rack (ToR) switches is proposed in \cite{Yan2015}. Additionally, SIC enables the aggregation of the OCS and the OPS within the AoD.
	
	The SDN-enabled OPS node for reconfigurable DCN is represented in \cite{Miao2015}. The OpenFlow protocol has been extended to support wavelength management, management of spatial and time switching elements, and flow management of the OPS node. The data plane of the OPS node is modular allowing the constant reconfiguration time (in the order of nanoseconds) regardless of the number of node ports. Optical flows generated by the ToRs contain an optical tag by which an FPGA-based controller performs an OPS table lookup and determines the OPS port to which flows are forwarded. Given the nature of statistical multiplexing, contention between input signals from the same ToR is possible. Therefore, between the FPGA-based OPS and ToR controller, bidirectional flow control with ACK and NACK signals has been established. To enable communication between OPS and OpenFlow controllers, an OpenFlow agent has been implemented.
	
	The SDN-based integration of time and wavelength division multiple access - passive optical metro networks (TWDM-PON) is demonstrated in \cite{Kondepu2015}. The proposed solution is based on a simple OpenFlow controller that runs on one node where network nodes are implemented using additional cards. The prototype, made of two optical network units and two optical service units of the 10Gbps Ethernet-based TWDM-PON, is implemented using two Altera FPGA Stratix IV GT chips. It has been experimentally demonstrated that such an architecture can achieve a reconfiguration time of the node in a data plane below 4 ms.
	
	The NEPHELE network architecture, featured in \cite{Bakopoulos2018}, has a scalable data plane built on verified commodity photonics technology. The NEPHELE data plane works in the time-division multiple access (TDMA) mode, where each slot is dynamically reserved for one communication on the rack to rack relation. Building blocks of the data plane are ToR and \textit{pod} switches. ToR switches connect devices within the datacenter rack and in their northern ports have specially-configured optical transceivers working in TDMA mode. \textit{Pod} switches, interconnected by the wavelength-division multiplexing (WDM) technology in the ring topology at the top of NEPHELE architecture, enable interconnection of all the ToR switches in the star topology. Within the \textit{pod} switches, switching is performed using the array waveguide grating routers. The integration of the data plane of NEPHELE architecture into the OpenFlow-based SDN architecture is enabled through the implementation of the support for three types of interaction:
	\begin{enumerate}
		\item{ability to advertise a device in the data plane (e.g., available wavelengths, active ports, available time slots),}
		\item{operational configuration of the device (e.g., adding records to the flow table, creating cross-connections),}
		\item{data plane monitoring, including asynchronous notifications and statistics download.}
	\end{enumerate}
	In a practical example, mentioned interactions are supported by implementing an SDN agent that acts as a proxy between the switches and the controller.
	
	\textit{\textbf{New data plane architecture}} --- called SplitArchitecture \cite{John2014}, allows flexible mapping of control plane layers to data plane layers in a hierarchically structured model. Additionally, SplitArchitecture separates the forwarding and processing functionalities of the data plane element. The ability to process data in the data plane enables the implementation of OAM functionality as support for tunneling in carrier networks (PPPoE, pseudo-wire emulation, etc.).
	
	A programmable and virtualizable all optic network infrastructure for DCN is presented in \cite{Saridis2015}. Its architecture is hybrid, and the data stream consists of SDN-enabled OPS and ToR switches connected to the SDN-enabled optical backplane with support for multicast circuit and packet switching. The optical backplane includes optical function blocks such as wavelength selective switches (WSS), OPSs and splitters, which can be dynamically configured to the arbitrary topology according to the requirements of a specific application. For intra-rack server interconnection, a programmable FPGA-based OPS/OCS hybrid NIC is used.
	
	Since OpenFlow specifications did not include optical layer constraints at the time, the use of hybrid switches was discussed in \cite{Elbers2016}. A typical core network transmits a combination of packet and circuit services. It is implemented as one of the following multilayer architectures:
	\begin{enumerate}
		\item{layered architecture in which the packet switching is positioned above the circuit switching,}
		\item{parallel architecture in which circuit and packet switching are located at the same level of the hierarchy,}
		\item{hybrid architecture in which a hybrid switch provides complete flexibility in the aggregation of circuit and packet switching on individual wavelengths.}
	\end{enumerate}
	This flexibility is enabled by the use of software-defined transceivers and the flexible network of ROADMs.
	
	One approach to the integration of optical networking devices into an OpenFlow-based SDN architecture is presented in \cite{Zhang2015,Xiong2018}. The architecture of the software-defined optical network (SDON) with a hierarchical data plane structure is proposed. The data plane is divided into four layers to achieve more accurate resource management. Integration was achieved by introducing transport controllers between data and control plane. The task of the transport controller is an abstraction of the optical network infrastructure to the unified controller as a set of virtual resources, using the techniques of inter-layer provisioning and resource adaptation.
	
	\textit{\textbf{Hardware abstraction layer (HAL)}} --- architecture, presented in \cite{Belter20141,Ogrodowczyk2014,Parniewicz2014}, consists of two parts: a cross-hardware platform (CHPL) and a hardware-specific layer (HSL). CHPL enables virtualization and realization of OpenFlow mechanisms, independent of the hardware platform below, with an efficient pipeline. The pipeline is processing packet abstractions rather than actual packets, where packet abstraction contains a reference to the actual packet stored in the network device memory. HSL is a set of hardware drivers which implement primitive network instructions, specific to different hardware platforms. Examples of HAL architecture adaptations are presented for three groups of network devices: (1) optical devices, (2) point to multi-point devices and (3) programmable platforms. Therefore, an example of the dense wavelength division multiplexing (DWDM) ROADM optical switch for optical devices is presented. In this example, the optical fiber specificities are abstracted by a cross-connection flow table containing information about the input port, the output port, and the wavelength at which the established communication is performed. 
	Additionally, an adaptation of the proposed HAL architecture for GPON technology is presented in \cite{Clegg2014}. 
	As an example one on multiple architectures, adaptation to the DOCSIS system is provided in \cite{Ogrodowczyk2014}. In this case, the entire data plane of the DOCSIS system consisting of the cable modem terminating system, cable modems and residential gateways is abstracted as a one aggregated switch. Since the DOCSIS platform is a closed system and configuration is only possible through standard interfaces, a DOCSIS proxy is introduced between the OpenFlow Controller and the DOCSIS system that performs described abstraction. Technical details of the DOCSIS proxy implementation are presented in \cite{Fuentes2014}. Adaptation of HAL architecture for programmable platforms is provided for an EZappliance platform, based on the EZchip NP-3 network processor, and the NetFPGA board.
	
	All-optical DCN architecture is presented in \cite{Kondepu2018}. The data plane of the proposed architecture consists of OCSs based on large-port-count fiber switches, above which the MCF switch is settled. MCF switch forwards traffic between data centers. Two layers of OCS combined with SMF and MCF devices, under SDN control, form a flat DCN architecture. Besides MCF switches, additional components (such as FPGA based high-speed TDM switches and NICs) are placed in the data plane to meet different requirements. Above the presented data plane, OpenStack-based virtualization has been established, which simplifies the process of provisioning and SDN-based network management.

\subsection{Network and network functions virtualization}
	The network functions virtualization (NFV) relies on host and network virtualization technologies, which enable the mapping of entire classes of network node functions to building blocks. Interconnecting (chaining) of building blocks creates complex communication services (e.g., firewalls, IDSs, traffic caches). Network functions virtualization complement the SDN so that it enables the organization of elements on the path through the data plane. Because of an unbreakable connection between NFV and SDN, research in the field of virtual networks and NFV has also implicated changes in the data plane of the SDN. An overview of these research is provided below.

	\textit{\textbf{Reconfigurable architecture}} --- for virtualized middlebox acceleration, called OpenANFV, is presented in \cite{Ge2014}.
	From top to bottom, OpenANFV's architecture consists of: (1) an OpenStack orchestration platform, (2) a virtual network function (VNF) controller, (3) a NFV infrastructure, and (4) a network functions acceleration platform (NFAP). In short, when a specific middlebox needs to be realized, each VNF that builds it is instantiated as a VM node. The VNF controller takes care of the resource management required by each VM. NFAP is implemented using a PCIe card that contains a FPGA chip with static and partially reconfigurable (PR) regions. Functionalities of packet classification and switching are implemented in the static region. When required, accelerators are implemented using PR and connect to a switch in a static region.
	
	The use of FPGAs as a platform for the NFV is proposed in \cite{Kachris2014}. In the proposed conceptual solution, FPGAs can be dynamically configured to provide hardware support to a specific application or another hardware. The platform should support hardware accelerators from different vendors, by defining a standard interface between these modules (NFV-IF). The NFV controller would perform the configuration of individual modules on the FPGA via the NFV configuration interface (NFV-Config-IF).

	\textit{\textbf{Data plane programming}} --- using P4 was used to create a portable virtualization platform HyPer4 \cite{Hancock2016}. HyPer4 consists of a P4 program, named \textit{persona}, running on a network device, a compiler and a data plane management unit. Persona is a generic P4 program that can be dynamically configured to mimic the functionality of other P4 programs through three phases:
	\begin{enumerate}
		\item{parsing and setup,}
		\item{match-action emulation,}
		\item{egress phase.}
	\end{enumerate}
	In the parsing and setup phase, a packet is received and prepared for the processing according to the specifications of the P4 program, which is emulated in the second stage. The output phase manages output-specific primitives and prepares the packet for sending. To be able to emulate arbitrary P4 programs, the persona supports:
	\begin{itemize}
		\item{programmable parsing,}
		\item{arbitrary definition of the packet field representation,}
		\item{matching to arbitrary fields,}
		\item{actions which can be complex collections of P4 primitives,}
		\item{virtual networking based on the recirculation of packets from one virtual device to another.}
	\end{itemize}
	As proof of the proposed concept, using HyPer4 are emulated L2 Ethernet switch, IPv4 router, ARP proxy and firewall.
	
	\textit{\textbf{New data plane architecture}} --- based on ForCES network elements with support for virtualization and programmability is proposed in \cite{Rong20162}. A virtualization support in ForCES network elements has been achieved by introducing virtual machine technology in CE and FE. The virtual ForCES router, called vForTER, is designed as a proof of the proposed concept. The vForTER architecture consists of virtual control elements (vCE), virtual forwarding elements (vFE) and a particular FE called switching element (SE). SE is in charge of managing the virtualization of CE and FEs and the internal scheduling of traffic between virtualized elements. Also, an FE algorithm for the dynamic allocation of FE resources is executed on the CE. In the proposed vForTER design, the data plane consists of one FE and two vFEs. vFE performs the processing of packets received from the SE, using functions provided by LFBs. The vForTER prototype was implemented using VMWare ESXi hypervisor and the Click modular router.
	
	A virtual filtering platform (VFP), based on a programmable virtual switch, is presented in \cite{Firestone2017}. The VFP is running on a Hyper-V extendable switch on the Microsoft Azure cloud platform and consists of match-action tables (MAT) and the packet processor. MATs are realized as layers that support configuration using a programming model and the operation with multiple controllers. The programming model is based on the configurable hierarchy of VFP objects: (1) ports on which the filtering policies are performed, (3) rules as MAT records, and (4) rule groups inside the single layer. By testing VFP on over a million hosts, over a period of 4 years, there have been several conclusions regarding data plane design:
	\begin{itemize}
		\item{the design should be conceived as a stateful from the beginning,}
		\item{a precise semantic of the forwarding table layering is needed,}
		\item{physical layer protocols need to be separated from the data plane,}
		\item{all operations should be modeled as actions (e.g., tunneling as encapsulation and decapsulation),}
		\item{forwarding should be kept simple,}
		\item{commodity NIC hardware is not ideal for the SDN.}
	\end{itemize}
	
	Virtual data planes, as described in \cite{Oguchi2017}, can be realized as software objects called virtual network objects (VNO), to which multiple network infrastructures can be mapped. The network slice can be created as needed by configuring the logical network via VNO, which liberates users from the need for expert knowledge in the field of virtual networks.
	
	\textit{\textbf{Hardware abstraction layer (HAL)}} --- model for data plane resources orchestration based on UNIFY BiS-BiS \cite{Sonkoly2015} is presented in \cite{Szabo2017}. In the proposed model, each hardware element which affects delay and bandwidth must be taken into account, e.g., processor cores, memory modules, physical or virtual network interfaces, etc. Fast and efficient resource orchestrator (FERO) generates an abstract model based on hardware infrastructure during a bootstrap process. Incoming network service requests are mapped to available resources using the generated graph model. The prototype of the proposed solution was implemented using Docker and various software switches with added support for DPDK.
	
	\begin{table*}[]
		\centering
		\caption{Correlation of treated problems and problem-solving approaches}
		\label{correlation}
		\begin{tabular}{|l|C{0.4cm}|C{0.4cm}|C{0.4cm}|C{0.4cm}|C{0.4cm}|C{0.4cm}|C{0.4cm}||L{4.5cm}|C{0.8cm}|}
			\hline
			\diagbox[width=5.5cm,height=5cm]{Problem-solving approach\\ \hskip 0pt}{\\Treated problem}  
			& \rotatebox{90}{\parbox{4.5cm}{\centering Performance}}
			& \rotatebox{90}{\parbox{4.5cm}{\centering Energy consumption}}
			& \rotatebox{90}{\parbox{4.5cm}{\centering Quality of service}}
			& \rotatebox{90}{\parbox{4.5cm}{\centering Measurement and monitoring}}
			& \rotatebox{90}{\parbox{4.5cm}{\centering Security and reliability}}
			& \rotatebox{90}{\parbox{4.5cm}{\centering Support for various\\network technologies}}
			& \rotatebox{90}{\parbox{4.5cm}{\centering Network and\\network functions virtualization}}
			& Key limitations of ForCES and OpenFlow data plane architectures
			& \rotatebox{90}{\parbox{4.5cm}{\centering Generalization of\\problem-solving approaches}}
			\\ \hline
			
			Data plane programming                     & & & & $\bullet$ & & & $\bullet$ & There is no ability to use the language to describe and program the data plane & A \\ \hline
			Stateful packet processing                 & $\bullet$ & & $\bullet$ & $\bullet$ & $\bullet$ & & & There is no ability for the state-aware or context-aware packet processing & B \\ \hline
			Reconfigurable architectures               & $\bullet$ & & & & & $\bullet$ & $\bullet$ & There is no ability for reconfiguration of functionality or data plane parameters & C \\ \hline
			Physical layer management                  & & $\bullet$ & & & & & & There is no ability for process management below the flow table lookup level & C \\ \hline
			New structures of flow tables              & $\bullet$ & $\bullet$ & $\bullet$ & & & $\bullet$ & & There is no ability to change the table flow structure (for example, when adding support for new protocols) & C/D \\ \hline
			New mechanisms for flow tables lookup      & $\bullet$ & $\bullet$ & & & & & & There is no ability to implement new algorithms for flow table lookup, or enhance existing ones & C/D \\ \hline
			New packet classification mechanisms       & $\bullet$ & & & & $\bullet$ & & & There is no ability for higher layer protocols' packets classification (e.g., TCP, HTTP, SIP) & B/C/D \\ \hline
			Hybrid architectures (SW-HW or HW-HW)      & $\bullet$ & & $\bullet$ & $\bullet$ & $\bullet$ & & & There is no ability to offload data plane processes from hardware to software or vice versa, and to accelerate processes with additional hardware & D \\ \hline
			New data plane architectures               & & & $\bullet$ & $\bullet$ & $\bullet$ & $\bullet$ & $\bullet$ & There are general limitations on the topology flexibility and data plane functionality & D \\ \hline
			Hardware abstraction layer                 & & $\bullet$ & $\bullet$ & & & $\bullet$ & $\bullet$ & There is no direct support for specific network technologies (e.g., DOCSIS, GPON, OCS/OPS, PPPoE) --- data plane implementations depend on the physical layer network technology or a target platform& D \\ \hline
		\end{tabular}
	\end{table*}

\subsection*{Key factors for data plane evolution}
	By reviewing the research which had addressed different problems by the categories listed in this section, we observed several common problem-solving approaches. Then we established a correlation between treated problems and problem-solving approaches. Afterwards, we identified the key limitations of ForCES and OpenFlow data plane architectures which have conditioned the specific problem-solving approach. Based on identified key limitations, we generalize the approaches to addressing the problem of programmability and flexibility of the SDN's data plane in four categories:
	\begin{enumerate}[label=\Alph*)]
		\item{data plane languages,}
		\item{stateful data plane,}
		\item{deeply programmable networks,}
		\item{new data plane architectures and abstractions.}
	\end{enumerate}
	Table~\ref{correlation} shows the correlation of treated problems and approaches to addressing an issue of programmability and flexibility of SDN's data plane, as well as a generalization of problem-solving approaches based on identified key limitations.

	From the motivations and results of the research presented in this section, we can conclude that the perceived limitations of ForCES- and OpenFlow-based data plane architectures cannot be relatively simply overcome. The programmability of the internal data plane structure is imperative for the future flexibility of the SDN from the aspects of flows, functions, resources and topology.
	
\section{Generic approaches to improve the data plane flexibility and programmability}
\label{sec:programmability}
	Before entering a critical review of the research which through the four generic approaches achieved the improvement of the programmability and flexibility of the SDN's data plane, we will make a brief recapitulation of the work done so far. In the first sections, an overview of the ForCES and OpenFlow data plane architectures, as well as its hardware- and software-based implementations, is provided. Afterwards, an overview of SDN-related research is presented, whose results implied the evolution of data plane, under which we meant a gradual deviation from the original architecture given by ForCES and OpenFlow specifications. By establishing a correlation between problems treated by SDN-related research and problem-solving approaches, we identified key limitations of ForCES and OpenFlow data plane architectures. Based on identified key constraints, we generalized approaches to addressing the problem of programmability and flexibility of the data plane, into four categories which represent the organizational units of this section.
		
\subsection{Data plane languages}
	By contemplating research which dealt with the definition of data plane language, we have recognized two categories:
	\begin{enumerate}
		\item{data plane description languages,}
		\item{data plane programming languages.}
	\end{enumerate}
	
	\subsubsection{Data plane description languages}
	Data plane description languages enable the description of the data plane structure and its components (e.g., an order of elements in the packet processing pipeline, parameters of elements). A review of the most relevant research from this category is given below.
	
	The high-level language for programming packet processors, called P4, is proposed in \cite{Bosshart2014}. When designing the P4 language, three goals have been set: (1) to enable on-demand reconfigurability of parsing and processing the packet, (2) to ensure protocol independence by enabling the specification of the packet header parsing and the match-action table, and (3) to achieve independence from the target platform. For the definition of P4 language, an abstract forwarding model was used in which the switch forwards packets to the pipeline through the programmable parser. The pipeline is composed of multiple match-action module stages, which can be arranged in a parallel, in a series or in a combination. The P4 program contains the following components:
	\begin{itemize}
		\item{packet header definition,}
		\item{parser definition,}
		\item{match-action tables,}
		\item{constructions of actions made of simple protocol-independent primitives,}
		\item{control programs which determine the order of application of the match-action tables on the packets.}
	\end{itemize}
	In short, by using the P4 language the programmer defines the packet processing mode in the data plane without taking into account the implementation details. The written P4 program is compiled into a table dependency graph which can be mapped to a specific software or hardware switch. The design of the P4 compiler for reconfigurable matching tables (RMT) \cite{Bosshart2013} and Intel FlexPipe programmable ASICs is presented in \cite{Jose2015}. A hardware abstraction, which is common for both chips, is defined to achieve the independence of the compiler from the target platform. The physical pipeline of the chip is modeled as a directed acyclical graph of processing stages, while the memory is abstracted with match tables which support records corresponding to the type of memory (e.g., TCAM as a ternary match table, SRAM as an exact match table).
	
	Considering existing P4 language constraints when describing the DCN data plane, the P4 language extension is proposed in \cite{Sivaraman20152}. Additional constructions were introduced: (1) cloning the packet, (2) rejecting the packet, (3) generating Digest, and (4) adding CRC-16 hash to a specific packet field. It has also been shown that existing P4 language constructions enable the description of a significant number of DCN switch functionalities.
	
	The PISCES, a P4 programmable software switch presented in \cite{Shahbaz2016}, is a modified version of OVS in which parsing, matching, and action execution codes are replaced with a C code generated by a P4 compiler. In order to customize the OVS function to the P4 principles, additional changes were made in the OPS: (1) support for arbitrary encapsulation and decapsulation was provided by adding two new primitives to manipulate the header of the packet, (2) support for conditioned action execution has been added, and (3) support to the generic verification and checksum update mechanism was provided. The PISCES compiler generates the source code of the OVS software switch based on the P4 program, which then needs to be compiled into executable files of the switch.
	
	 Since FPGA technology has become popular in prototype network hardware, the P4FPGA framework which enables compilation of the P4 program into the FPGA firmware is proposed in \cite{Wang2016,Wang20172}. P4FPGA generates the appropriate BSV code suitable both for simulation and synthesis, based on the P4 program. Generated BSV code contains a description of the P4 pipeline and additional supporting infrastructure. The supporting infrastructure includes FPGA memory management and the pipeline communication with other peripheral units on the target platform (e.g., PCIe interconnection, 10G Ethernet interface).
	
	In \cite{Wirbel2014}, Xilinx, one of the leading manufacturers of FPGA chips, has presented its contribution toward the implementation of reconfigurable network elements in both control and data plane in the form of the development environment SDNet. It allows the specification of packet switching node elements in a domain-specific language (e.g., P4), translation into hardware description languages (HDL), like VHDL and Verilog, and its synthesis and deployment on a broad range of FPGA and SoC chips. The basic features of SDNet environment include: 
	\begin{itemize}
		\item{generating custom hardware components to perform specific tasks (e.g., editing, searching, parsing),}
		\item{generating a specially customized data plane hardware subsystem according to the application or the user requirements,}
		\item{generating a particularly customized firmware for designed SDNet hardware architectures,}
		\item{generating testbench for validation and debugging.}
	\end{itemize}
	As an example of using the SDNet environment, the next generation NIC architecture, called the Softly Defined Line Card, has been presented. In the card architecture, the customized data plane consists of a packet processor and a programmable traffic controller. Also, IP cores for MAC, forward error correction (FEC), physical coding sublayer (PCS) and switching functionality can be used from the extensive network SmartCORE library. Other standard interfaces (e.g., external memory interfaces) are available from the LogiCORE IP library.
	
	An intermediate data plane language, called NetASM, is proposed in \cite{Shahbaz2015}. NetASM is conceived as a link between high-level languages (e.g., P4) and a diverse and growing set of hardware platforms. The instructions of NetASM languages are executed sequentially or concurrently on packet-related states, or on persistent states at the pipeline level. By introducing a packet-related state, it is possible to parallelize the pipeline. The NetASM language instruction set contains 23 instructions and enables: (1) data loading, (2) data storing, (3) calculation, (4) branching, (5) operations on the packet header, and (6) special operations such as hash and checksum.
	
	\subsubsection{Data plane programming languages}	
	On the other hand, data plane programming languages enable the implementation of new algorithms in the data plane (e.g., new packet scheduling mechanisms, new packet classification methods), and the most important representatives are Protocol-oblivious Forwarding (POF) \cite{Song2013}, and Domino \cite{Sivaraman20162}.
	
	POF does not need to know the packet format. Its only task is to extract the search keys from the header of the packet according to the controller's instructions, search the table, and then execute associated instructions. The instructions are given in the form of an executable compiled from the code written in the Flow Instruction Set (FIS) language. FIS instructions allow manipulation of packet headers, forwarding tables contents, and statistical counters. Thus, support for new protocols is provided without changes in the data plane.
	
	Domino is an imperative language with a syntax similar to C language and allows writing programs for packet processing using packet transactions. Packet transactions are inseparable sequential code blocks. The execution of Domino packet transactions is foreseen on the new machine model of the programmable line-rate switches, called Banzai. The Banzai machine architecture consists of an ingress and an egress pipeline. Parsing a packet is out of the scope of the Banzai model, and it is assumed those packet headers which are entering pipeline are already parsed. The pipeline consists of many stages that contain atoms, i.e., vectors of programmable units for packet header processing. The Domino compiler extracts the code fragments, which are executing atomically, from the description of the algorithm and maps them to the appropriate configuration of atoms in the Banzai machine.
	
	\subsubsection*{Remark on flexibility and programmability}
	In the first category --- data plane description languages --- P4 is particularly emphasized, which treated limited SDN's data plane flexibility by increasing the programmability of parsing, matching and action processes. Protocol independence was achieved by introducing reconfigurability in the packet parsing and processing, and target platform independence was achieved by hardware abstraction. However, in this way, some packet processing functionalities, such as queueing, scheduling and physical layer management, are neglected, limiting flexibility in term of forwarding function operation. As P4 does not envisage the use of hybrid architecture as a target platform, its flexibility in terms of function scaling and function placement is significantly limited.
	
	On the other hand, the alternative to the P4 language are hardware description languages, which provide high programmability but with a complicated and time-consuming development process. Although such a limitation can be partially overcome using compilers which translate domain-specific languages into HDL, domain-specific languages should retain sufficient details specific to target hardware which would allow the programmability of all data plane processes.
	
	The second category includes data plane programming languages in term of implementation of arbitrary packet processing algorithms directly in the data plane. Such languages achieve high flexibility in term of function operation through the programmability of all data plane processes. However, their dependency on the destination platform (specific packet processors or packet machines) limits the flexibility from the aspects of function scaling and placement.
	
	Therefore, as open problems which should be addressed by future research, we highlight the following:
	\begin{enumerate}
		\item{the lack of adequate constructs of current languages which would allow complete programmability of all data plane processes, and}
		\item{the inability of the existing languages to adequately abstract the details specific to the destination platform, which would ensure independence from the destination platform without losing the granularity in programming the data plane processes.}
	\end{enumerate}

\subsection{Stateful data plane}
	As already noted in the Section~\ref{sec:implications}, an approach based on the introduction of finite state machines into the data plane has been applied to solve a significant number of data plane problems. A review of the most important research aimed at generalizing the introduction of state-aware packet processing into the SDN's data plane is provided below.
	
	A stateful forwarding abstraction (SFA) in SDN's data plane is presented in \cite{Zhu2014}. The co-processor unit, called Forwarding Processor (FP), is implemented within the SDN switch using the CPU. Extended OpenFlow instructions are used to redirect packets or flows from flow tables to the FP, in which complex processing functionalities are implemented. This has enabled stateful network processing on the higher layers of the protocol stack.
	
	The new stateful datacentre architecture (SDPA) is proposed in \cite{Zhu2015} as a follow-up of the research presented in \cite{Zhu2014}. Unlike the standard match-action OpenFlow paradigm, a new "match-state-action" paradigm has been proposed in which state information is maintained within the data plane without the heavy involvement of the SDN controller. The SDN switch architecture, which supports SDPA in such a way that the FP, the state table and the policy module are added to the standard architecture of the SDN switch, is proposed. FP's task is to maintain the flow or the packet states. The policy module is used to customize and manage processing policies issued by the controller. Based on the proposed architecture, hardware and software prototypes of SDPA switches and application examples such as stateful firewall, defense against the domain name system (DNS) reflection attack and network address translation (NAT) functionality have been implemented.
	
	Although the OpenFlow architecture is limited regarding programmability within the switch, a non-trivial subclass of stateful control functions, which can be abstracted as Mealy's FSMs, and are already compatible with OpenFlow hardware from version 1.1 with minimal architectural changes, is advocated in \cite{Bianchi20142}. The proposed approach, OpenState \cite{Bianchi20141}, focuses on introducing programmable states and transition to the OpenFlow. The control logic uses packet-level events as triggers for the change of forwarding rules at wire speed inside the device itself. OpenState introduces the stateful block as an extension to a single flow table and can implement: (1) state table associated with flow identities, and (2) the XFSM table which performs a search based on the state label and packet header fields, and returns the associated forwarding action and the next state label. Stateful blocks can be chained into the pipeline with other stateful blocks as well as the classic OpenFlow tables. The proposed abstraction generalizes the OpenFlow match-action rules by using XFSMs which are directly executed within the switches, thus offloading controllers and creating abilities for complex control operations at the packet level in a fast data plane.
	
	The contribution to the improvement of the data plane programmability by introducing stateful packet processing within network switches is presented in \cite{Bianchi2016}. The proposed solution is called an Open Packet Processor (OPP), and its primary goal is to provide an ability for direct packet processing in the fast path, with efficient storage of flow state informations. OPP is based on the abstraction of the match-action phase of the OpenFlow using XFSM. The forwarding evolution is described by the FSM, where each state defines the forwarding policy, and the packet-level event initiates the transition to the next state.The workflow of the OPP architecture consists of four phases: (1) flow context table lookup, (2) conditions evaluation, (3) XFSMs execution, and (4) update of registers and flow context tables. The OPP prototype was implemented using the NetFPGA-SUME platform, where SRAM and TCAM were utilized for the flow context table and XFSM implementation, respectively. Later in \cite{Bonola20171}, the same authors have presented several cases of OPP usage in the implementation of advanced network functions:
	\begin{itemize}
		\item{packet forwarding based on load balancing,}
		\item{topology discovery based on L2 data,}
		\item{load balancing in the private network with static NAT.}
	\end{itemize}
	
	\subsubsection*{Remark on flexibility and programmability}
	The introduction of the match-state-action paradigm in the data plane has, through the increase in programmability of actions, improved the flexibility of the data plane from the aspect of the function operation. In the proposed solutions, the control logic is dependent on the state of the data plane and driven by packet-level events. Since the control plane does not have the ability to monitor and manage the state of the data plane, the flexibility from the aspect of the function placement is limited. Additionally, the lack of adequate synchronization of the data plane state reduces flexibility from the function scaling aspect.
	
	Although reviewed research has addressed the problem of introducing state-aware packet processing into the SDN's data plane adequately, we notice some key issues that have not been solved and which lead us back from the basic SDN idea --- managing the data plane from the control plane. Therefore, we highlight the following open issues:
	\begin{enumerate}
		\item{state monitoring and management,}
		\item{data plane state synchronization.}
	\end{enumerate}
	
\subsection{Deeply programmable networks}
	The deeply programmable network (DPN) is mentioned for the first time in \cite{Nakao2012}, and implies a step further to increase the data plane programmability in the SDN. The DPN's goal is to enable advanced packet processing functions such as caching, transcoding, and DPI, and to support new protocols, through increased data plane programmability below the flow table configuration level.
	
	\subsubsection{Original approach to the realization of DPN idea}
	Realization of the DPN idea has started through a VNode design which consists of slow and fast paths. The slow path represents a programming environment which consists of Intel Architecture servers, while the fast route performs network traffic processing using network processors. VNode relies on the generic router encapsulation (GRE) for tunneling of frames which come from different protocols. The extension of the VNode architecture with a physical node called Network ACommodation Equipment (NACE) is presented in \cite{Kanada2012}. NACE performs a dual role: (1) a gateway between a virtual network and an external Ethernet/virtual local area network (VLAN) network, and (2) a virtual switch between slices of the virtual network. To implement the data plane, NACE relies on 10Gbps Ethernet hardware including an L3 switch which supports VLAN and IP routing functions, and a service module card which performs conversion of packet formats between external and slice-internal formats. The internal format is based on GRE, according to VNode architecture, while the external format is based on the VLAN.
	Due to VNode's limitations regarding the complexity of fast path programming and the inability to handle other frames besides the conventional Ethernet frames, a new DPN node design inspired by the POF idea, called FLARE node, is proposed in \cite{Nakao2013}. FLARE can handle arbitrary protocol frames thanks to the PHY designed to support any frame. This means that there is no need for traffic tunneling which reduces overhead and increases network bandwidth and performance. The FLARE node architecture has supported the implementation of the multiple switching logic within a single node by dividing physical node resources into many fully programmable slivers, through the virtualization technology. Interfaces to physical ports of the node, a sliver management system, and a classification machine named packet slicer are implemented at the lowest level of architecture. The sliver management system enables dynamic installation and removal of slivers. The packet slicer performs fast packet scanning and its multiplexing or demultiplexing from or to slivers. A programmable control and data planes and virtual ports are available to the user within one sliver of the FLARE node. The data plane consists of a fast and slow path. The fast path of the FLARE node is implemented using a Click modular router, which runs as a multithreaded application on multi-core processors. The OpenFlow protocol support modules are implemented within a slow path of the FLARE node. Each sliver allows the implementation of arbitrary switch logic, e.g., one sliver can implement OpenFlow 1.0 and the other OpenFlow 1.3. The benefits of this approach are reflected in the abilities of instant replacement of the switch software, and the gradual upgrade of the network while maintaining compatibility with the legacy technologies. The general advantages of FLARE architecture are supporting the extension of SDN capabilities and the development of new protocols (e.g., non-IP protocols) in the research community.
	Later, improvements in VNode infrastructure using the FLARE node have been presented in \cite{Yamada2015,Nakao20153,Yamada2016,Minami2015}. Support for edge network virtualization and better resource management has been added. Different use cases of DPN-based NFV were also presented:
	\begin{itemize}
		\item{application-specific traffic control,}
		\item{smart M2M gateways,}
		\item{custom actions for OpenFlow switches,}
		\item{content/information-oriented networks.}
	\end{itemize}
	The use of the DPN in the application-specific slicing of the data plane of the mobile virtual network operators (MVNO) is shown in \cite{Nakao20152}. FLARE node is used for classification of traffic specific to the particular application or the device. In this way, it enables: (1) fine-grained QoS application, (2) traffic engineering based on the type of application, device, and other status information, (3) implementation of application-specific value added services, (4) support for intra-network security and parental control, and (5) improvement the bandwidth utilization based on statistical data on particular applications and devices. A context-aware IoT architecture based on the MVNO switch, as shown in \cite{Du2016}, has been built on the ground of this idea. In the proposed architecture, data collected through different sensors are transmitted over the IoT gateways to MVNO switches, which then forwards the received data to the central service controller for further processing. MVNO switches are realized as a slice of the FLARE node, and for packet forwarding are using information from the application layer content of the packet.
	As a fresh example of using the FLARE node, it is also worth mentioning the concept of network slicing in the 5G mobile network \cite{Nakao2017}. FLARE is used for the implementation of eNodeB and evolved packet core (EPC) nodes. eNodeB is running as a virtual machine instance within a FLARE slice. EPC is also implemented as a FLARE slice, where forwarding and processing of user data (e.g., Serving Gateway and Packet Data Network Gateway) are implemented in the data plane and signaling entities (e.g., Mobility Management Entity) in the control plane.
	
	The model of the L7 switch which forwards packets based on the application layer content using the regular expression, and offers a south-bound API for configuring regular expressions as needed during switch operation, is proposed in \cite{Ando2014}. The switch is realized by implementing two additional Click elements on the FLARE node: (1) L7Classifier and (2) L7Register. L7Classifier performs the identification of packet flows based on the IP address and TCP port number, and packet forwarding to the output interface or the L7Register element based on the content of the flow table. L7Register performs the matching of the packet contents of the particular flow with a regular expression. When there is matching with the regular expression in the L7Register element, an output interface is being updated in a flow table. By cascading multiple L7Register elements, it is possible to implement more complex processing of the application layer content.
	
	The TagFlow architecture for packet forwarding based on flow tags, which is implemented using a FLARE node, is proposed in \cite{Farhady2014}. TagFlow's architecture is based on a DCN which has two edges: the ingress edge and the egress edge. The ingress edge includes switches which connect the DCN's core with application servers and the egress edge is the point where packets leave DCN. The TagFlow working principle is as follows: (1) when the packet comes to the ingress edge, the edge switch performs the classification and tagging of the packet and forwards it to the next hop, (2) the packet is forwarded based on the tag to the egress edge, (2) the egress edge switch removes the tag from the packet and forwards it to the destination. The classification of the packet can be any application layer classification, where each traffic class can have its tag and be treated as a separate flow. The tag is added to the end of the packet (i.e., trailer tagging) to keep compatibility with current network technologies that are unaware that the tag exists. The tag-based packet forwarding is reduced to the forwarding table lookup on each core switch. The forwarding table contains ordered pairs <flow identifier, action> as records, with supported actions similar to those from the OpenFlow.
	
	The extension of TagFlow architectures with user-defined actions, designed to provide full flexibility and programmability of the SDN's data plane, is presented in \cite{Farhadi2014}. The network operator describes user-defined actions using high-level languages on commodity hardware. Implementations of user-defined actions on TagFlow and OpenFlow architectures were presented, and experimental comparison of its performance was performed. The TagFlow-based solution was 33\% faster than OpenFlow-based.
	
	The application of the DPN concept to improving the performance of DPI-based routing of applications transmitted via UDP traffic is presented in \cite{Nirasawa20161,Nirasawa20162}. The proposed solution is demonstrated on two UDP-based applications: (1) file reading and (2) database access. Unlike the standard methods which perform destination servers' load balancing using the round-robin method, the proposed solution conducts the forwarding of application requests to servers based on the content of the request. In the first case it is the file name, and in the second case, the database name. Later, an upgraded DPN switching node solution based on the FLARE node, for TCP-based applications, was presented in \cite{Nirasawa2017}. The demonstration was done on the HTTP application, where the DPN switching node mediated in communication between the client and the server. By analyzing the content of a client's request, when possible, the DPN node generates HTTP responses to the client without server engagement, thereby improving the performance of the application.
	
	\subsubsection{Other approaches to the realization of DPN idea}	
	Although the idea of DPN was developed by a group of researchers from the University of Tokyo, other researchers have also generated a significant contribution to this area.
	Thus SwitchBlade, a platform for the implementation of custom protocols on programmable hardware, is introduced in \cite{Anwer2010}. It enables the implementation of individual hardware modules on the fly without the need for hardware resynthesis. The SwitchBlade pipeline includes customizable hardware modules which implement the most common data plane functionalities. By combining these modules, it is possible to implement support for new protocols. For more complex tasks which can not be implemented on hardware, there is support for software processing based on packet rules or flow rules. Given the need to simultaneously run multiple protocols on the same hardware, the resource isolation mechanism was implemented by using separate forwarding tables. The prototype of the proposed solution is implemented on NetFPGA platform.
	
	Another example of DPN idea realization is a split SDN data plane (SSDP), presented in \cite{Narayanan2012} as a new switch architecture which combines non-flexible commodity switching chips and a deep programmable co-processor system to resolve the limitations of the SDN innovation potential. The SSDP architecture consists of two data paths: (1) a commodity switching chip which uses TCAM to forward aggregated flows, so-called macro-flows, and (2) NP units which perform micro-flows processing most often on L4 and higher layers. The SSDP prototype is implemented on the Dell PowerConnect 7024 platform which uses 24x1Gbps switching chips connected to the programmable subsystem (PS) via the XAUI interface. The PS is based on a 4-core microprocessor without interlocked pipeline stages (MIPS) running an OPS-based OpenFlow v1.0 software switch. The principle is that packets which do not match records in the TCAM of the switching chip are forwarded to the PS, where software flow tables are looked up. The OpenFlow controller determines where the particular flow will be written: in the TCAM on the hardware or in the software flow table.
	
	The solution presented in \cite{Risso2012} is based on the idea of an edge network node supporting user-driven data plane applications which can monitor and, if necessary, modify network traffic in transit. User-driven applications are intended to be executed within a network slice associated with the corresponding actor (e.g., network service provider, end user, content provider, etc.). The proposed edge network node architecture consists of the following components:
	\begin{itemize}
		\item{software switch,}
		\item{network hypervisor,}
		\item{controller,}
		\item{embedded web server,}
		\item{network gateway,}
		\item{management server.}
	\end{itemize}
	The software switch, implemented by OVS, performs packet forwarding based on the flow rules list. The FlowVisor-based network hypervisor performs virtualization and splitting of the network into so-called network slices, which allows connection of the node to multiple OpenFlow controllers. The controllers execute data plane applications on the encapsulated data obtained by the network hypervisor. Other components implement the functionalities of network node management and its connection to the rest of the network.
	
	The implementation of a customizable and programmable PC-based switch, called the NetOpen switch, which supports different traffic processing functionalities within a data plane, is presented in \cite{Kim2012}. The NetOpen switch architecture consists of interconnected processing modules and forwarding elements. The processing modules are working independently of the other modules, and by combining different modules, it is possible to create a data plane specially adapted for each traffic flow. Each processing module consists of a serially connected submodules: (1) pre-processing, (2) processing, and (3) post-processing. The forwarding element is responsible for the transfer of the packet between the network interfaces and the processing modules, following the service functionalities of the particular traffic flow. The topology and configuration of individual processing modules within the data plane are generated from the flow table. Hence, the flow table is used as a programmable interface between the data plane and the controller.
	
	The Scalable Programmable Packet Processing Platform (SP4) based on the software router is proposed in \cite{Gill2012}. The SP4 data plane architecture is a component-based pipeline supporting three types of components: (1) serial, in which packets are processed in strict FIFO order, (2) parallel, in which concurrent threads can process multiple packets, and (3) parallel, context-oriented components where packets are grouped by the context-key and within the group are processed according to the order of arrival in the system. The operation of the proposed solution is demonstrated in several use cases: analysis of simple mail transfer protocol (SMTP) traffic, interception of voice over IP (VoIP) calls, and DDoS attack detection.
	
	A new pipeline architecture of the switching chip based on the RMT is proposed in \cite{Bosshart2013}. RMT enables changes in the data plane without hardware modification. This is accomplished with the help of a minimal set of action primitives which specify how to handle the packet header in hardware through: (1) definition of arbitrary headers, (2) definition of arbitrary matching tables over arbitrary header fields, (3) definition of the packet header modification mode, and (4) maintenance of the states associated with packets. The architecture consists of a large number of stages of the physical pipeline, mapped by the logic stages of the RMT, and according to the resource needs of each logic stage. The components of the physical pipeline are the configurable parser, the configurable matching memory, and the configurable action machine. These components enable the implementation of arbitrary match-action processing of the packet.
	
	Inspired by the ideas of chemical reaction engineering, the realization of data plane functions using chemical algorithms (CA) was proposed in \cite{Monti2016,Monti2017}. It has been shown that CA can be easily and quickly modified and reprogrammed on FPGA hardware without the need for translation into an intermediate program or HDL code. By using CA, it is possible to give an expressive and straightforward representation of rule-based algorithms for network dynamics management, which is suitable for extending the functionalities of the SDN's data plane.
	
	\subsubsection*{Remark on flexibility and programmability}
	We believe that applying the DPN paradigm can solve a wide range of problems caused by insufficient programmability and flexibility of the SDN's data plane. For example, a deeply programmable node FLARE, using virtualization techniques, positively affects the flexibility from the aspect of function scaling. By introducing deep programmability into the Click-based software path, FLARE node achieves high flexibility in term of forwarding function operation.
	
	However, one group of researchers focused on the use of conventional servers eventually equipped with special network processors, while others decided to use FPGA technology. That partially limits the flexibility in term of the forwarding function placement.	Consequently, we claim that all reviewed DPN architectures have one common disadvantage --- they are not independent of the target platform, which remains an open problem for future research.
	
\subsection{New Data Plane Architectures and Abstractions}
	In the last category of generic approaches to improve the programmability and flexibility of the SDN's data plane, an overview of research which generated new architectures or abstractions of existing data plane architecture is given. Although approaches to creating new data plane level architectures are diverse, they share the same motivation --- overcoming limitations imposed by the original architecture of the OpenFlow-based SDN.
	
	\subsubsection{New data plane architectures}	
	The conceptual solution presented in \cite{Casado2012} is based on introducing a new component called \textit{network fabric}. Network fabric has been defined as a group of forwarding elements whose primary function is the transmission of the packet. The working principle of the proposed solution can be described in the following steps: (1) the source node sends packets to the input edge switch which, after providing network services, forwards packets to the network fabric, (2) the network fabric performs fast forwarding of packets to the egress edge switch, (3) the egress edge switch sends packets to the destination node. The network fabric is transparent to the end nodes. The management of edge switches and the network fabric is supported by the control plane.
		
	Several dynamic scenarios which illustrate the main challenges associated with the development of a framework for software-defined middleboxes are presented in \cite{Gember2012}. The representation of the middlebox state, the middlebox manipulation and the implementation of the control logic are listed as the main challenges. The abstraction which exploits the inherent mapping of the status to the value of the protocol header has been proposed for the representation of the middlebox state. The introduction of three basic operations has supported the manipulation of the middlebox state: (1) state retrieval, (2) state addition, and (3) state deletion. The control logic remains under the control of the SDN controller.
	
	In \cite{Fayazbakhsh2013} it is argued, through the use of practical application scenarios, that flow tracking is required to ensure the consistent implementation of network management policies in the presence of traffic dynamics. To achieve this, the extension of the SDN architecture has been proposed. In the proposed extension, middleboxes add tags to outgoing packets to provide the required context, and these tags are used by switches and other middleboxes for systematic implementation of the management policies. The context carried by the tag refers to middlebox-specific internal information which is critical for the implementation of management policies, such as the ratio of the number of hits and misses on the proxy or public-private NAT mapping. For the implementation of the proposed solution, switches must support the operations of inserting the tag into the header or the packet content and matching of the packet with the corresponding tag, while the purpose of the tag remains in charge of the controller.
	
	The SDN architecture which enables application-aware packet processing is presented in \cite{Mekky2014}. The basic idea is to intercept the packet before sending it to the controller and applying the application processing logic within the switch. The proposed solution has been implemented by extending the OVS with the support for special flow tables, so-called application tables, and application modules processing the traffic for which there is a matching within the application table. The pipeline of the OpenFlow data plane has been preserved with minor changes. Immediately after OpenFlow flow table lookup, packets for which there is no record in the flow table are forwarded to the additional application table lookup, instead of encapsulation and sending to the controller. Application modules can implement different functionalities such as firewall and context-aware load balancing.
	
	\subsubsection{Data plane abstractions}	
	On the other hand, the data plane abstraction motivation is established on the fact that a majority of data plane implementations depend on the physical layer network technology or the target hardware- or software-based platform. Some approaches to the data plane abstraction are presented below.
	
	New hardware abstraction, called Programmable Abstraction of Datapath (PAD), for various network devices such as access devices, network processors, and optical devices, is proposed in \cite{Belter20142}. PAD enabled data plane behavior programming through protocol and function header definitions, and by using generic byte-oriented operations. The PAD architecture consists of several functional elements: 
	\begin{itemize}
		\item{ports,}
		\item{search engines,}
		\item{search structures,}
		\item{execution engines,}
		\item{forwarding functions.}
	\end{itemize}
	The PAD's working principle is as follows: (1) the received packet is merged with metadata and forwarded to the search engine, (2) after a successful search, the packet metadata together with the search results are forwarded to the execution engine, (3) the execution engine calls the function on the packet based on the results of the preceding search, and finally (4) the packet is forwarded to the output port. Multiple packet processing through the above steps can be implemented using loopback ports. PAD configuration management is possible through the PAD API on the northbound interface. The PAD API functionalities are as follows: (1) retrieval of information about data plane capabilities supported by specific hardware, (2) managing search structures, functions, and network protocol definitions, and (3) adding and deleting records in search structures. Specially designed languages such as NetPDL and P4 can be used to define network protocols and operations running on the execution engine.
	
	The Network Abstraction Model (NAM) is proposed in \cite{Haleplidis20151} as a unique model for packet forwarding and network functionalities in SDN and NFV. The proposed model should enable the control, management, and orchestration of processes and network functionalities in the data plane, with the help of a unique protocol. It has been noted that the network device has a forwarding plane based on building blocks (BB) which together create required functions and the operating plane which is in charge of maintaining the state of the device. Based on the observed, ForCES was selected as the framework for the realization of the abstraction model. Additionally, the ForCES LFB-based model does not see the difference between a physical and a virtual device, which makes it convenient for NFV abstraction. The architecture of NAM is organized in such a way that BBs map to LFBs, and VNFs to one or more network devices.
	
	TableVisor, a transparent proxy layer which enables pipelined packet processing and extension of hardware flow tables using multiple hardware switches, is proposed in \cite{Geissler2017}. Pipelined packet processing is enabled by emulating single multi-table switch using multiple hardware switches. On the other hand, emulation of large hardware tables by combining TCAM memory from multiple switches is also supported.
	
	\subsubsection*{Remark on flexibility and programmability}
	
	New architectures, analyzed in this subsection, have tended in various ways to improve the flexibility of the data plane. In some research, a separation of the edge functions from the core forwarding functions influenced the flexibility from the aspect of function scaling. In other studies, by introducing a context into the packet processing process and using middleboxes, they improved flexibility in term of function operation.
	
	Also, different data plane abstractions have contributed to flexibility. Thus, hardware abstractions enabled greater flexibility in term of function scaling, while network functionalities abstractions had a positive effect on the flexibility from the aspect of the function operation.
	
	However, neither approach has given satisfactory improvements to all of the considered aspects of flexibility, which remains an open problem which should be addressed by future research.
	
	\Figure[t!](topskip=0pt, botskip=0pt, midskip=0pt)[clip, trim=0.5cm 7.5cm 0.5cm 7.2cm, width=1.0\textwidth]{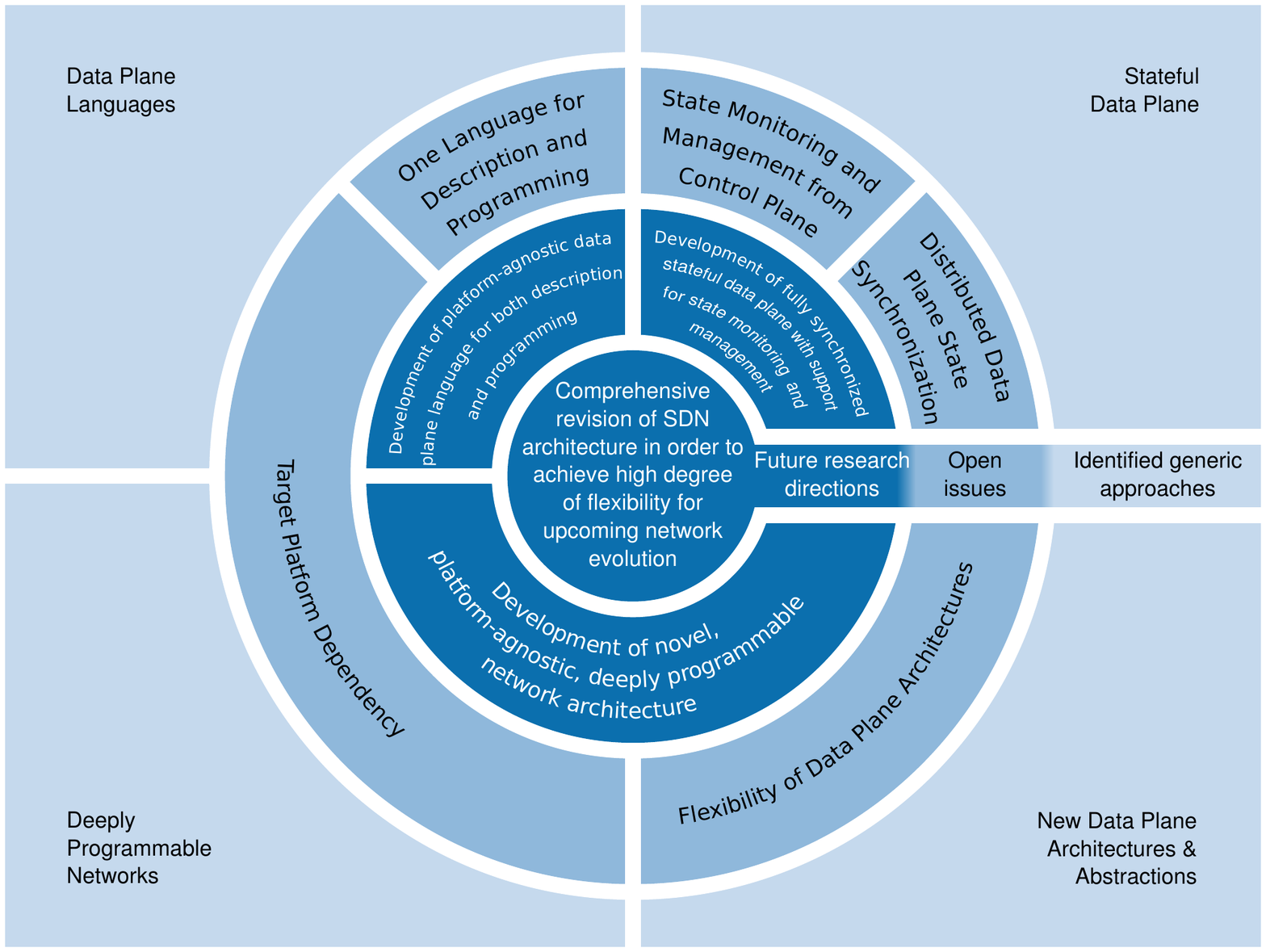}
	{The pursuit of novel data plane architecture through future research directions\label{fig:conclusion}}
	
\section{Future Research Directions and Conclusion}
\label{sec:futurework}
	Since previous surveys of the SDN-related research did not focus sufficiently on the data plane, this paper provides a comprehensive survey of the wired data plane in the SDN. The prerequisites for advancing the development of SDN's data plane are created through the proposal of future research directions which adequately address problems of programmability and flexibility. The complete process from identifying problems to the definition of future research directions, as shown in Figure~\ref{fig:conclusion}, is carried out through the following steps:
	\begin{enumerate}
		\item{An overview of actual SDN's data plane architectures is provided.}
		\item{An overview of software- and hardware-based supporting technologies which enabled data plane implementations is provided.}
		\item{A review of SDN-related research with the aim of identifying key factors influencing the data plane evolution is given.}
		\item{A critical review of generic approaches to improving the data plane flexibility and programmability is given.}
	\end{enumerate}
	
	In the realization of the first step, standardization of the SDN concept and architecture has been elaborated through the historical context. A particular attention is dedicated to the specification of ForCES architecture and to first attempts to realize SDN following that architecture such as NEon and Ethane. Since OpenFlow, albeit different from ForCES, emerged as a realization of the SDN idea, the most critical aspects of its architecture were compared with ForCES.
	
	Afterwards, an overview of the definitions of network flexibility and programmability and some general considerations of flexibility in other domains is given. Then, a review of the constraints of ForCES and OpenFlow-based data plane architectures, through the considered definitions and aspects of flexibility and programmability, is presented. Given that a lot of reviewed research is established on the experimental evaluation, an overview of hardware- and software-based technologies which served as good support for data plane implementation is given.
		
	To address problems of flexibility and programmability of the SDN's data plane, a lot of research generated solutions which have implicated the data plane evolution. The evolution of data plane means a gradual deviation from original data plane architectures given by ForCES and OpenFlow specifications. A comprehensive review of SDN-related research was made to identify the key factors influencing the data plane evolution. Then, the correlation between treated problems and problem-solving approaches is established and shown in the correlation table. Afterwards, key limitations of ForCES and OpenFlow data plane architectures are identified, which set the conditions for selecting a particular problem-solving approach. Based on identified key limitations and using subjective metric, approaches to address the problem of programmability and flexibility of SDN's data plane are generalized in four categories, as illustrated in the outer belt of the Figure~\ref{fig:conclusion}:
	\begin{enumerate}[label=\Alph*)]
		\item{data plane languages,}
		\item{stateful data plane,}
		\item{deeply programmable networks,}
		\item{new data plane architectures and abstractions.}
	\end{enumerate}
	Open issues are identified by a critical review of generic problem-solving approaches in terms of flexibility and programmability, as shown in the middle belt of Figure~\ref{fig:conclusion}. This established the ground for future research directions proposal (inner belt of the Figure~\ref{fig:conclusion}) as follows:
	\begin{itemize}
		\item{Development of platform-agnostic language for both description and programming of all processes in the data plane.}
		\item{Development of fully synchronized stateful data plane with support for state monitoring and management.}
		\item{Development of novel and platform-agnostic deeply programmable network architecture.}
	\end{itemize}
	In closing of this paper, it is important to emphasize that the research surveyed in this paper did not provide the complete solution to recognized problems. Since simple extensions cannot solve problems of programmability and flexibility of the existing data plane architectures, we advocate the idea of creating an entirely new SDN's data plane architecture which will provide a high degree of flexibility for the upcoming network evolution, as illustrated in the center of Figure~\ref{fig:conclusion}.

\section*{Acknowledgment}
	The authors are grateful to the Editor and anonymous reviewers that greatly helped to improve both the content and the readability of this paper.

\bibliography{Arxiv}{}

\begin{thebibliography}{100}
\providecommand{\url}[1]{#1}
\csname url@samestyle\endcsname
\providecommand{\newblock}{\relax}
\providecommand{\bibinfo}[2]{#2}
\providecommand{\BIBentrySTDinterwordspacing}{\spaceskip=0pt\relax}
\providecommand{\BIBentryALTinterwordstretchfactor}{4}
\providecommand{\BIBentryALTinterwordspacing}{\spaceskip=\fontdimen2\font plus
\BIBentryALTinterwordstretchfactor\fontdimen3\font minus
  \fontdimen4\font\relax}
\providecommand{\BIBforeignlanguage}[2]{{%
\expandafter\ifx\csname l@#1\endcsname\relax
\typeout{** WARNING: IEEEtran.bst: No hyphenation pattern has been}%
\typeout{** loaded for the language `#1'. Using the pattern for}%
\typeout{** the default language instead.}%
\else
\language=\csname l@#1\endcsname
\fi
#2}}
\providecommand{\BIBdecl}{\relax}
\BIBdecl

\bibitem{Leslie1991}
I.~Leslie and D.~McAuley, ``Fairisle: An {ATM} network for the local area,'' in
  \emph{Proceedings of the Conference on Communications Architecture \&Amp;
  Protocols}, ser. SIGCOMM '91.\hskip 1em plus 0.5em minus 0.4em\relax New
  York, NY, USA: ACM, 1991, pp. 327--.

\bibitem{Merwe1998}
J.~E. van~der Merwe, S.~Rooney, L.~Leslie, and S.~Crosby, ``The {Tempest} - a
  practical framework for network programmability,'' \emph{IEEE Network},
  vol.~12, no.~3, pp. 20--28, May 1998.

\bibitem{Tennenhouse1997}
D.~L. Tennenhouse, J.~M. Smith, W.~D. Sincoskie, D.~J. Wetherall, and G.~J.
  Minden, ``A survey of active network research,'' \emph{IEEE Communications
  Magazine}, vol.~35, no.~1, pp. 80--86, Jan 1997.

\bibitem{Campbell1999}
A.~T. Campbell, H.~G. De~Meer, M.~E. Kounavis, K.~Miki, J.~B. Vicente, and
  D.~Villela, ``A survey of programmable networks,'' \emph{SIGCOMM Comput.
  Commun. Rev.}, vol.~29, no.~2, pp. 7--23, Apr. 1999.

\bibitem{Feamster2014}
N.~Feamster, J.~Rexford, and E.~Zegura, ``The road to {SDN}: An intellectual
  history of programmable networks,'' \emph{SIGCOMM Comput. Commun. Rev.},
  vol.~44, no.~2, pp. 87--98, Apr. 2014.

\bibitem{Khosravi2003}
\BIBentryALTinterwordspacing
H.~M. Khosravi and T.~A. Anderson, ``Requirements for separation of {IP}
  control and forwarding,'' RFC 3654, Dec. 2003, accessed: 2018-10-12.
  [Online]. Available: \url{https://rfc-editor.org/rfc/rfc3654.txt}
\BIBentrySTDinterwordspacing

\bibitem{Yang2004}
\BIBentryALTinterwordspacing
L.~Yang, R.~Dantu, T.~Anderson, and R.~Gopal, ``Forwarding and control element
  separation ({ForCES}) framework,'' RFC 3746, Apr 2004, accessed: 2018-10-12.
  [Online]. Available: \url{https://rfc-editor.org/rfc/rfc3746.txt}
\BIBentrySTDinterwordspacing

\bibitem{McKeown2008}
N.~McKeown, T.~Anderson, H.~Balakrishnan, G.~Parulkar, L.~Peterson, J.~Rexford,
  S.~Shenker, and J.~Turner, ``{OpenFlow}: Enabling innovation in campus
  networks,'' \emph{SIGCOMM Comput. Commun. Rev.}, vol.~38, no.~2, pp. 69--74,
  Mar. 2008.

\bibitem{Dong2010}
\BIBentryALTinterwordspacing
L.~Dong, R.~Gopal, and J.~Halpern, ``Forwarding and control element separation
  ({ForCES}) protocol specification,'' RFC 5810, Mar 2010, accessed:
  2018-10-12. [Online]. Available: \url{https://rfc-editor.org/rfc/rfc5810.txt}
\BIBentrySTDinterwordspacing

\bibitem{Wang2012}
\BIBentryALTinterwordspacing
Z.~Wang, T.~Tsou, J.~Huang, X.~Shi, and X.~Yin, ``Analysis of comparisons
  between {OpenFlow} and {ForCES},'' Internet Draft, March 2012, accessed:
  2018-10-12. [Online]. Available:
  \url{https://tools.ietf.org/html/draft-wang-forces-compare-openflow-forces-01}
\BIBentrySTDinterwordspacing

\bibitem{Kreutz2015}
D.~Kreutz, F.~M.~V. Ramos, P.~E. Veríssimo, C.~E. Rothenberg, S.~Azodolmolky,
  and S.~Uhlig, ``Software-defined networking: A comprehensive survey,''
  \emph{Proceedings of the IEEE}, vol. 103, no.~1, pp. 14--76, Jan 2015.

\bibitem{Xia2015}
W.~Xia, Y.~Wen, C.~H. Foh, D.~Niyato, and H.~Xie, ``A survey on
  software-defined networking,'' \emph{IEEE Communications Surveys Tutorials},
  vol.~17, no.~1, pp. 27--51, 2015.

\bibitem{Farhady2015}
H.~Farhady, H.~Lee, and A.~Nakao, ``Software-defined networking: A survey,''
  \emph{Computer Networks}, vol.~81, no. Supplement C, pp. 79--95, 2015.

\bibitem{Masoudi2016}
R.~Masoudi and A.~Ghaffari, ``Software defined networks: A survey,''
  \emph{Journal of Network and Computer Applications}, vol.~67, no. Supplement
  C, pp. 1--25, 2016.

\bibitem{Cox2017}
J.~H. Cox, J.~Chung, S.~Donovan, J.~Ivey, R.~J. Clark, G.~Riley, and H.~L.
  Owen, ``Advancing software-defined networks: A survey,'' \emph{IEEE Access},
  vol.~5, pp. 25\,487--25\,526, 2017.

\bibitem{Akyildiz2014}
I.~F. Akyildiz, A.~Lee, P.~Wang, M.~Luo, and W.~Chou, ``A roadmap for traffic
  engineering in {SDN-OpenFlow} networks,'' \emph{Computer Networks}, vol.~71,
  no. Supplement C, pp. 1--30, 2014.

\bibitem{Akyildiz2016}
------, ``Research challenges for traffic engineering in software defined
  networks,'' \emph{IEEE Network}, vol.~30, no.~3, pp. 52--58, May 2016.

\bibitem{Karakus2017}
M.~Karakus and A.~Durresi, ``Quality of service ({QoS}) in software defined
  networking ({SDN}): A survey,'' \emph{Journal of Network and Computer
  Applications}, vol.~80, pp. 200--218, 2017.

\bibitem{Benzekki20161}
K.~Benzekki, A.~El~Fergougui, and A.~Elbelrhiti~Elalaoui, ``Software-defined
  networking ({SDN}): a survey,'' \emph{Security and Communication Networks},
  vol.~9, no.~18, pp. 5803--5833, 2016, sCN-16-0386.R1.

\bibitem{Bolla2011}
R.~Bolla, R.~Bruschi, F.~Davoli, and F.~Cucchietti, ``Energy efficiency in the
  future internet: A survey of existing approaches and trends in energy-aware
  fixed network infrastructures,'' \emph{IEEE Communications Surveys
  Tutorials}, vol.~13, no.~2, pp. 223--244, Second 2011.

\bibitem{Tuysuz2017}
M.~F. Tuysuz, Z.~K. Ankarali, and D.~Gözüpek, ``A survey on energy efficiency
  in software defined networks,'' \emph{Computer Networks}, vol. 113, no.
  Supplement C, pp. 188--204, 2017.

\bibitem{Anan2016}
M.~Anan, A.~Al-Fuqaha, N.~Nasser, T.-Y. Mu, and H.~Bustam, ``Empowering
  networking research and experimentation through software-defined
  networking,'' \emph{Journal of Network and Computer Applications}, vol.~70,
  no. Supplement C, pp. 140--155, 2016.

\bibitem{Kellerer2016}
W.~Kellerer, A.~Basta, and A.~Blenk, ``Using a flexibility measure for network
  design space analysis of sdn and nfv,'' in \emph{2016 IEEE Conference on
  Computer Communications Workshops (INFOCOM WKSHPS)}, April 2016, pp.
  423--428.

\bibitem{Kellerer2018}
W.~Kellerer, A.~Basta, P.~Babarczi, A.~Blenk, M.~He, M.~Klugel, and A.~M. Alba,
  ``How to measure network flexibility? a proposal for evaluating softwarized
  networks,'' \emph{IEEE Communications Magazine}, vol.~56, no.~10, pp.
  186--192, OCTOBER 2018.

\bibitem{He2019}
M.~He, A.~M. Alba, A.~Basta, A.~Blenk, and W.~Kellerer, ``Flexibility in
  softwarized networks: Classifications and research challenges,'' \emph{IEEE
  Communications Surveys Tutorials}, pp. 1--1, 2019.

\bibitem{Farhad2014}
H.~Farhad, H.~Lee, and A.~Nakao, ``Data plane programmability in {SDN},'' in
  \emph{2014 IEEE 22nd International Conference on Network Protocols}, Oct
  2014, pp. 583--588.

\bibitem{Nakao20151}
A.~NAKAO, ``Research and development on network virtualization technologies in
  japan,'' \emph{Journal of the National Institute of Information and
  Communications Technology}, vol.~62, no.~2, pp. 25--32, 2015.

\bibitem{Zilberman2015}
N.~Zilberman, P.~M. Watts, C.~Rotsos, and A.~W. Moore, ``Reconfigurable network
  systems and software-defined networking,'' \emph{Proceedings of the IEEE},
  vol. 103, no.~7, pp. 1102--1124, July 2015.

\bibitem{Dargahi2017}
T.~Dargahi, A.~Caponi, M.~Ambrosin, G.~Bianchi, and M.~Conti, ``A survey on the
  security of stateful {SDN} data planes,'' \emph{IEEE Communications Surveys
  Tutorials}, vol.~19, no.~3, pp. 1701--1725, thirdquarter 2017.

\bibitem{Bifulco2018}
R.~Bifulco and G.~R{\'e}tv{\'a}ri, ``A survey on the programmable data plane:
  Abstractions architectures and open problems,'' in \emph{Proc. IEEE HPSR},
  2018.

\bibitem{Cho2014}
H.~Cho, C.~Lai, T.~K. Shih, and H.~Chao, ``Integration of {SDR} and {SDN} for
  {5G},'' \emph{IEEE Access}, vol.~2, pp. 1196--1204, 2014.

\bibitem{Macedo2015}
D.~F. Macedo, D.~Guedes, L.~F.~M. Vieira, M.~A.~M. Vieira, and M.~Nogueira,
  ``Programmable networks - from software-defined radio to software-defined
  networking,'' \emph{IEEE Communications Surveys Tutorials}, vol.~17, no.~2,
  pp. 1102--1125, Secondquarter 2015.

\bibitem{Sun2015}
S.~Sun, M.~Kadoch, L.~Gong, and B.~Rong, ``Integrating network function
  virtualization with {SDR} and {SDN} for {4G/5G} networks,'' \emph{IEEE
  Network}, vol.~29, no.~3, pp. 54--59, May 2015.

\bibitem{Halpern2010}
\BIBentryALTinterwordspacing
J.~M. Halpern and J.~H. Salim, ``Forwarding and control element separation
  ({ForCES}) forwarding element model,'' RFC 5812, Mar. 2010, accessed:
  2018-10-12. [Online]. Available: \url{https://rfc-editor.org/rfc/rfc5812.txt}
\BIBentrySTDinterwordspacing

\bibitem{Dong2007}
L.~Dong, F.~Jia, and W.~Wang, ``Definition and implementation of logical
  function blocks compliant to {ForCES} specification,'' in \emph{2007 15th
  IEEE International Conference on Networks}, Nov 2007, pp. 531--536.

\bibitem{Ramirez2017}
P.~L.~G. Ram{\'\i}rez, J.~Lloret, S.~M. Cordero, and L.~C.~T. Arboleda,
  ``Design and implementation of {ForCES} protocol,'' \emph{Network Protocols
  and Algorithms}, vol.~9, no. 1-2, pp. 1--27, 2017.

\bibitem{Rong20161}
J.~Rong, H.~Xiong-xiong, and Z.~Guang-xin, ``A composition method of logical
  function block chain,'' \emph{International Journal of Future Generation
  Communication and Networking}, vol.~9, no.~3, pp. 57--68, 2016.

\bibitem{Haleplidis20152}
E.~Haleplidis, J.~H. Salim, J.~M. Halpern, S.~Hares, K.~Pentikousis, K.~Ogawa,
  W.~Wang, S.~Denazis, and O.~Koufopavlou, ``Network programmability with
  {ForCES},'' \emph{IEEE Communications Surveys Tutorials}, vol.~17, no.~3, pp.
  1423--1440, thirdquarter 2015.

\bibitem{Schuba2005}
C.~Schuba, J.~Goldschmidt, K.~Kalajan, and M.~F. Speer, ``Integrated network
  service processing using programmable network devices,'' Sun Microsystems,
  Mountain View, CA, USA, Tech. Rep., 2005.

\bibitem{Casado2007}
M.~Casado, M.~J. Freedman, J.~Pettit, J.~Luo, N.~McKeown, and S.~Shenker,
  ``Ethane: Taking control of the enterprise,'' \emph{SIGCOMM Comput. Commun.
  Rev.}, vol.~37, no.~4, pp. 1--12, Aug. 2007.

\bibitem{Haleplidis2012}
E.~Haleplidis, S.~Denazis, O.~Koufopavlou, J.~H. Salim, and J.~Halpern,
  ``Software-defined networking: Experimenting with the control to forwarding
  plane interface,'' in \emph{2012 European Workshop on Software Defined
  Networking}, Oct 2012, pp. 91--96.

\bibitem{Broniatowski2016}
D.~A. Broniatowski and J.~Moses, ``Measuring flexibility, descriptive
  complexity, and rework potential in generic system architectures,''
  \emph{Systems Engineering}, vol.~19, no.~3, pp. 207--221.

\bibitem{Nilchiani2007}
R.~Nilchiani and D.~E. Hastings, ``Measuring the value of flexibility in space
  systems: A six-element framework,'' \emph{Systems Engineering}, vol.~10,
  no.~1, pp. 26--44.

\bibitem{Broniatowski2017}
D.~A. Broniatowski, ``Flexibility due to abstraction and decomposition,''
  \emph{Systems Engineering}, vol.~20, no.~2, pp. 98--117.

\bibitem{Omer20092}
M.~Omer, R.~Nilchiani, and A.~Mostashari, ``Measuring the resilience of the
  trans-oceanic telecommunication cable system,'' \emph{IEEE Systems Journal},
  vol.~3, no.~3, pp. 295--303, Sept 2009.

\bibitem{Zobel2011}
C.~W. Zobel, ``Representing perceived tradeoffs in defining disaster
  resilience,'' \emph{Decision Support Systems}, vol.~50, no.~2, pp. 394 --
  403, 2011.

\bibitem{Filippini2014}
R.~Filippini and A.~Silva, ``A modeling framework for the resilience analysis
  of networked systems-of-systems based on functional dependencies,''
  \emph{Reliability Engineering \& System Safety}, vol. 125, pp. 82 -- 91,
  2014, special issue of selected articles from ESREL 2012.

\bibitem{Casado20051}
M.~Casado, G.~Watson, and N.~McKeown, ``Reconfigurable networking hardware: a
  classroom tool,'' in \emph{13th Symposium on High Performance Interconnects
  (HOTI'05)}, Aug 2005, pp. 151--157.

\bibitem{Casado20052}
------, ``Teaching networking hardware,'' in \emph{Proceedings of the 10th
  Annual SIGCSE Conference on Innovation and Technology in Computer Science
  Education}, ser. ITiCSE '05.\hskip 1em plus 0.5em minus 0.4em\relax New York,
  NY, USA: ACM, 2005, pp. 208--212.

\bibitem{Watson2006}
G.~Watson, N.~McKeown, and M.~Casado, ``{NetFPGA}: A tool for network research
  and education,'' in \emph{2nd workshop on Architectural Research using FPGA
  Platforms (WARFP)}, vol.~3, 2006.

\bibitem{Lockwood2007}
J.~W. Lockwood, N.~McKeown, G.~Watson, G.~Gibb, P.~Hartke, J.~Naous,
  R.~Raghuraman, and J.~Luo, ``{NetFPGA} - an open platform for gigabit-rate
  network switching and routing,'' in \emph{2007 IEEE International Conference
  on Microelectronic Systems Education (MSE'07)}, June 2007, pp. 160--161.

\bibitem{Luo2007}
J.~Luo, J.~Pettit, M.~Casado, J.~Lockwood, and N.~McKeown, ``Prototyping fast,
  simple, secure switches for ethane,'' in \emph{15th Annual IEEE Symposium on
  High-Performance Interconnects (HOTI 2007)}, Aug 2007, pp. 73--82.

\bibitem{Blott2010}
M.~Blott, J.~Ellithorpe, N.~McKeown, K.~Vissers, and H.~Zeng, ``{FPGA} research
  design platform fuels network advances.''

\bibitem{Zilberman2014}
N.~Zilberman, Y.~Audzevich, G.~A. Covington, and A.~W. Moore, ``{NetFPGA SUME}:
  Toward 100 {Gbps} as research commodity,'' \emph{IEEE Micro}, vol.~34, no.~5,
  pp. 32--41, Sept 2014.

\bibitem{Cao2015}
J.~Cao, X.~Zheng, L.~Sun, and J.~Jin, ``The development status and trend of
  {NetFPGA},'' in \emph{2015 International Conference on Network and
  Information Systems for Computers}, Jan 2015, pp. 101--105.

\bibitem{Varga2015}
P.~Varga, L.~Kovács, T.~Tóthfalusi, and P.~Orosz, ``{C-GEP}: 100 {Gbit/s}
  capable, {FPGA}-based, reconfigurable networking equipment,'' in \emph{2015
  IEEE 16th International Conference on High Performance Switching and Routing
  (HPSR)}, July 2015, pp. 1--6.

\bibitem{Naous20081}
J.~Naous, D.~Erickson, G.~A. Covington, G.~Appenzeller, and N.~McKeown,
  ``Implementing an {OpenFlow} switch on the {NetFPGA} platform,'' in
  \emph{Proceedings of the 4th ACM/IEEE Symposium on Architectures for
  Networking and Communications Systems}, ser. ANCS '08.\hskip 1em plus 0.5em
  minus 0.4em\relax New York, NY, USA: ACM, 2008, pp. 1--9.

\bibitem{Gibb2008}
G.~Gibb, J.~W. Lockwood, J.~Naous, P.~Hartke, and N.~McKeown, ``{NetFPGA} - an
  open platform for teaching how to build gigabit-rate network switches and
  routers,'' \emph{IEEE Transactions on Education}, vol.~51, no.~3, pp.
  364--369, Aug 2008.

\bibitem{Naous20082}
J.~Naous, G.~Gibb, S.~Bolouki, and N.~McKeown, ``{NetFPGA}: Reusable router
  architecture for experimental research,'' in \emph{Proceedings of the ACM
  Workshop on Programmable Routers for Extensible Services of Tomorrow}, ser.
  PRESTO '08.\hskip 1em plus 0.5em minus 0.4em\relax New York, NY, USA: ACM,
  2008, pp. 1--7.

\bibitem{Khan2013}
A.~Khan and N.~Dave, ``Enabling hardware exploration in software-defined
  networking: A flexible, portable {OpenFlow} switch,'' in \emph{2013 IEEE 21st
  Annual International Symposium on Field-Programmable Custom Computing
  Machines}, April 2013, pp. 145--148.

\bibitem{Gibb2010}
G.~Gibb and N.~McKeown, ``{OpenPipes}: Making distributed hardware systems
  easier,'' in \emph{2010 International Conference on Field-Programmable
  Technology}, Dec 2010, pp. 381--384.

\bibitem{Weerasinghe2015}
J.~Weerasinghe, F.~Abel, C.~Hagleitner, and A.~Herkersdorf, ``Enabling {FPGAs}
  in hyperscale data centers,'' in \emph{2015 IEEE 12th Intl Conf on Ubiquitous
  Intelligence and Computing and 2015 IEEE 12th Intl Conf on Autonomic and
  Trusted Computing and 2015 IEEE 15th Intl Conf on Scalable Computing and
  Communications and Its Associated Workshops (UIC-ATC-ScalCom)}, Aug 2015, pp.
  1078--1086.

\bibitem{Bitar2015}
A.~Bitar, M.~S. Abdelfattah, and V.~Betz, ``Bringing programmability to the
  data plane: Packet processing with a {NoC}-enhanced {FPGA},'' in \emph{2015
  International Conference on Field Programmable Technology (FPT)}, Dec 2015,
  pp. 24--31.

\bibitem{Antichi2013}
G.~Antichi, M.~Shahbaz, S.~Giordano, and A.~Moore, ``From {1G} to {10G}: Code
  reuse in action,'' in \emph{Proceedings of the First Edition Workshop on High
  Performance and Programmable Networking}, ser. HPPN '13.\hskip 1em plus 0.5em
  minus 0.4em\relax New York, NY, USA: ACM, 2013, pp. 31--38.

\bibitem{Chengchen2014}
C.~Hu, J.~Yang, H.~Zhao, and J.~Lu, ``Design of all programable innovation
  platform for software defined networking,'' in \emph{Presented as part of the
  Open Networking Summit 2014 ({ONS} 2014)}.\hskip 1em plus 0.5em minus
  0.4em\relax Santa Clara, CA: {USENIX}, 2014.

\bibitem{Zhou2014}
S.~Zhou, W.~Jiang, and V.~Prasanna, ``A programmable and scalable {OpenFlow}
  switch using heterogeneous {SoC} platforms,'' in \emph{Proceedings of the
  Third Workshop on Hot Topics in Software Defined Networking}, ser. HotSDN
  '14.\hskip 1em plus 0.5em minus 0.4em\relax New York, NY, USA: ACM, 2014, pp.
  239--240.

\bibitem{Lu2011}
G.~Lu, C.~Guo, Y.~Li, Z.~Zhou, T.~Yuan, H.~Wu, Y.~Xiong, R.~Gao, and Y.~Zhang,
  ``{ServerSwitch}: A programmable and high performance platform for data
  center networks,'' in \emph{Proceedings of the 8th USENIX Conference on
  Networked Systems Design and Implementation}, ser. NSDI'11.\hskip 1em plus
  0.5em minus 0.4em\relax Berkeley, CA, USA: USENIX Association, 2011, pp.
  15--28.

\bibitem{Wang2007}
W.~Wang, L.~Dong, B.~Zhuge, M.~Gao, F.~Jia, R.~Jin, J.~Yu, and X.~Wu, ``Design
  and implementation of an open programmable router compliant to {IETF}
  {ForCES} specifications,'' in \emph{Networking, 2007. ICN '07. Sixth
  International Conference on}, April 2007, pp. 82--82.

\bibitem{Rostami2012}
A.~Rostami, T.~Jungel, A.~Koepsel, H.~Woesner, and A.~Wolisz, ``{ORAN}:
  {OpenFlow} routers for academic networks,'' in \emph{2012 IEEE 13th
  International Conference on High Performance Switching and Routing}, June
  2012, pp. 216--222.

\bibitem{Kohler2000}
E.~Kohler, R.~Morris, B.~Chen, J.~Jannotti, and M.~F. Kaashoek, ``The {Click}
  modular router,'' \emph{ACM Trans. Comput. Syst.}, vol.~18, no.~3, pp.
  263--297, Aug. 2000.

\bibitem{Handley2003}
M.~Handley, O.~Hodson, and E.~Kohler, ``{XORP}: An open platform for network
  research,'' \emph{SIGCOMM Comput. Commun. Rev.}, vol.~33, no.~1, pp. 53--57,
  Jan. 2003.

\bibitem{Dobrescu2009}
M.~Dobrescu, N.~Egi, K.~Argyraki, B.-G. Chun, K.~Fall, G.~Iannaccone, A.~Knies,
  M.~Manesh, and S.~Ratnasamy, ``{RouteBricks}: Exploiting parallelism to scale
  software routers,'' in \emph{Proceedings of the ACM SIGOPS 22Nd Symposium on
  Operating Systems Principles}, ser. SOSP '09.\hskip 1em plus 0.5em minus
  0.4em\relax New York, NY, USA: ACM, 2009, pp. 15--28.

\bibitem{Mundada2009}
Y.~Mundada, R.~Sherwood, and N.~Feamster, ``An {OpenFlow} switch element for
  {Click},'' in \emph{in Symposium on Click Modular Router}, 2009.

\bibitem{Fernandes2014}
E.~L. Fernandes and C.~E. Rothenberg, ``{OpenFlow} 1.3 software switch,''
  \emph{Salao de Ferramentas do XXXII Simp{\'o}sio Brasileiro de Redes de
  Computadores e Sistemas Distribu{\i}dos SBRC}, pp. 1021--1028, 2014.

\bibitem{Risso2006}
F.~Risso and M.~Baldi, ``{NetPDL}: An extensible {XML}-based language for
  packet header description,'' \emph{Computer Networks}, vol.~50, no.~5, pp.
  688--706, 2006.

\bibitem{Kim2014}
H.~Kim, J.~Kim, and Y.~B. Ko, ``Developing a cost-effective {OpenFlow} testbed
  for small-scale software defined networking,'' in \emph{16th International
  Conference on Advanced Communication Technology}, Feb 2014, pp. 758--761.

\bibitem{OFREFIMPL}
\BIBentryALTinterwordspacing
``{OpenFlow} reference implementation,'' 2009, accessed: 2018-10-12. [Online].
  Available: \url{https://github.com/mininet/openflow}
\BIBentrySTDinterwordspacing

\bibitem{INDIGO}
\BIBentryALTinterwordspacing
``Indigo,'' 2013, accessed: 2018-10-12. [Online]. Available:
  \url{http://www.projectfloodlight.org/indigo/}
\BIBentrySTDinterwordspacing

\bibitem{PANTOU}
\BIBentryALTinterwordspacing
``Pantou,'' 2013, accessed: 2018-10-12. [Online]. Available:
  \url{https://github.com/CPqD/openflow-openwrt}
\BIBentrySTDinterwordspacing

\bibitem{OPENFAUCET}
\BIBentryALTinterwordspacing
``{OpenFaucet},'' 2012, accessed: 2018-10-12. [Online]. Available:
  \url{https://github.com/rlenglet/openfaucet}
\BIBentrySTDinterwordspacing

\bibitem{OFJAVA}
\BIBentryALTinterwordspacing
``{OpenFlow Java},'' 2013, accessed: 2018-10-12. [Online]. Available:
  \url{https://bitbucket.org/openflowj/openflowj/}
\BIBentrySTDinterwordspacing

\bibitem{OFLIBNODE}
\BIBentryALTinterwordspacing
``oflib-node,'' 2011, accessed: 2018-10-12. [Online]. Available:
  \url{https://github.com/TrafficLab/oflib-node}
\BIBentrySTDinterwordspacing

\bibitem{Fernandes2018}
E.~L. Fernandes, G.~Antichi, I.~Castro, and S.~Uhlig, ``An {SDN}-inspired model
  for faster network experimentation,'' in \emph{Proceedings of the 2018 ACM
  SIGSIM Conference on Principles of Advanced Discrete Simulation}, ser.
  SIGSIM-PADS '18.\hskip 1em plus 0.5em minus 0.4em\relax New York, NY, USA:
  ACM, 2018, pp. 29--32.

\bibitem{Lantz2010}
B.~Lantz, B.~Heller, and N.~McKeown, ``A network in a laptop: Rapid prototyping
  for software-defined networks,'' in \emph{Proceedings of the 9th ACM SIGCOMM
  Workshop on Hot Topics in Networks}, ser. Hotnets-IX.\hskip 1em plus 0.5em
  minus 0.4em\relax New York, NY, USA: ACM, 2010, pp. 19:1--19:6.

\bibitem{Ahmed2012}
M.~Ahmed, F.~Huici, and A.~Jahanpanah, ``Enabling dynamic network processing
  with {clickOS},'' in \emph{Proceedings of the ACM SIGCOMM 2012 Conference on
  Applications, Technologies, Architectures, and Protocols for Computer
  Communication}, ser. SIGCOMM '12.\hskip 1em plus 0.5em minus 0.4em\relax New
  York, NY, USA: ACM, 2012, pp. 293--294.

\bibitem{Rizzo20121}
L.~Rizzo, ``Netmap: A novel framework for fast packet i/o,'' in
  \emph{Proceedings of the 2012 USENIX Conference on Annual Technical
  Conference}, ser. USENIX ATC'12.\hskip 1em plus 0.5em minus 0.4em\relax
  Berkeley, CA, USA: USENIX Association, 2012, pp. 9--9.

\bibitem{Honda2015}
M.~Honda, F.~Huici, G.~Lettieri, and L.~Rizzo, ``{mSwitch}: A highly-scalable,
  modular software switch,'' in \emph{Proceedings of the 1st ACM SIGCOMM
  Symposium on Software Defined Networking Research}, ser. SOSR '15.\hskip 1em
  plus 0.5em minus 0.4em\relax New York, NY, USA: ACM, 2015, pp. 1:1--1:13.

\bibitem{Pfaff2015}
B.~Pfaff, J.~Pettit, T.~Koponen, E.~J. Jackson, A.~Zhou, J.~Rajahalme,
  J.~Gross, A.~Wang, J.~Stringer, P.~Shelar, K.~Amidon, and M.~Casado, ``The
  design and implementation of {Open vSwitch},'' in \emph{Proceedings of the
  12th USENIX Conference on Networked Systems Design and Implementation}, ser.
  NSDI'15.\hskip 1em plus 0.5em minus 0.4em\relax Berkeley, CA, USA: USENIX
  Association, 2015, pp. 117--130.

\bibitem{Han2010}
S.~Han, K.~Jang, K.~Park, and S.~Moon, ``{PacketShader}: A {GPU}-accelerated
  software router,'' \emph{SIGCOMM Comput. Commun. Rev.}, vol.~40, no.~4, pp.
  195--206, Aug. 2010.

\bibitem{Zhou2013}
D.~Zhou, B.~Fan, H.~Lim, M.~Kaminsky, and D.~G. Andersen, ``Scalable, high
  performance ethernet forwarding with {CuckooSwitch},'' in \emph{Proceedings
  of the Ninth ACM Conference on Emerging Networking Experiments and
  Technologies}, ser. CoNEXT '13.\hskip 1em plus 0.5em minus 0.4em\relax New
  York, NY, USA: ACM, 2013, pp. 97--108.

\bibitem{DPDK}
\BIBentryALTinterwordspacing
Intel, ``Data plane development kit,'' 2010, accessed: 2018-10-12. [Online].
  Available: \url{https://software.intel.com/en-us/networking/dpdk}
\BIBentrySTDinterwordspacing

\bibitem{Pongracz20131}
G.~Pongrácz, L.~Molnár, and Z.~L. Kis, ``Removing roadblocks from {SDN}:
  {OpenFlow} software switch performance on {Intel DPDK},'' in \emph{2013
  Second European Workshop on Software Defined Networks}, Oct 2013, pp. 62--67.

\bibitem{Keinanen2009}
J.~Kein{\"a}nen, P.~Jokela, and K.~Slavov, ``Implementing {zFilter} based
  forwarding node on a {NetFPGA},'' in \emph{Proc. of NetFPGA Developers
  Workshop}, 2009.

\bibitem{Antichi2011}
G.~Antichi, A.~D. Pietro, S.~Giordano, G.~Procissi, and D.~Ficara, ``Design and
  development of an {OpenFlow} compliant smart gigabit switch,'' in \emph{2011
  IEEE Global Telecommunications Conference - GLOBECOM 2011}, Dec 2011, pp.
  1--5.

\bibitem{Curtis2011}
A.~R. Curtis, J.~C. Mogul, J.~Tourrilhes, P.~Yalagandula, P.~Sharma, and
  S.~Banerjee, ``{DevoFlow}: Scaling flow management for high-performance
  networks,'' \emph{SIGCOMM Comput. Commun. Rev.}, vol.~41, no.~4, pp.
  254--265, Aug. 2011.

\bibitem{Ferkouss2011}
O.~E. Ferkouss, I.~Snaiki, O.~Mounaouar, H.~Dahmouni, R.~B. Ali, Y.~Lemieux,
  and C.~Omar, ``A {100Gig} network processor platform for {OpenFlow},'' in
  \emph{2011 7th International Conference on Network and Service Management},
  Oct 2011, pp. 1--4.

\bibitem{Qu2013}
Y.~R. Qu, S.~Zhou, and V.~K. Prasanna, ``High-performance architecture for
  dynamically updatable packet classification on {FPGA},'' in
  \emph{Architectures for Networking and Communications Systems}, Oct 2013, pp.
  125--136.

\bibitem{Martinello2014}
M.~Martinello, M.~R.~N. Ribeiro, R.~E.~Z. de~Oliveira, and R.~de~Angelis~Vitoi,
  ``{KeyFlow}: A prototype for evolving {SDN} toward core network fabrics,''
  \emph{IEEE Network}, vol.~28, no.~2, pp. 12--19, March 2014.

\bibitem{Perez20141}
K.~G. Pérez, X.~Yang, S.~Scott-Hayward, and S.~Sezer, ``A configurable packet
  classification architecture for software-defined networking,'' in \emph{2014
  27th IEEE International System-on-Chip Conference (SOCC)}, Sept 2014, pp.
  353--358.

\bibitem{Perez20142}
------, ``Optimized packet classification for software-defined networking,'' in
  \emph{2014 IEEE International Conference on Communications (ICC)}, June 2014,
  pp. 859--864.

\bibitem{Yan2014}
B.~Yan, Y.~Xu, H.~Xing, K.~Xi, and H.~J. Chao, ``{CAB}: A reactive wildcard
  rule caching system for software-defined networks,'' in \emph{Proceedings of
  the Third Workshop on Hot Topics in Software Defined Networking}, ser. HotSDN
  '14.\hskip 1em plus 0.5em minus 0.4em\relax New York, NY, USA: ACM, 2014, pp.
  163--168.

\bibitem{Yanbiao2014}
Y.~Li, D.~Zhang, K.~Huang, D.~He, and W.~Long, ``A memory-efficient parallel
  routing lookup model with fast updates,'' \emph{Computer Communications},
  vol.~38, no. Supplement C, pp. 60--71, 2014.

\bibitem{Kalyaev2015}
A.~Kalyaev, I.~Korovin, M.~Khisamutdinov, G.~Schaefer, and M.~A.~R. Ahad, ``A
  hardware approach for organisation of software defined network switches based
  on {FPGA},'' in \emph{2015 International Conference on Informatics,
  Electronics Vision (ICIEV)}, June 2015, pp. 1--4.

\bibitem{Ciesla2009}
M.~Ciesla, V.~Sivaraman, and A.~Seneviratne, ``{URL} extraction on the
  {NetFPGA} reference router,'' in \emph{NetFPGA Developers Workshop}, 2009.

\bibitem{Luo2009}
Y.~Luo, P.~Cascon, E.~Murray, and J.~Ortega, ``Accelerating {OpenFlow}
  switching with network processors,'' in \emph{Proceedings of the 5th ACM/IEEE
  Symposium on Architectures for Networking and Communications Systems}, ser.
  ANCS '09.\hskip 1em plus 0.5em minus 0.4em\relax New York, NY, USA: ACM,
  2009, pp. 70--71.

\bibitem{Ram2010}
K.~K. Ram, J.~Mudigonda, A.~L. Cox, S.~Rixner, P.~Ranganathan, and J.~R.
  Santos, ``{sNICh}: Efficient last hop networking in the data center,'' in
  \emph{Proceedings of the 6th ACM/IEEE Symposium on Architectures for
  Networking and Communications Systems}, ser. ANCS '10.\hskip 1em plus 0.5em
  minus 0.4em\relax New York, NY, USA: ACM, 2010, pp. 26:1--26:12.

\bibitem{Tanyingyong2010}
V.~Tanyingyong, M.~Hidell, and P.~Sj\"{o}din, ``Improving {PC}-based {OpenFlow}
  switching performance,'' in \emph{Proceedings of the 6th ACM/IEEE Symposium
  on Architectures for Networking and Communications Systems}, ser. ANCS
  '10.\hskip 1em plus 0.5em minus 0.4em\relax New York, NY, USA: ACM, 2010, pp.
  13:1--13:2.

\bibitem{Tanyingyong2011}
V.~Tanyingyong, M.~Hidell, and P.~Sjödin, ``Using hardware classification to
  improve {PC}-based {OpenFlow} switching,'' in \emph{2011 IEEE 12th
  International Conference on High Performance Switching and Routing}, July
  2011, pp. 215--221.

\bibitem{Gao2012}
S.~Gao, S.~Shimizu, S.~Okamoto, and N.~Yamanaka, ``A high-speed routing engine
  for software defined network,'' \emph{Journal of Selected Areas in
  Telecommunications (JSAT)}, pp. 1--7, 2012.

\bibitem{Lu2012}
G.~Lu, R.~Miao, Y.~Xiong, and C.~Guo, ``Using {CPU} as a traffic co-processing
  unit in commodity switches,'' in \emph{Proceedings of the First Workshop on
  Hot Topics in Software Defined Networks}, ser. HotSDN '12.\hskip 1em plus
  0.5em minus 0.4em\relax New York, NY, USA: ACM, 2012, pp. 31--36.

\bibitem{Katta2014}
N.~Katta, O.~Alipourfard, J.~Rexford, and D.~Walker, ``Infinite {CacheFlow} in
  software-defined networks,'' in \emph{Proceedings of the Third Workshop on
  Hot Topics in Software Defined Networking}, ser. HotSDN '14.\hskip 1em plus
  0.5em minus 0.4em\relax New York, NY, USA: ACM, 2014, pp. 175--180.

\bibitem{Vencioneck2014}
R.~D. Vencioneck, G.~Vassoler, M.~Martinello, M.~R.~N. Ribeiro, and
  C.~Marcondes, ``{FlexForward}: Enabling an {SDN} manageable forwarding engine
  in {Open vSwitch},'' in \emph{10th International Conference on Network and
  Service Management (CNSM) and Workshop}, Nov 2014, pp. 296--299.

\bibitem{Dang2015}
H.~T. Dang, D.~Sciascia, M.~Canini, F.~Pedone, and R.~Soul{\'e}, ``{NetPaxos}:
  Consensus at network speed,'' in \emph{Proceedings of the 1st ACM SIGCOMM
  Symposium on Software Defined Networking Research}, ser. SOSR '15.\hskip 1em
  plus 0.5em minus 0.4em\relax New York, NY, USA: ACM, 2015, pp. 5:1--5:7.

\bibitem{Bifulco2015}
R.~Bifulco and A.~Matsiuk, ``Towards scalable {SDN} switches: Enabling faster
  flow table entries installation,'' in \emph{Proceedings of the 2015 ACM
  Conference on Special Interest Group on Data Communication}, ser. SIGCOMM
  '15.\hskip 1em plus 0.5em minus 0.4em\relax New York, NY, USA: ACM, 2015, pp.
  343--344.

\bibitem{Sanvito2017}
D.~Sanvito, D.~Moro, and A.~Capone, ``Towards traffic classification offloading
  to stateful {SDN} data planes,'' in \emph{2017 IEEE Conference on Network
  Softwarization (NetSoft)}, July 2017, pp. 1--4.

\bibitem{Bhuyan1984}
L.~N. Bhuyan and D.~P. Agrawal, ``Generalized hypercube and hyperbus structures
  for a computer network,'' \emph{IEEE Trans. Comput.}, vol.~33, no.~4, pp.
  323--333, Apr. 1984.

\bibitem{Kannan2013}
K.~Kannan and S.~Banerjee, \emph{Compact {TCAM}: Flow Entry Compaction in
  {TCAM} for Power Aware {SDN}}.\hskip 1em plus 0.5em minus 0.4em\relax Berlin,
  Heidelberg: Springer Berlin Heidelberg, 2013, pp. 439--444.

\bibitem{Banerjee2014}
S.~Banerjee and K.~Kannan, ``{Tag-In-Tag}: Efficient flow table management in
  {SDN} switches,'' in \emph{10th International Conference on Network and
  Service Management (CNSM) and Workshop}, Nov 2014, pp. 109--117.

\bibitem{Congdon2014}
P.~T. Congdon, P.~Mohapatra, M.~Farrens, and V.~Akella, ``Simultaneously
  reducing latency and power consumption in {OpenFlow} switches,''
  \emph{IEEE/ACM Trans. Netw.}, vol.~22, no.~3, pp. 1007--1020, Jun. 2014.

\bibitem{Tran2012}
T.~H. Vu, P.~N. Nam, T.~Thanh, L.~T. Hung, L.~A. Van, N.~D. Linh, T.~D. Thien,
  and N.~H. Thanh, ``Power aware {OpenFlow} switch extension for energy saving
  in data centers,'' in \emph{The 2012 International Conference on Advanced
  Technologies for Communications}, Oct 2012, pp. 309--313.

\bibitem{Bolla2013}
R.~Bolla, R.~Bruschi, F.~Davoli, L.~D. Gregorio, P.~Donadio, L.~Fialho,
  M.~Collier, A.~Lombardo, D.~R. Recupero, and T.~Szemethy, ``The green
  abstraction layer: A standard power-management interface for next-generation
  network devices,'' \emph{IEEE Internet Computing}, vol.~17, no.~2, pp.
  82--86, March 2013.

\bibitem{NamSeok2013}
N.-S. KO, H.~HEO, J.-D. PARK, and H.-S. PARK, ``{OpenQFlow}: Scalable
  {OpenFlow} with flow-based {QoS},'' \emph{IEICE Transactions on
  Communications}, vol. E96.B, no.~2, pp. 479--488, 2013.

\bibitem{Sonkoly2012}
B.~Sonkoly, A.~Gulyás, F.~Németh, J.~Czentye, K.~Kurucz, B.~Novák, and
  G.~Vaszkun, ``On {QoS} support to {Ofelia} and {OpenFlow},'' in \emph{2012
  European Workshop on Software Defined Networking}, Oct 2012, pp. 109--113.

\bibitem{Melazzi2012}
N.~B. Melazzi, A.~Detti, G.~Mazza, G.~Morabito, S.~Salsano, and L.~Veltri, ``An
  {OpenFlow}-based testbed for information centric networking,'' in \emph{2012
  Future Network Mobile Summit (FutureNetw)}, July 2012, pp. 1--9.

\bibitem{OFCONFIG}
\BIBentryALTinterwordspacing
``{OpenFlow} configuration and managementgement protocol {OF-CONFIG} 1.0,''
  June 2012, accessed: 2018-10-12. [Online]. Available:
  \url{https://www.opennetworking.org/wp-content/uploads/2013/02/of-config1dot0-final.pdf}
\BIBentrySTDinterwordspacing

\bibitem{Enns2011}
\BIBentryALTinterwordspacing
R.~Enns, M.~Björklund, A.~Bierman, and J.~Schönwälder, ``Network
  configuration protocol ({NETCONF}),'' RFC 6241, Jun. 2011, accessed:
  2018-10-12. [Online]. Available: \url{https://rfc-editor.org/rfc/rfc6241.txt}
\BIBentrySTDinterwordspacing

\bibitem{Jain2013}
S.~Jain, A.~Kumar, S.~Mandal, J.~Ong, L.~Poutievski, A.~Singh, S.~Venkata,
  J.~Wanderer, J.~Zhou, M.~Zhu, J.~Zolla, U.~H\"{o}lzle, S.~Stuart, and
  A.~Vahdat, ``B4: Experience with a globally-deployed software defined wan,''
  \emph{SIGCOMM Comput. Commun. Rev.}, vol.~43, no.~4, pp. 3--14, Aug. 2013.

\bibitem{Palma2014}
D.~Palma, J.~Gonçalves, B.~Sousa, L.~Cordeiro, P.~Simoes, S.~Sharma, and
  D.~Staessens, ``The {QueuePusher}: Enabling queue management in {OpenFlow},''
  in \emph{2014 Third European Workshop on Software Defined Networks}, Sept
  2014, pp. 125--126.

\bibitem{Caba2015}
C.~Caba and J.~Soler, ``{APIs} for {QoS} configuration in software defined
  networks,'' in \emph{Proceedings of the 2015 1st IEEE Conference on Network
  Softwarization (NetSoft)}, April 2015, pp. 1--5.

\bibitem{Sivaraman2013}
A.~Sivaraman, K.~Winstein, S.~Subramanian, and H.~Balakrishnan, ``No silver
  bullet: Extending {SDN} to the data plane,'' in \emph{Proceedings of the
  Twelfth ACM Workshop on Hot Topics in Networks}, ser. HotNets-XII.\hskip 1em
  plus 0.5em minus 0.4em\relax New York, NY, USA: ACM, 2013, pp. 19:1--19:7.

\bibitem{Szymanski2015}
T.~H. Szymanski and M.~Rezaee, ``An {FPGA} controller for deterministic
  guaranteed-rate optical packet switching,'' in \emph{2015 IFIP/IEEE
  International Symposium on Integrated Network Management (IM)}, May 2015, pp.
  1177--1183.

\bibitem{Wang2014}
W.~Wang, Q.~Qi, X.~Gong, Y.~Hu, and X.~Que, ``Autonomic {QoS} management
  mechanism in software defined network,'' \emph{China Communications},
  vol.~11, no.~7, pp. 13--23, July 2014.

\bibitem{Wang2015}
W.~Wang, Y.~T. X. G.~Q. Qi, and Y.~Hu, ``Software defined autonomic {QoS} model
  for future internet,'' \emph{Journal of Systems and Software}, vol. 110, no.
  Supplement C, pp. 122--135, 2015.

\bibitem{Wette2013}
P.~Wette and H.~Karl, ``Which flows are hiding behind my wildcard rule?: Adding
  packet sampling to {OpenFlow},'' \emph{SIGCOMM Comput. Commun. Rev.},
  vol.~43, no.~4, pp. 541--542, Aug. 2013.

\bibitem{Handigol2014}
N.~Handigol, B.~Heller, V.~Jeyakumar, D.~Mazi{\`e}res, and N.~McKeown, ``I know
  what your packet did last hop: Using packet histories to troubleshoot
  networks,'' in \emph{11th {USENIX} Symposium on Networked Systems Design and
  Implementation ({NSDI} 14)}.\hskip 1em plus 0.5em minus 0.4em\relax Seattle,
  WA: {USENIX} Association, 2014, pp. 71--85.

\bibitem{Zuo2014}
Q.~Zuo, M.~Chen, K.~Ding, and B.~Xu, ``On generality of the data plane and
  scalability of the control plane in software-defined networking,''
  \emph{China Communications}, vol.~11, no.~2, pp. 55--64, Feb 2014.

\bibitem{Bonola20172}
M.~Bonola, G.~Bianchi, G.~Picierro, S.~Pontarelli, and M.~Monaci, ``{StreaMon}:
  A data-plane programming abstraction for software-defined stream
  monitoring,'' \emph{IEEE Transactions on Dependable and Secure Computing},
  vol.~14, no.~6, pp. 664--678, Nov 2017.

\bibitem{Mogul2012}
J.~C. Mogul and P.~Congdon, ``Hey, you darned counters!: Get off my {ASIC}!''
  in \emph{Proceedings of the First Workshop on Hot Topics in Software Defined
  Networks}, ser. HotSDN '12.\hskip 1em plus 0.5em minus 0.4em\relax New York,
  NY, USA: ACM, 2012, pp. 25--30.

\bibitem{Yu2013}
M.~Yu, L.~Jose, and R.~Miao, ``Software defined traffic measurement with
  {OpenSketch},'' in \emph{Proceedings of the 10th USENIX Conference on
  Networked Systems Design and Implementation}, ser. nsdi'13.\hskip 1em plus
  0.5em minus 0.4em\relax Berkeley, CA, USA: USENIX Association, 2013, pp.
  29--42.

\bibitem{Jeyakumar2013}
V.~Jeyakumar, M.~Alizadeh, C.~Kim, and D.~Mazi\`{e}res, ``Tiny packet programs
  for low-latency network control and monitoring,'' in \emph{Proceedings of the
  Twelfth ACM Workshop on Hot Topics in Networks}, ser. HotNets-XII.\hskip 1em
  plus 0.5em minus 0.4em\relax New York, NY, USA: ACM, 2013, pp. 8:1--8:7.

\bibitem{Choi2010}
Y.~Choi, ``Implementation of content-oriented networking architecture ({CONA}):
  a focus on {DDoS} countermeasure,'' in \emph{Proc of 1st European NetFPGA
  Developers Workshop}, 2010.

\bibitem{Benzekki20162}
K.~Benzekki, A.~El~Fergougui, and A.~El~Belrhiti El~Alaoui, ``Devolving {IEEE
  802.1X} authentication capability to data plane in software-defined
  networking ({SDN}) architecture,'' \emph{Security and Communication
  Networks}, vol.~9, no.~17, pp. 4369--4377, 2016, sec.1613.

\bibitem{Fiessler2016}
A.~Fiessler, S.~Hager, B.~Scheuermann, and A.~W. Moore, ``{HyPaFilter} - a
  versatile hybrid {FPGA} packet filter,'' in \emph{2016 ACM/IEEE Symposium on
  Architectures for Networking and Communications Systems (ANCS)}, March 2016,
  pp. 25--36.

\bibitem{Han2016}
W.~Han, H.~Hu, Z.~Zhao, A.~Doup{\'e}, G.-J. Ahn, K.-C. Wang, and J.~Deng,
  ``State-aware network access management for software-defined networks,'' in
  \emph{Proceedings of the 21st ACM on Symposium on Access Control Models and
  Technologies}, ser. SACMAT '16.\hskip 1em plus 0.5em minus 0.4em\relax New
  York, NY, USA: ACM, 2016, pp. 1--11.

\bibitem{Kempf2012}
J.~Kempf, E.~Bellagamba, A.~Kern, D.~Jocha, A.~Takacs, and P.~Sköldström,
  ``Scalable fault management for {OpenFlow},'' in \emph{2012 IEEE
  International Conference on Communications (ICC)}, June 2012, pp. 6606--6610.

\bibitem{Capone2015}
A.~Capone, C.~Cascone, A.~Q.~T. Nguyen, and B.~Sansò, ``Detour planning for
  fast and reliable failure recovery in {SDN} with {OpenState},'' in \emph{2015
  11th International Conference on the Design of Reliable Communication
  Networks (DRCN)}, March 2015, pp. 25--32.

\bibitem{Cascone20172}
C.~Cascone, D.~Sanvito, L.~Pollini, A.~Capone, and B.~Sansò, ``Fast failure
  detection and recovery in {SDN} with stateful data plane,''
  \emph{International Journal of Network Management}, vol.~27, no.~2, 2017,
  e1957 nem.1957.

\bibitem{Petrucci2017}
L.~Petrucci, M.~Bonola, S.~Pontarelli, G.~Bianchi, and R.~Bifulco,
  ``Implementing iptables using a programmable stateful data plane abstraction:
  Demo,'' in \emph{Proceedings of the Symposium on SDN Research}, ser. SOSR
  '17.\hskip 1em plus 0.5em minus 0.4em\relax New York, NY, USA: ACM, 2017, pp.
  193--194.

\bibitem{Bianchi20141}
G.~Bianchi, M.~Bonola, A.~Capone, and C.~Cascone, ``{OpenState}: Programming
  platform-independent stateful {OpenFlow} applications inside the switch,''
  \emph{SIGCOMM Comput. Commun. Rev.}, vol.~44, no.~2, pp. 44--51, Apr. 2014.

\bibitem{Kempf2011}
J.~Kempf, S.~Whyte, J.~Ellithorpe, P.~Kazemian, M.~Haitjema, N.~Beheshti,
  S.~Stuart, and H.~Green, ``{OpenFlow} {MPLS} and the open source label
  switched router,'' in \emph{Proceedings of the 23rd International Teletraffic
  Congress}, ser. ITC '11.\hskip 1em plus 0.5em minus 0.4em\relax International
  Teletraffic Congress, 2011, pp. 8--14.

\bibitem{Shirazipour2012}
M.~Shirazipour, W.~John, J.~Kempf, H.~Green, and M.~Tatipamula, ``Realizing
  packet-optical integration with {SDN} and {OpenFlow} 1.1 extensions,'' in
  \emph{2012 IEEE International Conference on Communications (ICC)}, June 2012,
  pp. 6633--6637.

\bibitem{Sadasivarao2013}
A.~Sadasivarao, S.~Syed, P.~Pan, C.~Liou, A.~Lake, C.~Guok, and I.~Monga,
  ``{Open Transport Switch}: A software defined networking architecture for
  transport networks,'' in \emph{Proceedings of the Second ACM SIGCOMM Workshop
  on Hot Topics in Software Defined Networking}, ser. HotSDN '13.\hskip 1em
  plus 0.5em minus 0.4em\relax New York, NY, USA: ACM, 2013, pp. 115--120.

\bibitem{Amaya2013}
N.~Amaya, S.~Yan, M.~Channegowda, B.~R. Rofoee, Y.~Shu, M.~Rashidi, Y.~Ou,
  G.~Zervas, R.~Nejabati, D.~Simeonidou, B.~J. Puttnam, W.~Klaus, J.~Sakaguchi,
  T.~Miyazawa, Y.~Awaji, H.~Harai, and N.~Wada, ``First demonstration of
  software defined networking ({SDN}) over space division multiplexing ({SDM})
  optical networks,'' in \emph{39th European Conference and Exhibition on
  Optical Communication (ECOC 2013)}, Sept 2013, pp. 1--3.

\bibitem{Yan2015}
Y.~Yan, Y.~Shu, G.~M. Saridis, B.~R. Rofoee, G.~Zervas, and D.~Simeonidou,
  ``{FPGA}-based optical programmable switch and interface card for
  disaggregated {OPS/OCS} data centre networks,'' in \emph{2015 European
  Conference on Optical Communication (ECOC)}, Sept 2015, pp. 1--3.

\bibitem{Miao2015}
W.~Miao, F.~Agraz, S.~Peng, S.~Spadaro, G.~Bernini, J.~Perell\'{o}, G.~Zervas,
  R.~Nejabati, N.~Ciulli, D.~Simeonidou, H.~Dorren, and N.~Calabretta,
  ``{SDN}-enabled {OPS} with {QoS} guarantee for reconfigurable virtual data
  center networks,'' \emph{J. Opt. Commun. Netw.}, vol.~7, no.~7, pp. 634--643,
  Jul 2015.

\bibitem{Kondepu2015}
K.~Kondepu, A.~Sgambelluri, L.~Valcarenghi, F.~Cugini, and P.~Castoldi, ``An
  {SDN}-based integration of green {TWDM-PONs} and metro networks preserving
  end-to-end delay,'' in \emph{2015 Optical Fiber Communications Conference and
  Exhibition (OFC)}, March 2015, pp. 1--3.

\bibitem{Bakopoulos2018}
P.~Bakopoulos, K.~Christodoulopoulos, G.~Landi, M.~Aziz, E.~Zahavi, D.~Gallico,
  R.~Pitwon, K.~Tokas, I.~Patronas, M.~Capitani, C.~Spatharakis,
  K.~Yiannopoulos, K.~Wang, K.~Kontodimas, I.~Lazarou, P.~Wieder, D.~I. Reisis,
  E.~M. Varvarigos, M.~Biancani, and H.~Avramopoulos, ``{NEPHELE}: An
  end-to-end scalable and dynamically reconfigurable optical architecture for
  application-aware {SDN} cloud data centers,'' \emph{IEEE Communications
  Magazine}, vol.~56, no.~2, pp. 178--188, Feb 2018.

\bibitem{John2014}
W.~John, A.~Kern, M.~Kind, P.~Skoldstrom, D.~Staessens, and H.~Woesner,
  ``{SplitArchitecture}: {SDN} for the carrier domain,'' \emph{IEEE
  Communications Magazine}, vol.~52, no.~10, pp. 146--152, October 2014.

\bibitem{Saridis2015}
G.~M. Saridis, S.~Peng, Y.~Yan, A.~Aguado, B.~Guo, M.~Arslan, C.~Jackson,
  W.~Miao, N.~Calabretta, F.~Agraz, S.~Spadaro, G.~Bernini, N.~Ciulli,
  G.~Zervas, R.~Nejabati, and D.~Simeonidou, ``{LIGHTNESS}: A
  deeply-programmable {SDN}-enabled data centre network with {OCS/OPS}
  multicast/unicast switch-over,'' in \emph{2015 European Conference on Optical
  Communication (ECOC)}, Sept 2015, pp. 1--3.

\bibitem{Elbers2016}
J.~P. Elbers and A.~Autenrieth, ``From static to software-defined optical
  networks,'' in \emph{2012 16th International Conference on Optical Network
  Design and Modelling (ONDM)}, April 2012, pp. 1--4.

\bibitem{Zhang2015}
D.~Zhang, X.~Song, C.~Chen, H.~Guo, J.~Wu, and Y.~Xia, ``Software defined
  synergistic {IP}+optical resilient transport networks [invited],''
  \emph{IEEE/OSA Journal of Optical Communications and Networking}, vol.~7,
  no.~2, pp. A209--A217, February 2015.

\bibitem{Xiong2018}
Y.~Xiong, Z.~Li, B.~Zhou, and X.~Dong, ``Cross-layer shared protection strategy
  towards data plane in software defined optical networks,'' \emph{Optics
  Communications}, vol. 412, pp. 66--73, 2018.

\bibitem{Belter20141}
B.~Belter, D.~Parniewicz, L.~Ogrodowczyk, A.~Binczewski, M.~Stroiñski,
  V.~Fuentes, J.~Matias, M.~Huarte, and E.~Jacob, ``Hardware abstraction layer
  as an {SDN}-enabler for non-{OpenFlow} network equipment,'' in \emph{2014
  Third European Workshop on Software Defined Networks}, Sept 2014, pp.
  117--118.

\bibitem{Ogrodowczyk2014}
L.~Ogrodowczyk, B.~Belter, A.~Binczewski, K.~Dombek, A.~Juszczyk, I.~Olszewski,
  D.~Parniewicz, R.~D. Corin, M.~Gerola, E.~Salvadori \emph{et~al.}, ``Hardware
  abstraction layer for non-{OpenFlow} capable devices,'' in \emph{TERENA
  Networking Conference}, 2014, pp. 1--15.

\bibitem{Parniewicz2014}
D.~Parniewicz, R.~Doriguzzi~Corin, L.~Ogrodowczyk, M.~Rashidi~Fard, J.~Matias,
  M.~Gerola, V.~Fuentes, U.~Toseef, A.~Zaalouk, B.~Belter, E.~Jacob, and
  K.~Pentikousis, ``Design and implementation of an {OpenFlow} hardware
  abstraction layer,'' in \emph{Proceedings of the 2014 ACM SIGCOMM Workshop on
  Distributed Cloud Computing}, ser. DCC '14.\hskip 1em plus 0.5em minus
  0.4em\relax New York, NY, USA: ACM, 2014, pp. 71--76.

\bibitem{Clegg2014}
R.~G. Clegg, J.~Spencer, R.~Landa, M.~Thakur, J.~Mitchell, and M.~Rio,
  ``Pushing software defined networking to the access,'' in \emph{2014 Third
  European Workshop on Software Defined Networks}, Sept 2014, pp. 31--36.

\bibitem{Fuentes2014}
V.~Fuentes, J.~Matias, A.~Mendiola, M.~Huarte, J.~Unzilla, and E.~Jacob,
  ``Integrating complex legacy systems under {OpenFlow} control: The {DOCSIS}
  use case,'' in \emph{2014 Third European Workshop on Software Defined
  Networks}, Sept 2014, pp. 37--42.

\bibitem{Kondepu2018}
K.~Kondepu, C.~Jackson, Y.~Ou, A.~Beldachi, A.~Pag\`{e}s, F.~Agraz,
  F.~Moscatelli, W.~Miao, V.~Kamchevska, N.~Calabretta, G.~Landi, S.~Spadaro,
  S.~Yan, D.~Simeonidou, and R.~Nejabati, ``Fully {SDN}-enabled all-optical
  architecture for data center virtualization with time and space
  multiplexing,'' \emph{J. Opt. Commun. Netw.}, vol.~10, no.~7, pp. B90--B101,
  Jul 2018.

\bibitem{Ge2014}
X.~Ge, Y.~Liu, D.~H. Du, L.~Zhang, H.~Guan, J.~Chen, Y.~Zhao, and X.~Hu,
  ``{OpenANFV}: Accelerating network function virtualization with a
  consolidated framework in {OpenStack},'' in \emph{Proceedings of the 2014 ACM
  Conference on SIGCOMM}, ser. SIGCOMM '14.\hskip 1em plus 0.5em minus
  0.4em\relax New York, NY, USA: ACM, 2014, pp. 353--354.

\bibitem{Kachris2014}
C.~Kachris, G.~C. Sirakoulis, and D.~Soudris, ``Network function virtualization
  based on {FPGAs}: A framework for all-programmable network devices,''
  \emph{CoRR}, vol. abs/1406.0309, 2014.

\bibitem{Hancock2016}
D.~Hancock and J.~{Van Der Merwe}, \emph{{HyPer4}: Using {P4} to virtualize the
  programmable data plane}.\hskip 1em plus 0.5em minus 0.4em\relax Association
  for Computing Machinery, Inc, 12 2016, pp. 35--49.

\bibitem{Rong20162}
R.~Jin, X.~He, and M.~Gao, ``A method of {ForCES} virtualization,''
  \emph{International Journal of Future Generation Communication and
  Networking}, vol.~9, no.~1, pp. 271--284, 2016.

\bibitem{Firestone2017}
D.~Firestone, ``{VFP}: A virtual switch platform for host {SDN} in the public
  cloud,'' in \emph{14th USENIX Symposium on Networked Systems Design and
  Implementation (NSDI 17)}.\hskip 1em plus 0.5em minus 0.4em\relax Boston, MA:
  USENIX Association, 2017, pp. 315--328.

\bibitem{Oguchi2017}
N.~Oguchi and M.~Sekiya, ``Virtual data planes for easy creation and operation
  of end-to-end virtual networks,'' in \emph{2017 25th International Conference
  on Software, Telecommunications and Computer Networks (SoftCOM)}, Sept 2017,
  pp. 1--6.

\bibitem{Sonkoly2015}
B.~Sonkoly, R.~Szabo, D.~Jocha, J.~Czentye, M.~Kind, and F.~Westphal,
  ``{UNIFYing} cloud and carrier network resources: An architectural view,'' in
  \emph{2015 IEEE Global Communications Conference (GLOBECOM)}, Dec 2015, pp.
  1--7.

\bibitem{Szabo2017}
M.~Szabo, A.~Majdan, G.~Pongracz, L.~Toka, and B.~Sonkoly, ``Making the data
  plane ready for {NFV}: An effective way of handling resources,'' in
  \emph{Proceedings of the SIGCOMM Posters and Demos}, ser. SIGCOMM Posters and
  Demos '17.\hskip 1em plus 0.5em minus 0.4em\relax New York, NY, USA: ACM,
  2017, pp. 97--99.

\bibitem{Bosshart2014}
P.~Bosshart, D.~Daly, G.~Gibb, M.~Izzard, N.~McKeown, J.~Rexford,
  C.~Schlesinger, D.~Talayco, A.~Vahdat, G.~Varghese, and D.~Walker, ``P4:
  Programming protocol-independent packet processors,'' \emph{SIGCOMM Comput.
  Commun. Rev.}, vol.~44, no.~3, pp. 87--95, Jul. 2014.

\bibitem{Bosshart2013}
P.~Bosshart, G.~Gibb, H.-S. Kim, G.~Varghese, N.~McKeown, M.~Izzard, F.~Mujica,
  and M.~Horowitz, ``Forwarding metamorphosis: Fast programmable match-action
  processing in hardware for {SDN},'' in \emph{Proceedings of the ACM SIGCOMM
  2013 Conference on SIGCOMM}, ser. SIGCOMM '13.\hskip 1em plus 0.5em minus
  0.4em\relax New York, NY, USA: ACM, 2013, pp. 99--110.

\bibitem{Jose2015}
L.~Jose, L.~Yan, G.~Varghese, and N.~McKeown, ``Compiling packet programs to
  reconfigurable switches,'' in \emph{Proceedings of the 12th USENIX Conference
  on Networked Systems Design and Implementation}, ser. NSDI'15.\hskip 1em plus
  0.5em minus 0.4em\relax Berkeley, CA, USA: USENIX Association, 2015, pp.
  103--115.

\bibitem{Sivaraman20152}
A.~Sivaraman, C.~Kim, R.~Krishnamoorthy, A.~Dixit, and M.~Budiu, ``{DC.p4}:
  Programming the forwarding plane of a data-center switch,'' in
  \emph{Proceedings of the 1st ACM SIGCOMM Symposium on Software Defined
  Networking Research}, ser. SOSR '15.\hskip 1em plus 0.5em minus 0.4em\relax
  New York, NY, USA: ACM, 2015, pp. 2:1--2:8.

\bibitem{Shahbaz2016}
M.~Shahbaz, S.~Choi, B.~Pfaff, C.~Kim, N.~Feamster, N.~McKeown, and J.~Rexford,
  ``{PISCES}: A programmable, protocol-independent software switch,'' in
  \emph{Proceedings of the 2016 ACM SIGCOMM Conference}, ser. SIGCOMM
  '16.\hskip 1em plus 0.5em minus 0.4em\relax New York, NY, USA: ACM, 2016, pp.
  525--538.

\bibitem{Wang2016}
H.~Wang, K.~S. Lee, V.~Shrivastav, and H.~Weatherspoon, ``{P4FPGA}: High level
  synthesis for networking.''

\bibitem{Wang20172}
H.~Wang, R.~Soul{\'e}, H.~T. Dang, K.~S. Lee, V.~Shrivastav, N.~Foster, and
  H.~Weatherspoon, ``{P4FPGA}: A rapid prototyping framework for {P4},'' in
  \emph{Proceedings of the Symposium on SDN Research}, ser. SOSR '17.\hskip 1em
  plus 0.5em minus 0.4em\relax New York, NY, USA: ACM, 2017, pp. 122--135.

\bibitem{Wirbel2014}
L.~Wirbel, ``Xilinx {SDNet}: A new way to specify network hardware,'' The
  Linley Group, Tech. Rep., 2014.

\bibitem{Shahbaz2015}
M.~Shahbaz and N.~Feamster, ``The case for an intermediate representation for
  programmable data planes,'' in \emph{Proceedings of the 1st ACM SIGCOMM
  Symposium on Software Defined Networking Research}, ser. SOSR '15.\hskip 1em
  plus 0.5em minus 0.4em\relax New York, NY, USA: ACM, 2015, pp. 3:1--3:6.

\bibitem{Song2013}
H.~Song, ``{Protocol-Oblivious Forwarding}: Unleash the power of {SDN} through
  a future-proof forwarding plane,'' in \emph{Proceedings of the Second ACM
  SIGCOMM Workshop on Hot Topics in Software Defined Networking}, ser. HotSDN
  '13.\hskip 1em plus 0.5em minus 0.4em\relax New York, NY, USA: ACM, 2013, pp.
  127--132.

\bibitem{Sivaraman20162}
A.~Sivaraman, A.~Cheung, M.~Budiu, C.~Kim, M.~Alizadeh, H.~Balakrishnan,
  G.~Varghese, N.~McKeown, and S.~Licking, ``Packet transactions: High-level
  programming for line-rate switches,'' in \emph{Proceedings of the 2016 ACM
  SIGCOMM Conference}, ser. SIGCOMM '16.\hskip 1em plus 0.5em minus 0.4em\relax
  New York, NY, USA: ACM, 2016, pp. 15--28.

\bibitem{Zhu2014}
S.~Zhu, J.~Bi, and C.~Sun, ``{SFA}: Stateful forwarding abstraction in {SDN}
  data plane,'' in \emph{Presented as part of the Open Networking Summit 2014
  (ONS 2014)}.\hskip 1em plus 0.5em minus 0.4em\relax Santa Clara, CA: USENIX,
  2014.

\bibitem{Zhu2015}
S.~Zhu, J.~Bi, C.~Sun, C.~Wu, and H.~Hu, ``{SDPA}: Enhancing stateful
  forwarding for software-defined networking,'' in \emph{2015 IEEE 23rd
  International Conference on Network Protocols (ICNP)}, Nov 2015, pp.
  323--333.

\bibitem{Bianchi20142}
G.~{Bianchi}, M.~{Bonola}, A.~{Capone}, C.~{Cascone}, and S.~{Pontarelli},
  ``Towards wire-speed platform-agnostic control of {OpenFlow} switches,''
  \emph{ArXiv e-prints}, Aug. 2014.

\bibitem{Bianchi2016}
G.~Bianchi, M.~Bonola, S.~Pontarelli, D.~Sanvito, A.~Capone, and C.~Cascone,
  ``{Open Packet Processor}: a programmable architecture for wire speed
  platform-independent stateful in-network processing,'' \emph{CoRR}, vol.
  abs/1605.01977, 2016.

\bibitem{Bonola20171}
M.~Bonola, R.~Bifulco, L.~Petrucci, S.~Pontarelli, A.~Tulumello, and
  G.~Bianchi, ``Implementing advanced network functions for datacenters with
  stateful programmable data planes,'' in \emph{2017 IEEE International
  Symposium on Local and Metropolitan Area Networks (LANMAN)}, June 2017, pp.
  1--6.

\bibitem{Nakao2012}
A.~Nakao, ``{VNode}: A deeply programmable network testbed through network
  virtualization,'' \emph{3rd IEICE Technical Committee on Network
  Virtualization}, 2012.

\bibitem{Kanada2012}
Y.~Kanada, K.~Shiraishi, and A.~Nakao, ``High-performance network accommodation
  and intra-slice switching using a type of virtualization node.''\hskip 1em
  plus 0.5em minus 0.4em\relax INFOCOM, 2012.

\bibitem{Nakao2013}
A.~Nakao, ``Deeply programmable network; emerging technologies for network
  virtualization and software defined network ({SDN}),'' in \emph{2013
  Proceedings of ITU Kaleidoscope: Building Sustainable Communities}, April
  2013, pp. 1--1.

\bibitem{Yamada2015}
K.~Yamada, Y.~Kanada, K.~Amemiya, A.~Nakao, and Y.~Saida, ``{VNode}
  infrastucture enhancement — deeply programmable network virtualization,''
  in \emph{2015 21st Asia-Pacific Conference on Communications (APCC)}, Oct
  2015, pp. 244--249.

\bibitem{Nakao20153}
A.~NAKAO, ``Software-defined data plane enhancing {SDN} and {NFV},''
  \emph{IEICE Transactions on Communications}, vol. E98.B, no.~1, pp. 12--19,
  2015.

\bibitem{Yamada2016}
K.~YAMADA, A.~NAKAO, Y.~KANADA, Y.~SAIDA, K.~AMEMIYA, and Y.~MINAMI, ``Design
  and deployment of enhanced {VNode} infrastructure — deeply programmable
  network virtualization,'' \emph{IEICE Transactions on Communications},
  vol.~99, no.~8, pp. 1629--1637, 2016.

\bibitem{Minami2015}
Y.~Minami and K.~Yamada, ``Novel applications and experiments on programmable
  infrastructures,'' in \emph{2015 24th International Conference on Computer
  Communication and Networks (ICCCN)}, Aug 2015, pp. 1--6.

\bibitem{Nakao20152}
A.~NAKAO, P.~DU, and T.~IWAI, ``Application specific slicing for {MVNO} through
  software-defined data plane enhancing {SDN},'' \emph{IEICE Transactions on
  Communications}, vol. E98.B, no.~11, pp. 2111--2120, 2015.

\bibitem{Du2016}
P.~Du, P.~Putra, S.~Yamamoto, and A.~Nakao, ``A context-aware {IoT}
  architecture through software-defined data plane,'' in \emph{2016 IEEE Region
  10 Symposium (TENSYMP)}, May 2016, pp. 315--320.

\bibitem{Nakao2017}
A.~Nakao, P.~Du, Y.~Kiriha, F.~Granelli, A.~A. Gebremariam, T.~Taleb, and
  M.~Bagaa, ``End-to-end network slicing for {5G} mobile networks,''
  \emph{Journal of Information Processing}, vol.~25, pp. 153--163, 2017.

\bibitem{Ando2014}
S.~Ando and A.~Nakao, ``L7 packet switch: packet switch applying regular
  expression to packet payload,'' in \emph{2014 IEEE International Workshop
  Technical Committee on Communications Quality and Reliability (CQR)}, May
  2014, pp. 1--6.

\bibitem{Farhady2014}
H.~Farhady and A.~Nakao, ``{TagFlow}: Efficient flow classification in {SDN},''
  \emph{IEICE Transactions on Communications}, vol.~97, pp. 2302--2310, 2014.

\bibitem{Farhadi2014}
H.~Farhadi, P.~Du, and A.~Nakao, ``User-defined actions for {SDN},'' in
  \emph{Proceedings of The Ninth International Conference on Future Internet
  Technologies}, ser. CFI '14.\hskip 1em plus 0.5em minus 0.4em\relax New York,
  NY, USA: ACM, 2014, pp. 3:1--3:6.

\bibitem{Nirasawa20161}
S.~Nirasawa, M.~Hara, S.~Yamaguchi, M.~Oguchi, A.~Nakao, and S.~Yamamoto,
  ``Application performance improvement with application aware {DPN}
  switches,'' in \emph{2016 18th Asia-Pacific Network Operations and Management
  Symposium (APNOMS)}, Oct 2016, pp. 1--4.

\bibitem{Nirasawa20162}
S.~Nirasawa, M.~Hara, A.~Nakao, M.~Oguchi, S.~Yamamoto, and S.~Yamaguchi,
  ``Network application performance improvement with deeply programmable
  switch,'' in \emph{Adjunct Proceedings of the 13th International Conference
  on Mobile and Ubiquitous Systems: Computing Networking and Services}, ser.
  MOBIQUITOUS 2016.\hskip 1em plus 0.5em minus 0.4em\relax New York, NY, USA:
  ACM, 2016, pp. 263--267.

\bibitem{Nirasawa2017}
S.~Nirasawa, A.~Nakao, S.~Yamamoto, M.~Hara, M.~Oguchi, and S.~Yamaguchi,
  ``Application switch using {DPN} for improving {TCP} based data center
  applications,'' in \emph{2017 IFIP/IEEE Symposium on Integrated Network and
  Service Management (IM)}, May 2017, pp. 995--998.

\bibitem{Anwer2010}
M.~B. Anwer, M.~Motiwala, M.~b. Tariq, and N.~Feamster, ``{SwitchBlade}: A
  platform for rapid deployment of network protocols on programmable
  hardware,'' \emph{SIGCOMM Comput. Commun. Rev.}, vol.~40, no.~4, pp.
  183--194, Aug. 2010.

\bibitem{Narayanan2012}
R.~Narayanan, S.~Kotha, G.~Lin, A.~Khan, S.~Rizvi, W.~Javed, H.~Khan, and S.~A.
  Khayam, ``Macroflows and microflows: Enabling rapid network innovation
  through a split {SDN} data plane,'' in \emph{2012 European Workshop on
  Software Defined Networking}, Oct 2012, pp. 79--84.

\bibitem{Risso2012}
F.~Risso and I.~Cerrato, ``Customizing data-plane processing in edge routers,''
  in \emph{2012 European Workshop on Software Defined Networking}, Oct 2012,
  pp. 114--120.

\bibitem{Kim2012}
N.~Kim, J.~Y. Yoo, N.~L. Kim, and J.~Kim, ``A programmable networking switch
  node with in-network processing support,'' in \emph{2012 IEEE International
  Conference on Communications (ICC)}, June 2012, pp. 6611--6615.

\bibitem{Gill2012}
H.~Gill, D.~Lin, L.~Sarna, R.~Mead, K.~C. Lee, and B.~T. Loo, ``{SP4}: Scalable
  programmable packet processing platform,'' \emph{SIGCOMM Comput. Commun.
  Rev.}, vol.~42, no.~4, pp. 75--76, Aug. 2012.

\bibitem{Monti2016}
M.~Monti, M.~Sifalakis, C.~F. Tschudin, and M.~Luise, ``Towards programmable
  network dynamics: A chemistry-inspired abstraction for hardware design,''
  \emph{CoRR}, vol. abs/1601.05356, 2016.

\bibitem{Monti2017}
------, ``On hardware programmable network dynamics with a chemistry-inspired
  abstraction,'' \emph{IEEE/ACM Trans. Netw.}, vol.~25, no.~4, pp. 2054--2067,
  Aug. 2017.

\bibitem{Casado2012}
M.~Casado, T.~Koponen, S.~Shenker, and A.~Tootoonchian, ``Fabric: A
  retrospective on evolving {SDN},'' in \emph{Proceedings of the First Workshop
  on Hot Topics in Software Defined Networks}, ser. HotSDN '12.\hskip 1em plus
  0.5em minus 0.4em\relax New York, NY, USA: ACM, 2012, pp. 85--90.

\bibitem{Gember2012}
A.~Gember, P.~Prabhu, Z.~Ghadiyali, and A.~Akella, ``Toward software-defined
  middlebox networking,'' in \emph{Proceedings of the 11th ACM Workshop on Hot
  Topics in Networks}, ser. HotNets-XI.\hskip 1em plus 0.5em minus 0.4em\relax
  New York, NY, USA: ACM, 2012, pp. 7--12.

\bibitem{Fayazbakhsh2013}
S.~K. Fayazbakhsh, V.~Sekar, M.~Yu, and J.~C. Mogul, ``{FlowTags}: Enforcing
  network-wide policies in the presence of dynamic middlebox actions,'' in
  \emph{Proceedings of the Second ACM SIGCOMM Workshop on Hot Topics in
  Software Defined Networking}, ser. HotSDN '13.\hskip 1em plus 0.5em minus
  0.4em\relax New York, NY, USA: ACM, 2013, pp. 19--24.

\bibitem{Mekky2014}
H.~Mekky, F.~Hao, S.~Mukherjee, Z.-L. Zhang, and T.~Lakshman,
  ``Application-aware data plane processing in {SDN},'' in \emph{Proceedings of
  the Third Workshop on Hot Topics in Software Defined Networking}, ser. HotSDN
  '14.\hskip 1em plus 0.5em minus 0.4em\relax New York, NY, USA: ACM, 2014, pp.
  13--18.

\bibitem{Belter20142}
B.~Belter, A.~Binczewski, K.~Dombek, A.~Juszczyk, L.~Ogrodowczyk,
  D.~Parniewicz, M.~Stroiñski, and I.~Olszewski, ``Programmable abstraction of
  datapath,'' in \emph{2014 Third European Workshop on Software Defined
  Networks}, Sept 2014, pp. 7--12.

\bibitem{Haleplidis20151}
E.~Haleplidis, J.~Hadi~Salim, S.~Denazis, and O.~Koufopavlou, ``Towards a
  network abstraction model for {SDN},'' \emph{Journal of Network and Systems
  Management}, vol.~23, no.~2, pp. 309--327, Apr 2015.

\bibitem{Geissler2017}
S.~Geissler, S.~Herrnleben, R.~Bauer, S.~Gebert, T.~Zinner, and M.~Jarschel,
  ``{TableVisor} 2.0: Towards full-featured, scalable and hardware-independent
  multi table processing,'' in \emph{2017 IEEE Conference on Network
  Softwarization (NetSoft)}, July 2017, pp. 1--8.

\end{thebibliography}
\bibliographystyle{IEEEtran}

\begin{IEEEbiography}[{\includegraphics[width=1in,height=1.25in,clip,keepaspectratio]{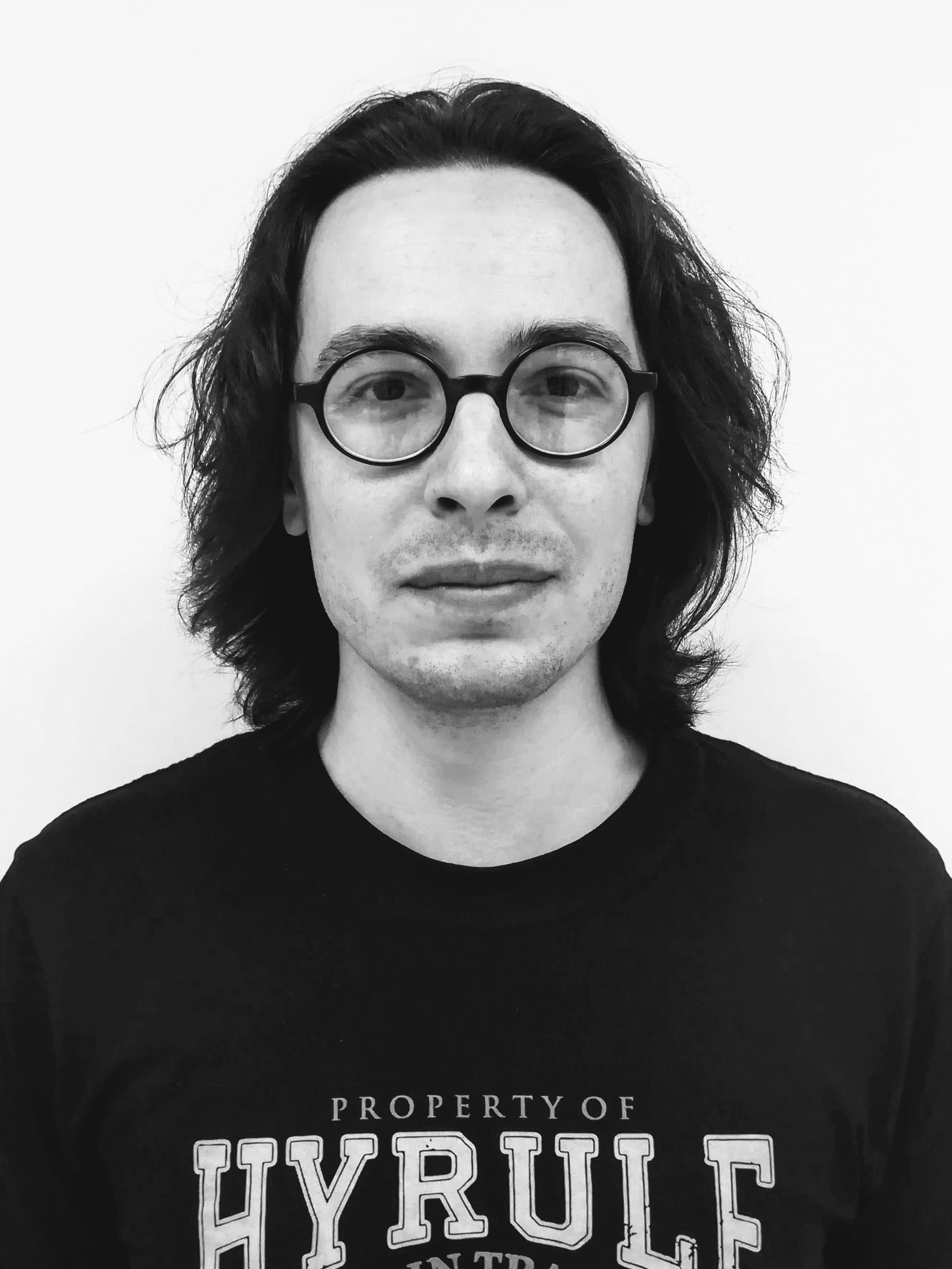}}]{\textbf{Enio Kaljic}}
	received Bachelor's and Master's degrees in telecommunication engineering from the Faculty of Electrical Engineering, University of Sarajevo, Bosnia and Herzegovina, in 2008 and 2010, respectively. From 2010 to 2014 he worked as a Teaching Assistant at the Department of Telecommunications at the Faculty of Electrical Engineering of the University of Sarajevo. Since 2014 he is a Senior Teaching Assistant at the same department, where he is~also currently pursuing his Ph.D. His research interests include software-defined networking and packet switching architectures for wireless and wired networking.
\end{IEEEbiography}

\begin{IEEEbiography}[{\includegraphics[width=1in,height=1.25in,clip,keepaspectratio]{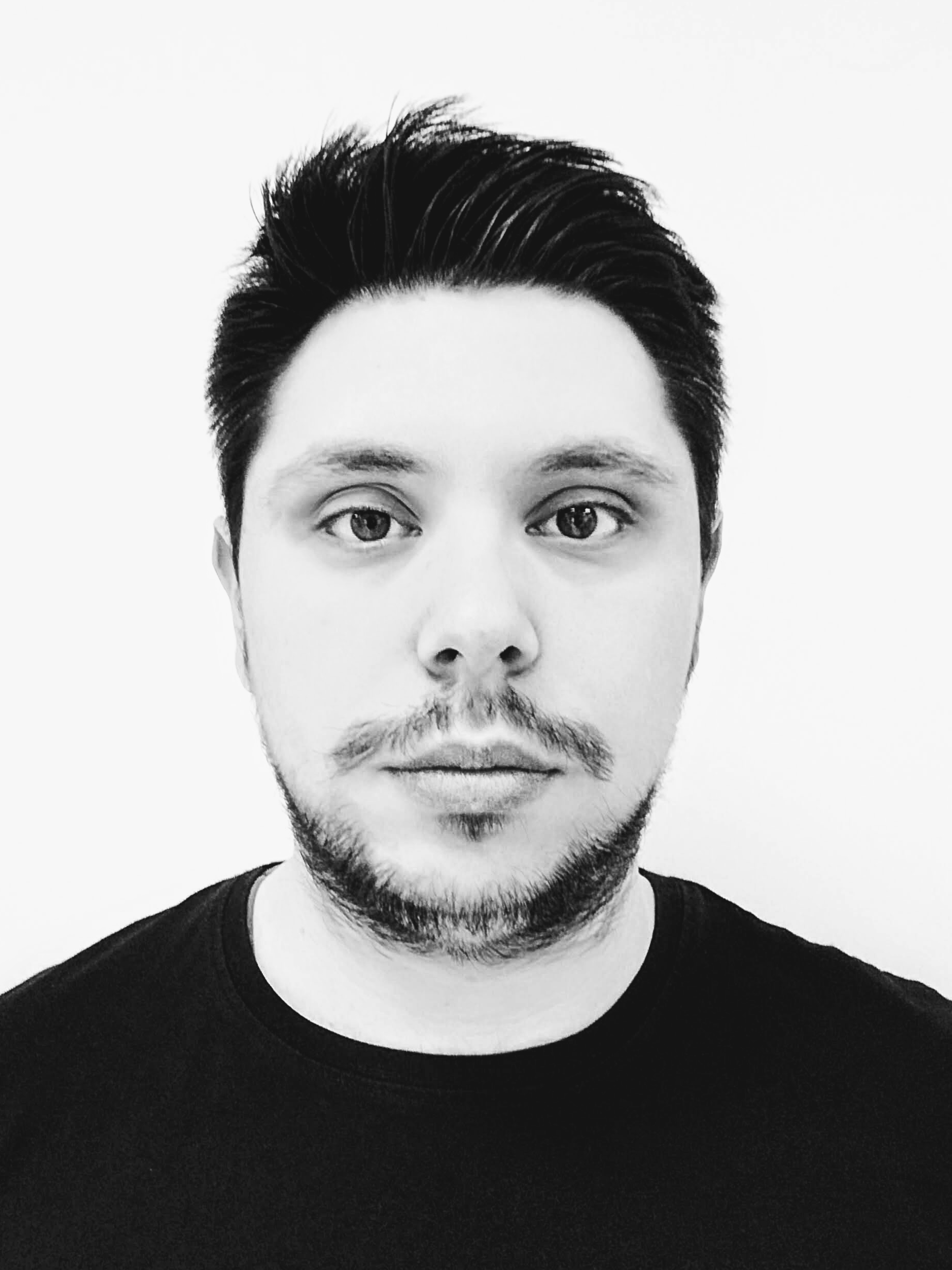}}]{\textbf{Almir Maric}}
	received Bachelor's and Master's degrees in electrical engineering from the University of Sarajevo, Bosnia and Herzegovina, in 2010 and 2013, respectively. He is currently working toward the Ph.D. degree with the Department of Telecommunications at the Faculty of Electrical Engineering of the University of Sarajevo. He is also working as a Teaching Assistant at the same department. His current research area includes physical channel modeling, communication channel characterization, fading channel and network simulators. 
\end{IEEEbiography}

\begin{IEEEbiography}[{\includegraphics[width=1in,height=1.25in,clip,keepaspectratio]{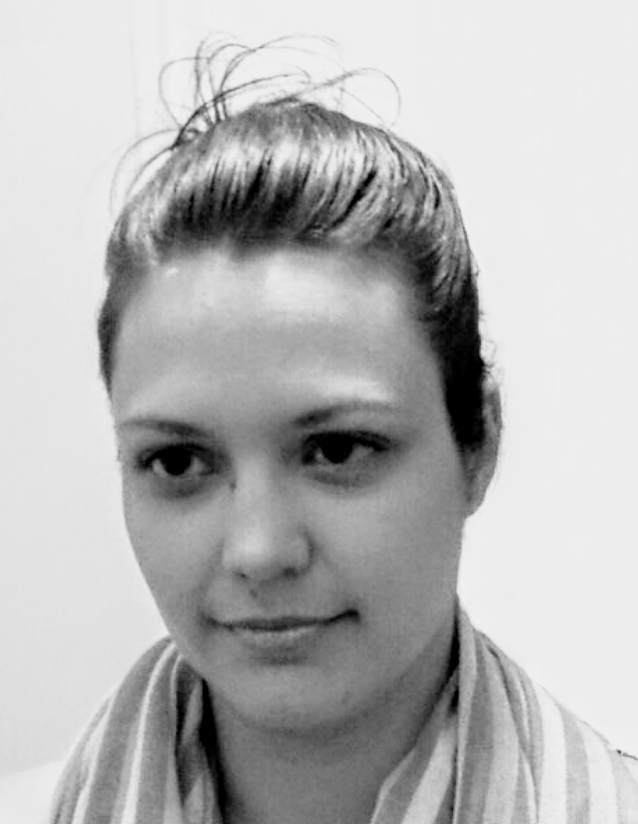}}]{\textbf{Pamela Njemcevic}}
	graduated at the Faculty of Electrical Engineering, University of Sarajevo in 2006, received M.Sc. degree in 2011 and Ph.D. in 2016 from the same University. Her professional carrier began in 2006. From that date, Pamela has been working as a teaching assistant and an assistant professor at the	Faculty of Electrical Engineering and the Faculty of Transport, Traffic and Communications, University of Sarajevo. She was involved in implementation of many research projects, related mostly to wireless
	communications and wireless propagation mechanisms. For two years, she was working for EUPM BH, as a Junior EU PPU CARDS expert for TETRA systems.
\end{IEEEbiography}

\begin{IEEEbiography}[{\includegraphics[width=1in,height=1.25in,clip,keepaspectratio]{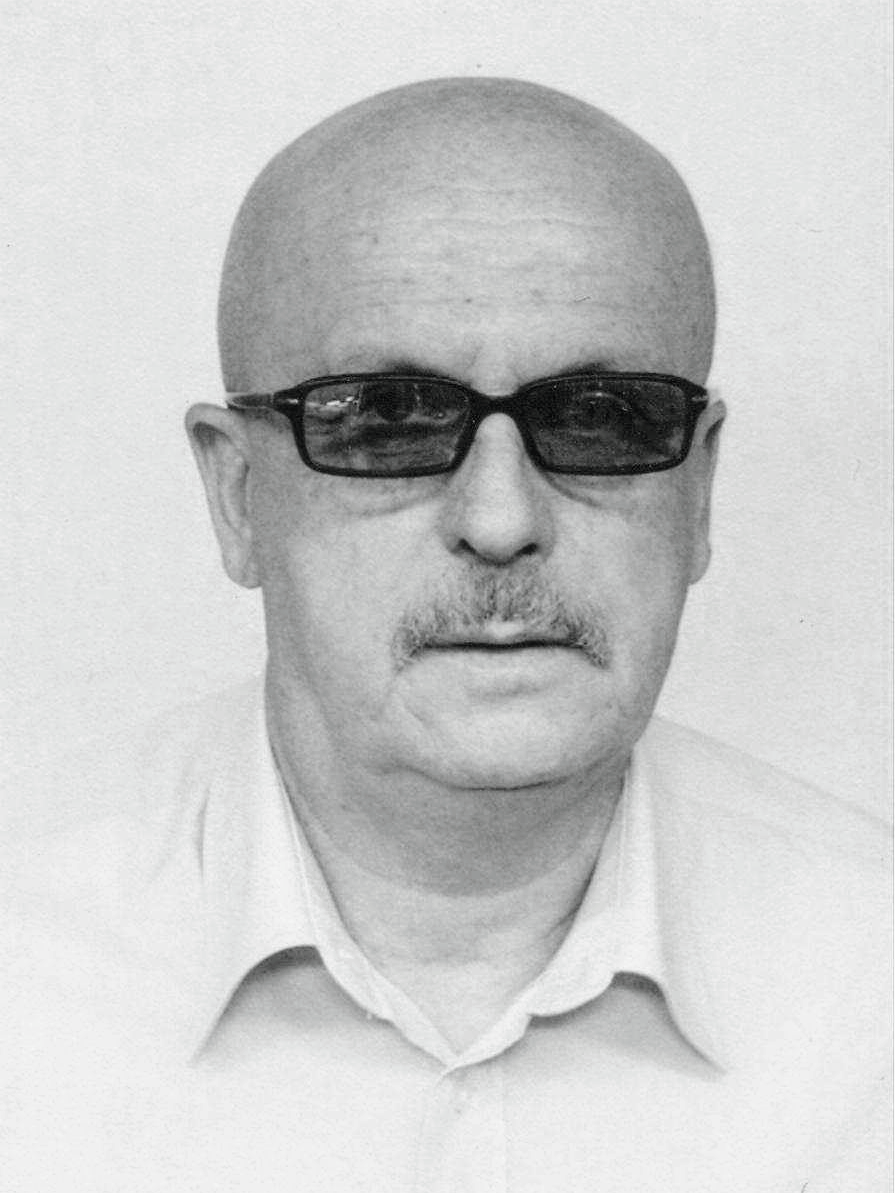}}]{\textbf{Mesud Hadzialic}}
	(deceased) received the Dipl.Ing., M.S.E.E., and Ph.D. degrees from the Faculty of Electrical Engineering, University of Sarajevo, Sarajevo, in 1978, 1986, and 2001, respectively. Since 2015, he has been a Full Professor with the Department of Telecommunications, Faculty of Electrical Engineering, University of Sarajevo. Since 2009, he has been the Vice Dean of science and research and the Head of the Department of Telecommunications. He has authored or co-authored five textbook of which three books university textbooks, several papers in journals, and over 50 papers at conferences. He led and participated in five scientific projects in the field of telecommunications supported by the Federal Ministry of Science of Federation of Bosnia and Herzegovina and over ten local and international projects in the domain of network simulations. He was the BiH project leader in projects co-funded by the European Union (South East Europe Transnational Cooperation Program). His research and teaching interests were in the general area of telecommunication techniques, theory and practice in the information networks, simulation methods, and techniques in telecommunication channels and networks.
\end{IEEEbiography}

\EOD

\end{document}